\documentclass[12pt]{iopart}

\usepackage{iopams}  

\usepackage{graphicx}
\usepackage{pslatex}
\usepackage{xspace}
\usepackage{subfigure}
\usepackage{eurosym}

\newcommand{\degree}{$^{\circ}$}
\newcommand{\pp}{\overline{pp}}
\newcommand{\Wp}{{W$_\textrm{p}$}\xspace}

\usepackage{hyperref}

\begin{document}

\review{Polymer--Fullerene Bulk Heterojunction Solar Cells}

\author{Carsten Deibel$^1$, Vladimir Dyakonov$^{1,2}$ }
\address{$^1$ Experimental Physics VI, Julius-Maximilians-University of W{\"u}rzburg, 97074 W{\"u}rzburg, Germany}
\address{$^2$ Energy Technology, Bavarian Centre for Applied Energy Research (ZAE Bayern), 97074 W{\"u}rzburg, Germany}
\ead{deibel@physik.uni-wuerzburg.de, dyakonov@physik.uni-wuerzburg.de}

\date{\today}

\begin{abstract}
Organic solar cells have the potential to be low-cost and efficient solar energy converters, with a promising energy balance. They are made from carbon-based semiconductors, which exhibit favourable light absorption and charge generation properties, and can be manufactured by low temperature processes such as printing from solvent-based inks, which are compatible with flexible plastic substrates or even paper. In this review, we will present an overview of the physical function of organic solar cells, their state-of-the-art performance and limitations, as well as novel concepts to achieve a better material stability and higher power conversion efficiencies. We will also briefly review processing and cost in view of the market potential.\\[2ex]
{\bf \href{http://dx.doi.org/10.1088/0034-4885/73/9/096401}{Rep.\ Prog.\ Phys. 73, 096401 (2010)}}

\end{abstract}


\maketitle

\tableofcontents

\clearpage

\section{Introduction}

The interest in organic solar cells has risen strongly in recent years, due to their interesting properties in terms of light incoupling and photocurrent generation, combined with the prospect of high throughput and low-cost processing~\cite{brabec2008book,deibel2010review3}. Many organic semiconductors exhibit very high absorption coefficients, making them promising  compounds for photovoltaic devices. However, compared to silicon solar cells which dominate the market these days---mostly in form of single crystals and polycrystalline material---the organic semiconductors which are of interest for photovoltaic applications are usually amorphous or polycrystalline, although single crystals do exist as well. The advantage of allowing a higher degree of  disorder is an easier and cheaper processing---in terms of cost as well as energy. Organic semiconductors are thermally evaporated at low temperatures as compared to inorganic crystals, or are processed by printing or coating at room temperature from solution. Inorganic semiconductors usually latter require high-temperature processing steps often exceeding 1000\degree{}C, although recent developments for solution-processing have been pursued, also showing promising efficiencies~\cite{gur2005,hou2009a,todorov2010}. Coming back to organic materials, although limited by the lower solar cell efficiencies, the cost structure as well as the energy balance can be very favourable. Unfortunately, other properties such as charge transport are more limited: crystals allow electrons to move almost freely, whereas in disordered matter the lack of long-range order leads to electrons hopping from one localised state to the next. Nevertheless, we will see that the resulting low charge carrier mobility is not a major limiting factor for organic solar cells; other properties are more critical but can be overcome in the majority of cases.

Generally, it is appropriate to distinguish between organic and inorganic materials, as well as ordered and disordered matter. In this review, the focus will be on rather disordered organic semiconductors, because these offer the most interesting properties in view of solar cell processing, while still being suitable in terms of charge generation and transport. The photoactive layer of efficient organic solar cells always consist of two compounds, a so-called electron donor and an electron accepting material, in order to be able to separate the photogenerated electron--hole pairs with a high yield. The research and development of organic solar cells focusses mostly on two concepts: either soluble blends of conjugated polymers with fullerene derivatives, or the combination of small molecular donor and acceptor materials, a material combination which can be thermally evaporated. Many aspects of the physical function are similar, and although the focus of this review will be on solution processable organic solar cells, we will usually include the small molecular materials as well.

This review is organised as follows: in the introductory section,  the milestones in the historical development of organic solar cells will be recounted. In this context, the basics of semiconductivity in organics as well as an overview of the physical principle behind the photovoltaic action will be given, with a focus on brevity rather than detail. In the second section, device physics, the working principle of organic solar cells will be discussed in detail. Also, the current limitations concerning the power conversion efficiency will be explained. Most of these limitations can be circumvented by applying novel concepts, as outlined in section~\ref{sec:concepts}: light trapping, tandem and hybrid solar cells as well as material engineering will be covered. In the fourth section, the processing methods and market potential of organic solar cells will be briefly considered. Finally, the conclusions will give an outlook on the future opportunities.

\subsection{Brief history}\label{sec:history}

The basis for organic solar cells was the finding of dark conductivity in halogen doped organic compounds in 1954, although many of them were not stable~\cite{akamatsu1954}. In the following years, a lot of systematic research was done on the charge transport properties of small molecules~\cite{pope1999book}. In the late seventies, the conductivity of the polymer polyacetylene---again by doping with halogens---was discovered~\cite{shirakawa1977}, for which its three main contributors Shirakawa, Heeger, and MacDiarmid were awarded the Nobel Price in Chemistry in 2000. 

The conductivity in carbon based semiconductors is due to conjugation, the alternation of single and double bonds between the carbon atoms~\cite{pope1999book}. The ground state of a carbon atom is in the $1s^{2}2s^{2}2p^{2}$ configuration. In organic semiconductors, the $s$ and $p$ orbitals form 3 $sp^{2}$ orbitals, the $\sigma$ bonds. The fourth orbital, $p_{z}$, is perpendicular to the plane spanned by the $sp^{2}$ orbitals. The overlap of the $p_{z}$ electron wavefunctions leads to delocalisation of the charges, which is the origin of the conductivity in these organic compounds. Due to the Peierls instability two delocalised energy bands are formed, the bonding and antibonding $\pi$ and $\pi^*$ orbitals, also-called highest occupied molecular orbital (HOMO) and lowest unoccupied molecular orbital (LUMO), respectively. HOMO and LUMO are separated by a bandgap in the order of typically one to three electron volts, making such an organic compound a semiconductor. The transition between these two levels can be excited by light in the visible spectrum. The semiconducting and light absorbing properties make conjugated organics a very interesting choice for photovoltaics.

The first organic solar cells were based on an active layer made of a single material, sandwiched between two electrodes of different work functions. By the absorption of light, strongly Coulomb-bound electron--hole pairs are created, so-called singlet excitons. As their binding energy in organic semiconductors, inhibiting much lower effective dielectric constants as their inorganic counterparts, is usually between 0.5 and 1~eV, the excitons have to be separated to finally generate a photocurrent. In order to overcome the exciton binding energy, one has  either to rely on the thermal energy, or dissociate the exciton at the contacts~\cite{hertel2008review}. Unfortunately, both processes have a rather low efficiency: under the operating conditions of solar cells, the temperature is not high enough, and the sample thickness is much thicker than the exciton diffusion length. The consequence is that not all excitons are dissociated, but can as well recombine radiatively by photoluminescence. Consequently, the single layer organic solar cells had power conversion efficiencies of far below 1\% for the solar spectrum~\cite{chamberlain1983review}.

\begin{figure}[bt]
	\centering\includegraphics[width=12cm]{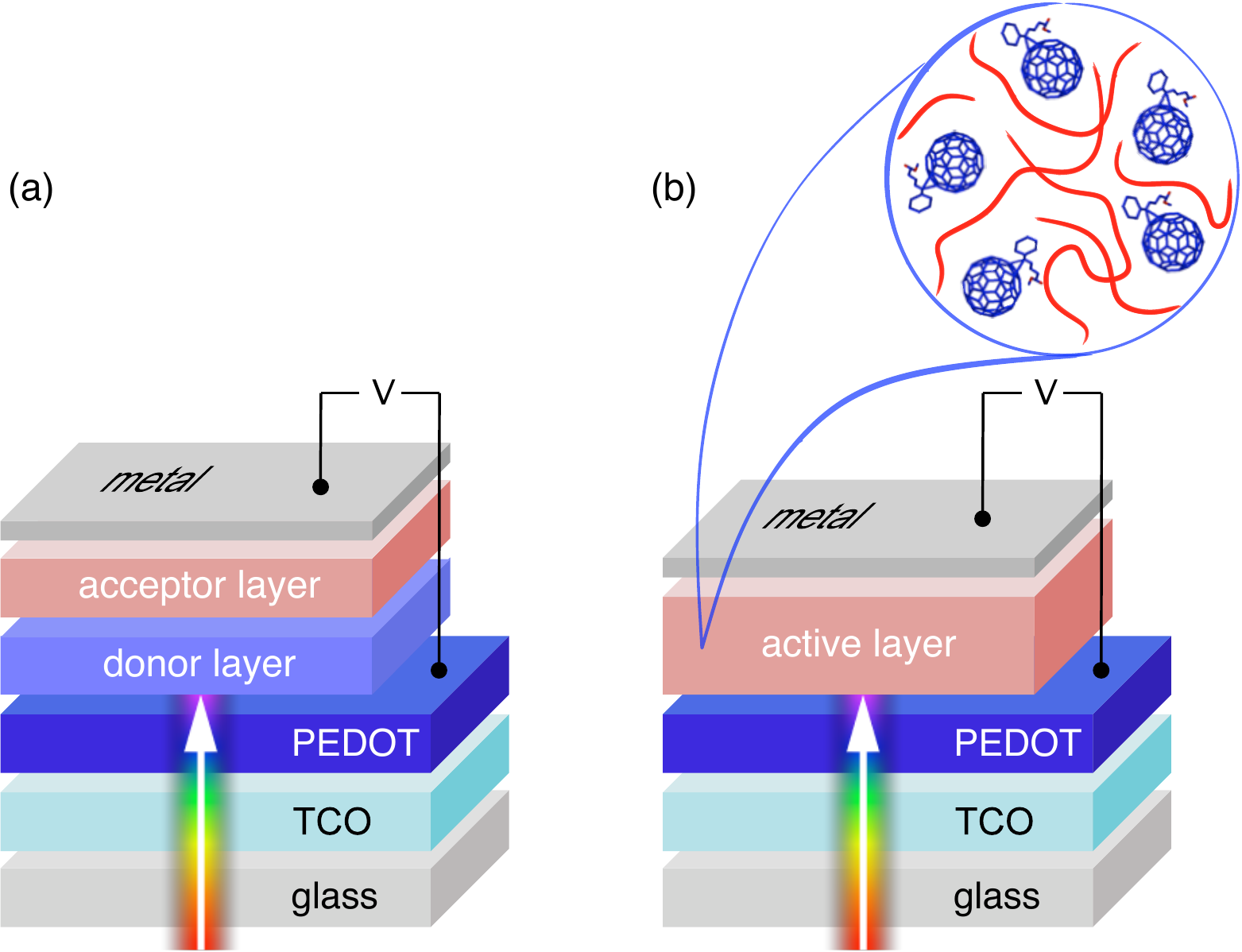}
	\caption{Typical device configurations of organic solar cells: (a) bilayer device with planar heterojunction, (b) bulk heterojunction device consisting of a blend of conjugated polymer with a fullerene derivative. On top of the glass substrate, a transparent conductive oxide (TCO) such as indium tin oxide acts as anode, a poly(3,4-ethylendioxythiophen):polystyrolsulfonate (PEDOT) interlayer helps to avoid local shunts. The active layer consists of either the bilayer or the blend of organic semiconductors. On top, a metallic electrode acts as cathode.%
	\label{fig:osc-bilayer+bhj}}
\end{figure}

The introduction of a second organic semiconductor layer was a quantum leap in terms of power conversion efficiency---though still on a low level. The first organic bilayer solar cells was presented by C.~W.~Tang in the mid eighties~\cite{tang1986}, although it is said that such devices were already developed in the late seventies. The typical device configuration is shown in Figure~\ref{fig:osc-bilayer+bhj}(a). The light is usually absorbed in the so-called donor material, a hole conducting small molecule such as copper phthalocyanine. In bilayer devices, the photogenerated singlet excitons could diffuse within the donor towards the planar interface to the second material, the acceptor, which is usually chosen to be strongly electronegative. The acceptor material provides the energy needed for the singlet exciton to be separated, as the electron can go to a state of much lower energy within the acceptor. This charge transfer dissociates the exciton, the electron moving to the acceptor material, whereas the hole remains on the donor. A prominent example for an electron acceptor material is the buckminsterfullerene (C$_{60}$)~\cite{smalley1999review}.

Roughly speaking, the difference between the electron energy on the donor and the corresponding acceptor level has to be larger than the exciton binding energy, in order to initiate a charge transfer from donor to acceptor material. If the exciton reaches this donor--acceptor heterointerface by diffusion---as it is neutral---it is energetically favourable for the electron to be transferred to the acceptor molecule. This charge transfer, or electron transfer, is reported to be very fast. Indeed, was found to be faster than 100~fs in polymer--fullerene systems, and very efficient, as the alternative loss mechanisms are much slower~\cite{sariciftci1992}. Thus, the exciton is dissociated and the resulting charge carriers are spatially separated. 

Even though the electron and the hole now reside on two separate materials, they are still Coulomb bound due to the weak screening of the electric field in organic semiconductors. Therefore, a further step is needed for the final charge pair dissociation, for instance initiated by an electric field or the energetic disorder of the organic semiconductors. Thus, the mutual Coulomb attraction is overcome. This dependence becomes manifest in the strongly field and temperature dependent photocurrent of organic solar cells, which also influences fill factor and short circuit current: only if this charge carrier separation is successfull, can electron and hole hop towards their respective contacts, in order to generate a photocurrent. 

The organic bilayer solar cells invented by C.~W.~Tang were made of two conjugated small molecules, and achieved a power conversion efficiency of about 1~\%~\cite{tang1986}. The limiting factor in this concept is that for a full absorption of the incident light, a layer thickness of the absorbing material has to be of the order of the absorption length, approx.\ 100~nm. This is much more than the diffusion length of the excitons, about 10nm in disordered and semicrystalline polymers and small molecules. In this example, maybe 100~\% of the incoming photons (within the absorption band) can be absorbed, but only 10~\% of these could reach the donor-acceptor interface and be dissociated to charge carrier pairs. As mostly the exciton diffusion length is much lower than the absorption length, the potential of the bilayer solar cell is difficult to exploit. 

In the beginning nineties, a novel concept was introduced, accounting for the low exciton diffusion length in disordered organic semiconductors, as well as the required thickness for a sufficient light absorption: the so-called bulk heterojunction solar cell~\cite{yu1995}. This approach---shown in Figure~\ref{fig:osc-bilayer+bhj}(b)---features a distributed junction between donor and acceptor material: both components interpenetrate one another, so that the interface between them is not planar any more, but spatially distributed. This concept is implemented by spincoating a polymer--fullerene blend, or by coevaporation of conjugated small molecules. Bulk heterojunctions have the advantage of being able to dissociate excitons very efficiently over the whole extent of the solar cell, and thus generating electron--hole pairs throughout in the film. The disadvantages are that it is somewhat more difficult to separate these still strongly Coulomb bound charge carrier pairs due to the increased disorder, and that percolation to the contacts is not always given in the disordered material mixtures. Also, it is more likely that trapped charge carriers recombine with mobile ones. However, the positive effects on the device performance outweigh the drawbacks. 

For an efficient bulk heterojunction solar cell, a good control of the morphology is very important. Rather simple methods of optimisation have been successfully performed in the last decade. The choice of solvents~\cite{shaheen2001} as well as the annealing of the solution processed polymer:fullerene solar cells~\cite{padinger2003}  both lead to a more favourable inner structure in view of the dissociation of bound electron--hole pairs and the subsequent charge transport. Thus, the power conversion efficiency was increased manyfold, in case of the annealing from a bare half percent to above 3~\%. Indeed, optimisation by novel routes is an ongoing process, and within the last five years, further steps in improving the power conversion efficiency have been made. Coevaporated copper phthalocyanine/fullerene solar cells have reached 5.0~\% efficiency using a concept called planar-mixed heterojunction~\cite{xue2005}, and solution processed polythiophene:fullerene cells achieved between 6~and~8\% efficiency by the use of novel materials as well as additives optimising the phase separation~\cite{peet2007,park2009,green2010review}.

\subsection{Current--voltage characteristics and solar cell parameters}

\begin{figure}
	\subfigure{\includegraphics[height=7cm]{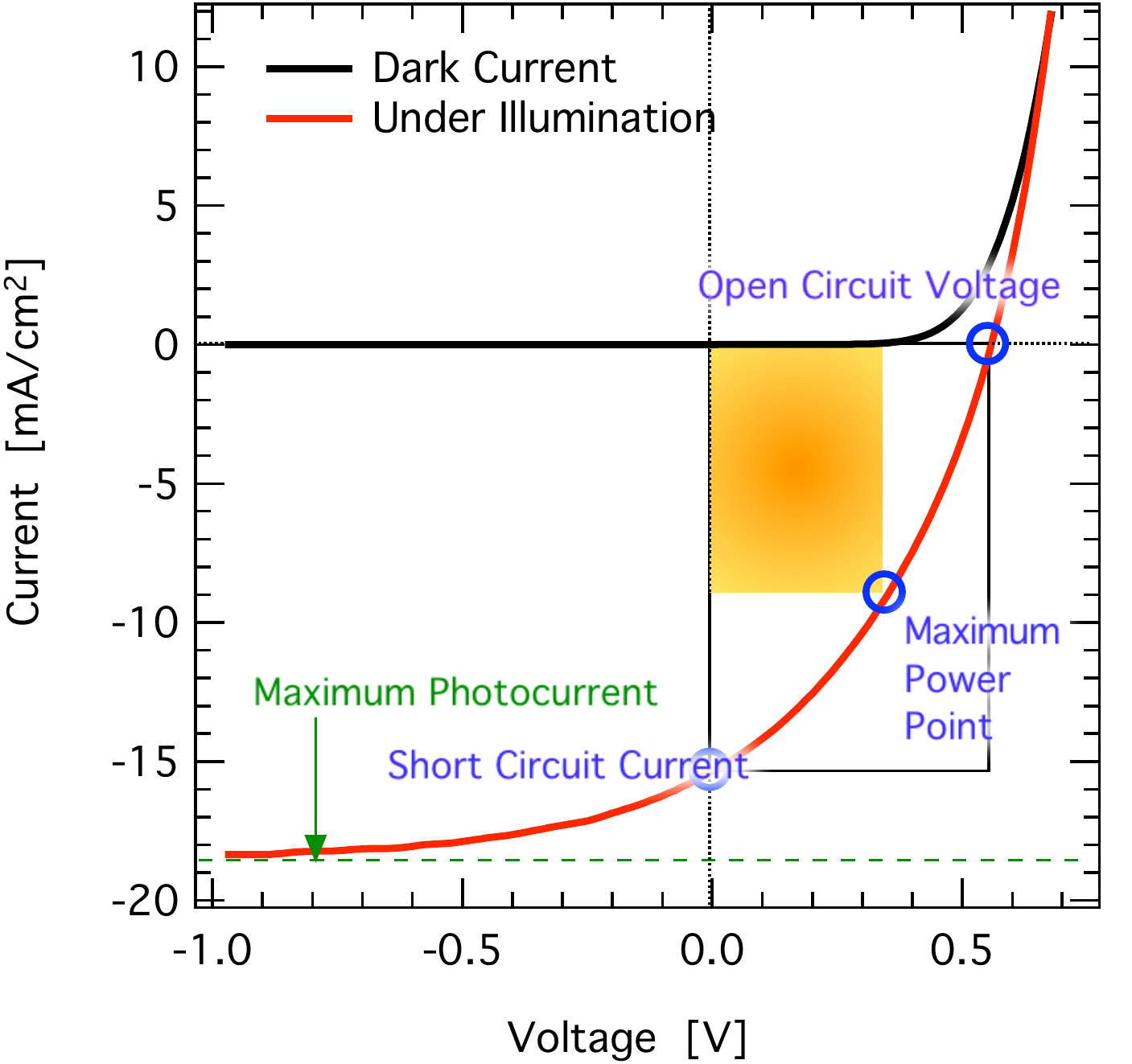}}\quad
	\subfigure{\includegraphics[height=7cm]{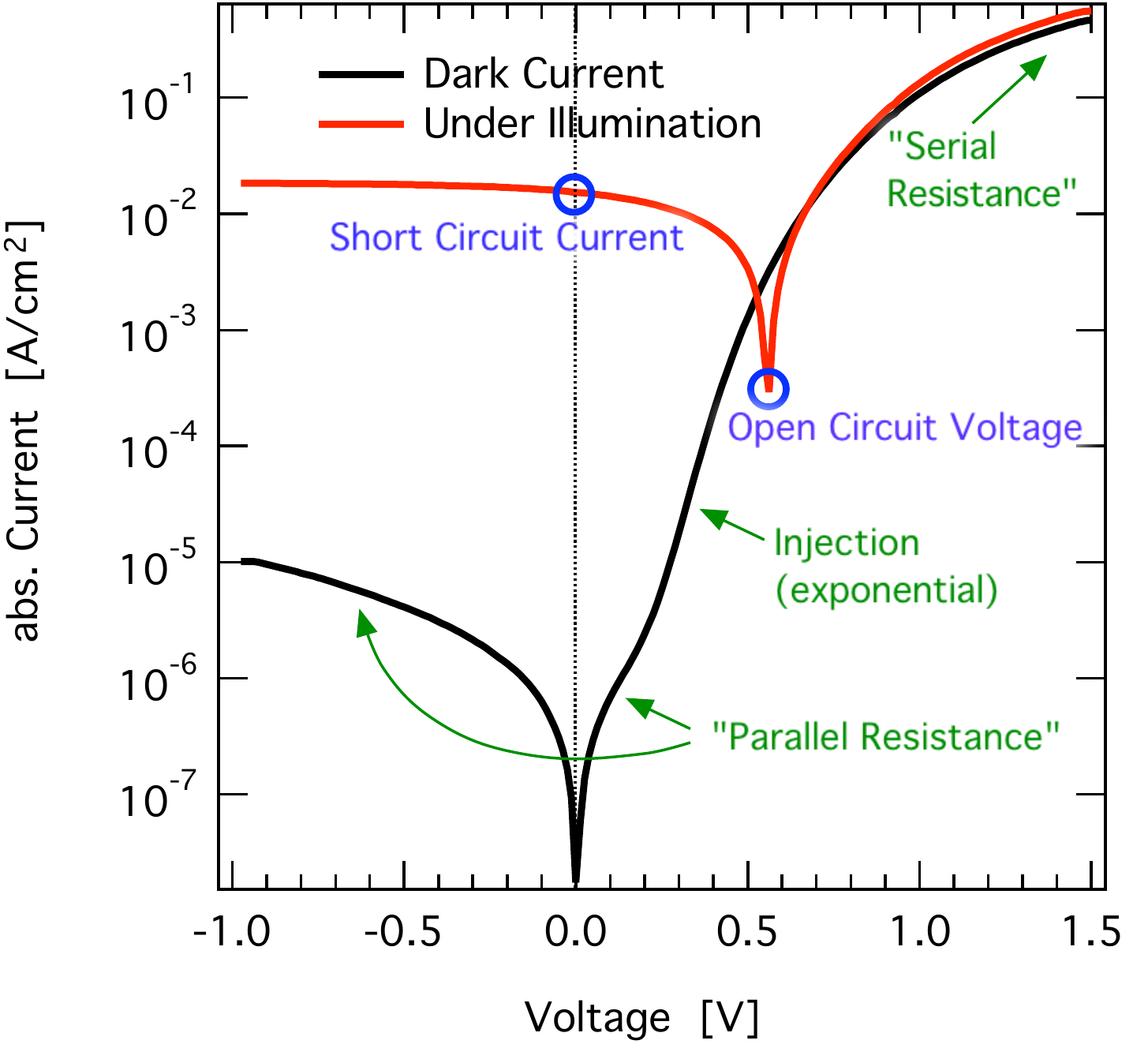}}%
	\caption{Schematic current--voltage characteristics of bulk heterojunction solar cells in (a) linear and (b) semilogarithmic representation. As the photocurrent in organic photovoltaic devices is usually not constant, but voltage dependent, does the photocurrent reach saturation not necessarily at short circuit conditions, but negative voltage bias. This reduces also the fill factor, and finally the solar cell efficiency.  When applying the enhanced Shockley equation, Eqn.~(\ref{eqn:shockley}), to describe the current--voltage characteristics of organic solar cells, parallel and series resistance values have to be assumed to be voltage dependent. This allows a fit of the data, although the physical processes responsible for the apparent voltage dependence, voltage-dependent photocurrent and space charge limited currents, are not considered.  %
	\label{fig:IV}}
\end{figure}

As indicated in the previous section, the most important figure of merit describing the performance of a solar cell ist the power conversion efficiency $\eta$. It is based on the parameters open circuit voltage $V_{oc}$, short circuit current density $j_{sc}$, and fill factor $FF$. They are highlighted in the current--voltage characteristics of a bulk heterojunction solar cell shown in Figure~\ref{fig:IV}. The accurate, reproducible and standard-conform measurement of the current--voltage characteristics of organic solar cells is essential, and has been described in literature~\cite{shrotriya2006}. 

The fill factor is given by the quotient of maximum power (yellow rectangle in Figure~\ref{fig:IV}) and the product of open circuit voltage and short circuit current (white rectangle); it therefore decribes the "squareness" of the solar cell's current--voltage characteristics.  The efficiency is the ratio of maximum power to incident radiant power $P_L$---typically radiated by the sun,
\begin{equation}
	\eta = \frac{FF \, j_{sc} \, V_{oc}}{P_L} .
\end{equation}
The maximum attainable efficiency is given by the detailed balance limit for inorganic p--n junction solar cells, published by Shockley and Queisser in 1961~\cite{shockley1961}, which will be discussed in Section~\ref{sec:concepts}. The famous Shockley diode equation is used in that publication as well, and although derived for inorganic devices, it has been applied by many researchers in the field of organic semiconductors to describe or fit the current--voltage characteristics~\cite{schilinsky2004,riedel2005}. It is based on an exponential term, as a positive voltage bias leads to the injection of charge carriers into the solar cell. The current increases exponentially, leading to a rectifying behaviour in the ideal case. In order to account for real solar cells, however, the ideal Shockley equation was extended by two resistors. The so-called series resistance $R_s$---in series with the ideal diode---was included to describe contact resistances such as injection barriers and sheet resistances. In contrast, the parallel resistance $R_p$ covers the influence of local shunts between the two electrodes, i.e., additional current paths circumventing the diode. Thus, for inorganic devices, the current $j$ after the Shockley equation is given as a function of voltage $V$,
\begin{equation}
	j(V) = j_0 \left( \exp\left(\frac{q(V-j R_s)}{nkT} \right) -1 \right) - \frac{V-jR_s}{R_p} - j_{ph} ,
	\label{eqn:shockley}
\end{equation}
where $j_0$ is the saturation current density,  $q$ is the elementary charge, $kT$ the thermal energy, and $n$ the diode ideality factor. The optional photocurrent $j_{ph}$ is included by a parallel shift of the current-voltage curve down the current axis. 

The current--voltage characteristics of a bulk heterojunction organic solar cell shown in Figure~\ref{fig:IV}, however, can hardly be described by Eqn.~(\ref{eqn:shockley})---the strongly field dependent photocurrent in the third quadrant is already striking (Figure~\ref{fig:IV}(a)). In some devices, it happens that the maximum photocurrent is not reached under short circuit conditions, but only at more negative bias, corresponding to a higher internal field.  Also, the crossing point of dark and illuminated curve at approx.\ 700~mV (Figure~\ref{fig:IV}(b)) cannot be explained by the diode equation. Trying to press the experimental data into the shape of the Shockley diode equation consequently leads to define a voltage and light intensity dependent parallel resistance $R_p$ and series resistance $R_s$ without physically justified foundation. This does not invalidate the Shockley diode equation in itself, it just shows that its extension by an equivalent circuit without accounting for a field-dependent photocurrent and other mechanisms is not sufficient to describe organic photovoltaic devices.

Thus, organic solar cells show some differences as compared to their inorganic counterparts. Highlighting just some of these aspects, most of which will be discussed in the following sections: the field dependent photocurrent is due to the separation of photogenerated, Coulomb bound electron--hole pairs (Section~\ref{sec:pp-diss}), and the extraction of the resulting free charges (Section~\ref{sec:extraction}). It leads to the apparent field dependent parallel resistance $R_p$, together with shunt resistances which certainly also can occur in organic devices. The series resistance can partly be due to organic semiconductors being less conductive as inorganics, so that at higher voltages space charge builds up, leading to space charge limited currents. As the charge transport is due to hopping between localised states, the charge carrier mobility is electric field and carrier concentration dependent (Section~\ref{sec:transport}). The relation between mobility and diffusion, the Einstein relation, is not valid in an as far range as in inorganics. Even the space charge limited currents do not strictly follow the well-known Mott--Gurney law with the current being proportional to V$^2$. Nevertheless, the expected series resistance seems to become voltage dependent.

In order to understand what sets organic solar cells apart, we will come to their detailed function, including advances as well as limitations of the state-of-the-art systems (Section~\ref{sec:device}), and to promising concepts for circumventing these limitations and improving the power conversion efficiency (Section~\ref{sec:concepts}).

\section{Device physics}\label{sec:device}

\begin{figure}[tb]
	\centering
	\subfigure{\includegraphics[height=6cm]{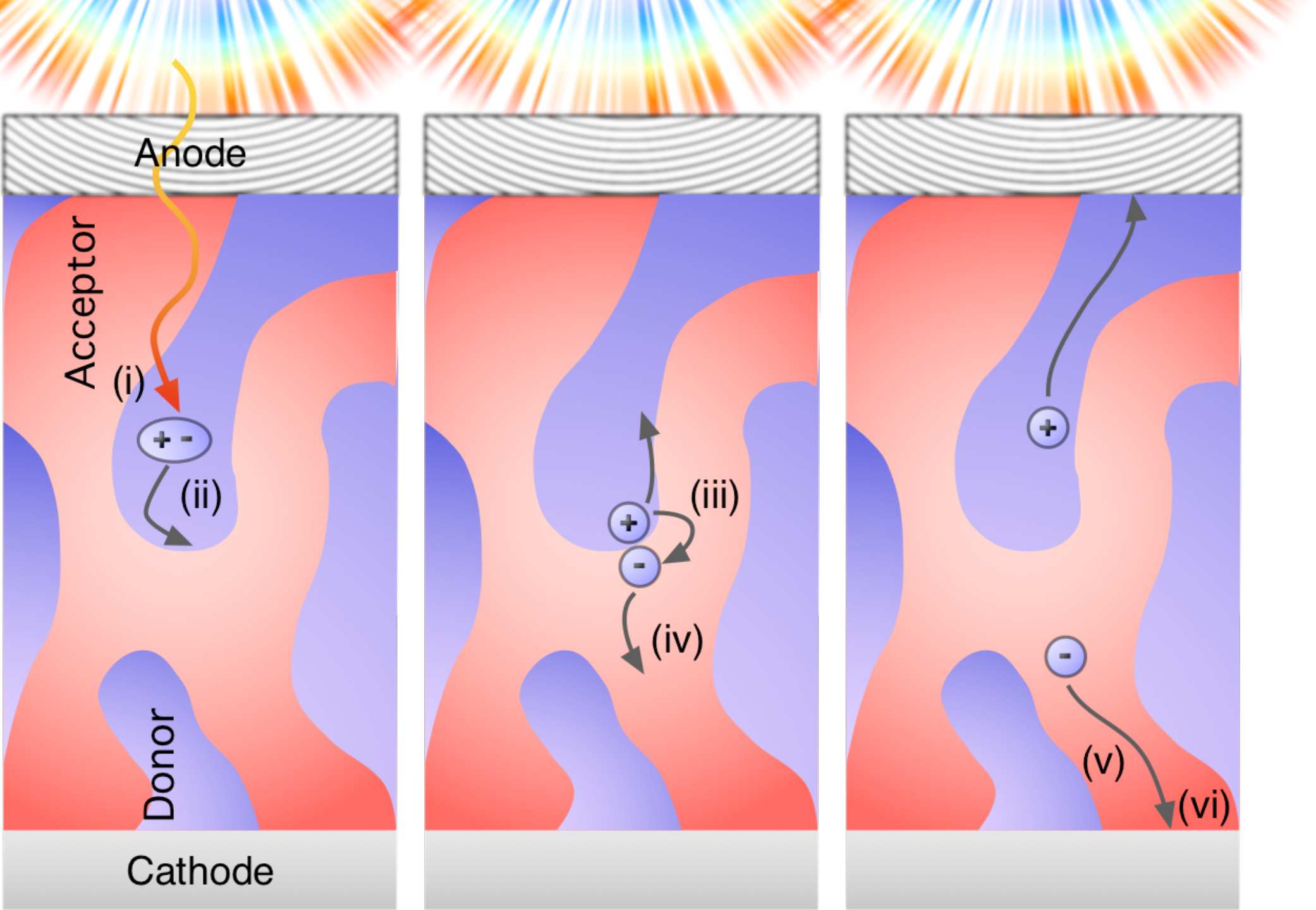}}\quad
	\subfigure{\includegraphics[height=6cm]{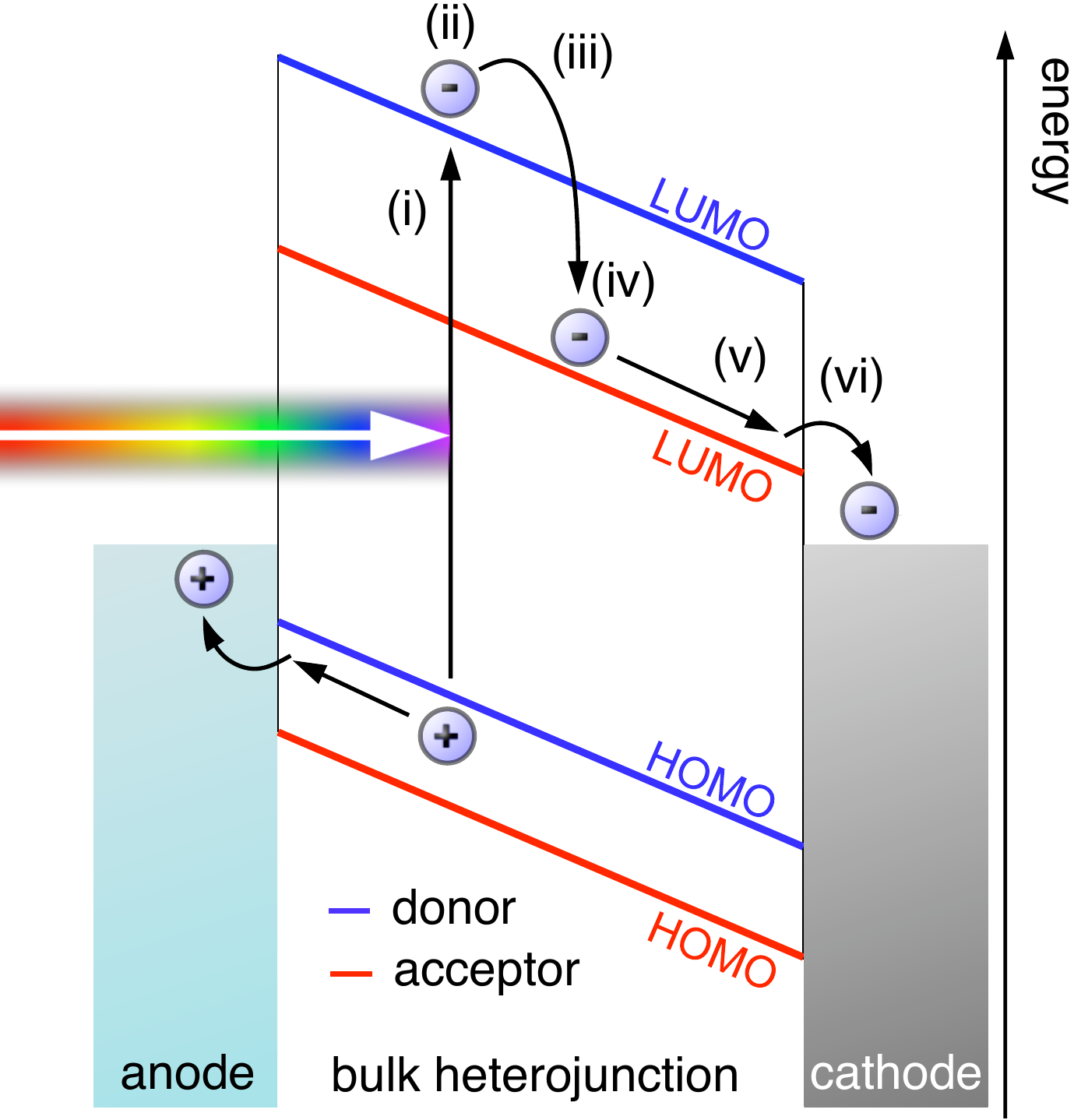}}%
	\caption{From light absorption to photocurrent in a bulk heterojunction solar cell. Left: from a kinetic point of view, right: simplified energy diagram (binding energies for excitons and polaron pairs are not shown). (i) singlet exciton generation from an absorbed photon in the donor material. (ii) exciton diffusion to the acceptor interface. (iii) exciton dissociation by electron transfer to the electronegative acceptor molecules. (iv) separation of the still Coulomb bound electron--hole pair due to electric field and material disorder. (v) charge transport of electron resp.\ hole by hopping between localised states. At this state, nongeminate recombination between independently generated electrons and holes can occur. (vi) extraction of the charges: photocurrent.%
	\label{fig:osc-bhj-morph}}
\end{figure}

In this section, the function of an organic solar cell will be described in more detail as compared to the introduction, including loss mechanisms and other limitations. The state-of-the-art literature will be discussed, and controversial debates will not be concealed. The section is ordered according to the different steps from photon absorption to photocurrent generation, as shown in Figure~\ref{fig:osc-bhj-morph} for a bulk heterojunction solar cell from a kinetic and an energetic perspective. 

\subsection{Exciton generation and dissociation}\label{sec:exciton}

In an organic donor--acceptor solar cell, light is usually absorbed mostly in the donor material, e.g., a conjugated polymer. This process is shown in Figure~\ref{fig:osc-bhj-morph} (i). Organic semiconductors---in analogy to many organic dyes---often exhibit very high absorption coefficients above $10^7$~m$^{-1}$. Consequently, very low thicknesses between 100--300~nm are sufficient for a good absorption yield in organic photovoltaic devices. In contrast, solar cells based on the inorganic polycrystalline semiconductor CuInSe$_2$ need few micron thick active layer for good absorption, and crystalline silicon solar cells more than 100~microns. Thus, lower material amounts are needed for organic cells. Unfortunately, many organic materials have a rather narrow absorption width. Conjugated polymers  commonly used in organic solar cells typically cover the visible optical spectrum only~\cite{martin1999,hoppe2003,lioudakis2007}, although polymers with wider absorption bands exist~\cite{gavrilenko2008,schueppel2008}. In contrast, the inorganic semiconductors silicon and CuInSe$_2$ absorb across the whole visible spectrum of the sun light, and beyond to more than 1000~nm (optical bandgap 1.1~eV). Novel concepts described in Section~\ref{sec:concepts} can help to avoid this limitation.

From the absorbed light, electron--hole pairs are generated in the organic semiconductors. For a long time, a controversial debate about the primary photoexcitation in organic semiconductors was held~\cite{sariciftci1997book,bredas1996,hertel2008review}. The question was whether free electron--hole pairs~\cite{moses2000} or strongly bound excitons~\cite{kersting1994}---excited quasiparticles in a solid, which can be seen as a Coulomb-bound electron--hole pairs---are generated due to light absorption. This discussion was connected with the anticipated magnitude of the exciton binding energy, which has important consequences for organic photovoltaics, as discussed further below. For weak binding energies, the thermal energy at room temperature would be sufficient for dissociation. However, lately, it is largely agreed that singlet excitons indeed are the primary excitation following the absorption of photons, exhibiting a significant binding energy much larger than the thermal energy~\cite{hertel2008review}. As the dielectric constant is lower in organics, typically between 3 and 4, the screening length is larger. Considering the Coulomb attraction of an electron--hole pair as a rough estimate for the exciton binding energy, a separation by 1~nm in a material with a dielectric constant of 3, the Coulomb binding energy is 0.5~eV. Thus, the exciton binding energy exceeds the thermal energy at room temperature by far, and the name Frenkel exciton is commonly used; the weakly bound type is called Wannier--Mott~\cite{scholes2006review}. Some people might point out that an exciton is defined only in highly ordered materials with a periodic lattice, however, in literature the term is commonly used also regarding polycrystalline and disordered semiconductors. In organic materials, excitons usually resides on one molecule or along an extended polymer chain segment, intrachain excitons~\cite{beenken2009a}, although in some cases, interchain excitons residing on adjacent molecules are reported~\cite{osterbacka2000}. The  combined spin-state of the two charges forming an exciton have a high importance. Singlet excitons have a resulting spin of zero, and triplet excitons with a spin of 1---which is possible in three different combinations, thus the name. Singlet excitons can be generated upon illumination, which is due to specific selection rules. In contrast, both, singlet as well as triplet excitons can be formed due to interaction following charge injection. Theoretically, this follows a one-to-three ratio, i.e., only a quarter is of singlet type. Details about the spin state can be found elsewhere~\cite{dyakonov1997,baldo1998,dyakonov1998,bredas2004review}.

Excitons inhibit a certain lifetime, after which they recombine radiatively. For singlet excitons in organic semiconductors, the lifetime is around one nanosecond, with photoluminescence as subsequent decay path. Radiative recombination after the triplet excitons' longer lifetime of up to milliseconds---the transition actually being spin forbidden---occurs by phosphorescence. The emission is related to the triplet energy, which is typically 0.7~eV lower as compared to the singlet~\cite{kohler2004}. As a side note, phosphorescence can be applied in so-called triplet emitters, being an important concept for organic light emitting diodes~\cite{yersin2004}. For photovoltaics, triplets are not yet advantageous---though might be if exploited in novel concepts---and act as loss mechanisms instead.

As discussed, the exciton binding energy of singlets is much larger than the thermal energy in organic semiconductors. Thus, in view of photovoltaic current generation, another driving force is needed to dissociate them. As already outlined in the brief history, its lack is the reason why single layer organic solar cells do not work efficiently. A second organic semiconductor, the electronegative acceptor, has to be introduced either in a bilayer or bulk heterojunction device configuration (Figure~\ref{fig:osc-bilayer+bhj}) in order to yield an efficient exciton dissociation. The class of materials with the currently best acceptor properties are the (buckminster)fullerenes, C60 and its derivatives, e.g., [6,6]-phenyl-C61 butyric acid methyl ester. The reason for their success is probably due to a combination of the spherical shape and the favourable electronegativity~\cite{imahori2007}, which leads to the fullerene having no quadrupolar moment~\cite{verlaak2009}. Polymer acceptors~\cite{halls1999,mcneill2008,huang2008a} and non-fullerene small molecule acceptors~\cite{zhou2009} have shown a less efficient photogeneration of charges up to now, which seems to be due to the LUMO not being electronegative enough for efficient electron accepting properties and too low electron mobilities~\cite{mcneill2009review}.

In Figure~\ref{fig:osc-bhj-morph} (ii) and (iii), the processes of exciton diffusion to the exciton dissociating donor-acceptor interface and the subsequent charge transfer of the electron from donor to acceptor material are shown. Indeed, before singlet exciton dissociation at the donor--acceptor interface can take place, the neutral exciton has to diffuse towards this interface~\cite{barbour2008,collini2009}. In bilayer solar cells, the small exciton diffusion length limits the thickness of the absorbing donor layer~\cite{peumans2003review}. Excitons generated further away from the planar interface to the acceptor than the exciton diffusion length are lost by recombination---usually radiative in form of photoluminescence. Consequently, the dimensions of the donor layer should ideally be chosen accordingly. The bulk heterojunction concept, featuring a distributed donor--acceptor interface over the extent of the whole active layer, puts this guideline into practice. For small molecule solar cells, it is implemented by coevaporation~\cite{peumans2003review}, for solution processed blends it is implicit to the approach~\cite{yu1995}. In principle, just accomodating the requirements of exciton dissociation, the phase separation could be as small as possible. However, as in a solar cell the subsequent steps of electron--hole pair dissociation and charge transport to the electrodes both need phase separations as coarse as possible, the optimum dimensions are a trade-off between two oppositional requirements. Thus, for a given material, the best achievable phase separation for an organic solar cell will usually be to exploit the full exciton diffusion length, thus losing no excitons to radiative recombination, in the hope that the resulting charge separation and transport properties will be reasonably good.

Organic single crystals can have exciton diffusion lengths in excess of 100~nm~\cite{kurrle2008}, which have the potential as donor of efficient bilayer solar cells, but have not been exploited yet. Instead, rather disordered materials are typically used, such as the donor--acceptor combinations poly(3-hexylthiophene) (P3HT)--[6,6]-phenyl-C$_{61}$ butyric acid methyl ester (PCBM)  and copper phthalocyanine--C$_{60}$, for solution processed and thermally evaporated solar cells, respectively. Experimentally determined diffusion lengths in these materials are in the range of 3-30~nm for small molecule donor materials, and 40~nm for C$_{60}$~\cite{peumans2003review}. The solution processed conjugated polymer P3HT has a exciton diffusion length of only 4~nm~\cite{luer2004}, whereas polyphenylenevinylene is reported by another group to have 12~nm~\cite{stubinger2001}. In each of these materials, the value of the exciton diffusion length is much smaller than the absorption length, showing that the bulk heterojunction concept is most suitable for organic solar cells. Another study pointed out that irrespective of exciton diffusion, the charge transfer itself can have a range of 8~nm in polymer/fullerene bilayers and blends, which they interpreted as delocalisation of the excitation~\cite{vacar1997}.

The phase separation is thus limited to small dimensions, and it can be anticipated that the charge transport might be made difficult by these restraints. In accordance with the importance of the donor-acceptor morphology, several groups have studied the phase separation of polymer--fullerene systems, such as polyphenylenevinylene (PPV)--PCBM~\cite{hoppe2004} and P3HT--PCBM~\cite{chirvase2004,vanlaeke2006a,campoy-quiles2008}.

\begin{figure}[bt]
	\centering\includegraphics[width=15cm]{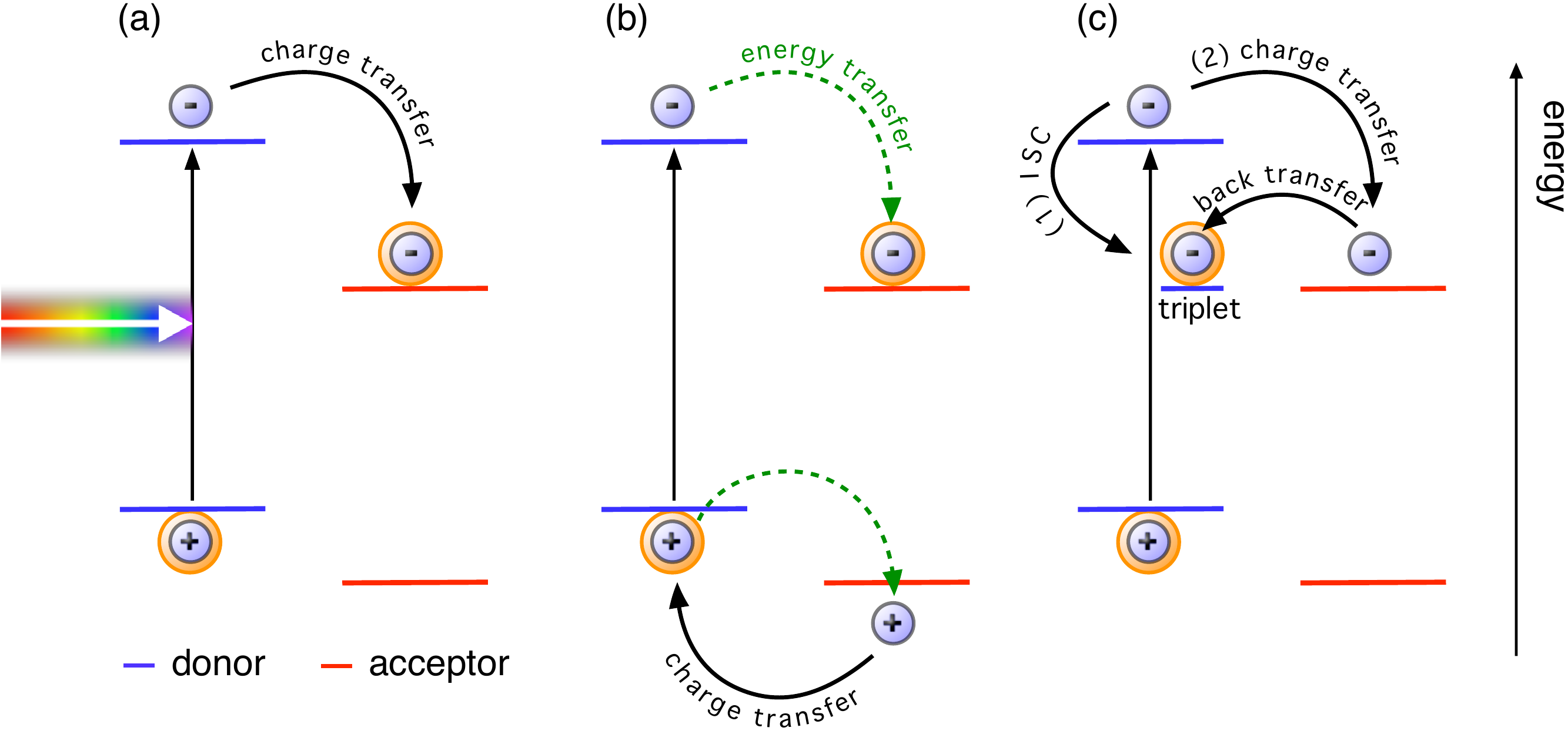}
	\caption{Typical charge transfer reactions in organic solar cells. (a) The singlet exciton on the donor material is dissociated by an electron transfer to the acceptor material. The final state is indicated by the orange frame of the charge carriers. (b) F{\"o}rster energy transfer of the exciton from donor to acceptor, followed by a hole transfer from acceptor to donor. The final state is similar to case (a). (c) Triplet excitons can act as loss mechanisms in two ways: (1) if intersystem crossing (ISC) is faster than the charge transfer, the singlet exciton is converted to a triplet, which decays subsequently. (2) if the electron--hole pair resulting from the electron transfer cannot be separated, an electron back transfer to a triplet exciton level in the donor can take place. In (1) and (2), no charge carriers are generated.%
	\label{fig:ct}}
\end{figure}

An exciton moving closer to the heterojunction of its host molecular material to the acceptor, it is influenced by the interface, as it experiences a different environment as compared to the bulk. This leads to a slightly red shifted photoluminescence. In order to consider the altered properties, the excited state is called exciplex (from excited complex) in order to distinguish it from the bulk exciton~\cite{morteani2005a,yin2007,deibel2010review2}.

Once the excited state---diffusing within the donor phase---has reached the interface to the acceptor material within its diffusion length, it can transfer its electron to the electronegative acceptor (Figure~\ref{fig:osc-bhj-morph}(iii) and Figure~\ref{fig:ct}(a)). This charge transfer is reported to be extremely fast, on the order of tens of femtoseconds~\cite{sariciftci1992,hwang2008}. 
Some groups claim that due to the prevalence of electron transfer in polymeric systems~\cite{farinha2008review}, the charge generation process consists of exciton generation mostly in the polymer, energy (not electron) transfer to the fullerene and subsequent hole transfer to the polymer~\cite{lloyd2008}. This process is indicated in Figure~\ref{fig:ct}(b). The intermediate step is ascribed to a F{\"o}rster resonant energy transfer, in which the whole excitation may move from polymer to fullerene. This discussion is not completely solved yet, although mostly a direct electron transfer from donor to acceptor is anticipated.

In suitable material combinations, the charge transfer is much faster than any competing loss processes such as photoluminescence or intersystem crossing (shown in Figure~\ref{fig:ct}(c)), the latter being associated with the transition from singlet to triplet exciton.  Excitons dissociate only at energetically favourable acceptor molecules such as the fullerenes, when the energy gain is larger than the exciton binding energy. This energy gain is actually not as straight forward as implied by some publications, where it is described as the energy offset between the LUMO of the donor and the LUMO of the acceptor. Such definition can act as a rough guide, but as LUMOs are energy levels of uncharged molecules, they do describe neither a charged molecule nor optical transitions~\cite{deibel2010}. Briefly, the exciton can dissociate when its energy is larger than the energy of the electron--hole pair after the electron transfer, often called polaron pair or charge transfer complex~\cite{ohkita2008,veldman2009,deibel2010review2}. Only then an electron or charge transfer takes place, dissociating the singlet exciton into an electron on the fullerene acceptor and a hole on the polymer.

The terms polaron and polaron pair are very common in the context of organic photovoltaics. A polaron is a charge, i.e., an electron or a hole, in combination with a distortion of the charge's environment. In a crystalline inorganic material, setting a charge onto a site does not change the surroundings significantly, as the crystal lattice is rigid. In contrast, in many organic semiconductors, putting a charge onto a certain molecular site can deform the whole molecule. The implication is that charge transport becomes more difficult, the charge carrier mobility becomes lower---which will be discussed later. A polaron pair is a Coulomb bound pair of a negative and a positive polaron, either situated in a single material---being less strongly bound than an exciton---or on different molecules, typically the positive polaron on the donor material of an organic solar cell, and the negative polaron on the acceptor. As outlined above, such polaron pairs are the intermediate step from an exciton to a pair of free polarons~\cite{frankevich1992}, and therefore important in order to understand photogeneration in organic semiconductors~\cite{clarke2010review, deibel2010review2}.

As pointed out in the beginning of this section, most of the sun light is absorbed in the donor material, such as a conjugated polymer, and is separated due to electron transfer to an electron accepting fullerene. However, light can also be absorbed by fullerenes, and although the exciton generation is much lower as compared to polymers, a subsequent charge transfer to the polymer from the fullerene has been observed experimentally~\cite{cook2009}. The terms donor for the polymer and acceptor for the fullerene seem to be incorrect with respect to this hole transfer, but the full notation is electron donor resp.\ electron acceptor which was defined for the dominant case of electron transfer from polymer to fullerene.

\subsection{Charge carrier pair dissociation} \label{sec:pp-diss}

\begin{figure}[bt]
	\centering
	\subfigure{\includegraphics[height=6.8cm]{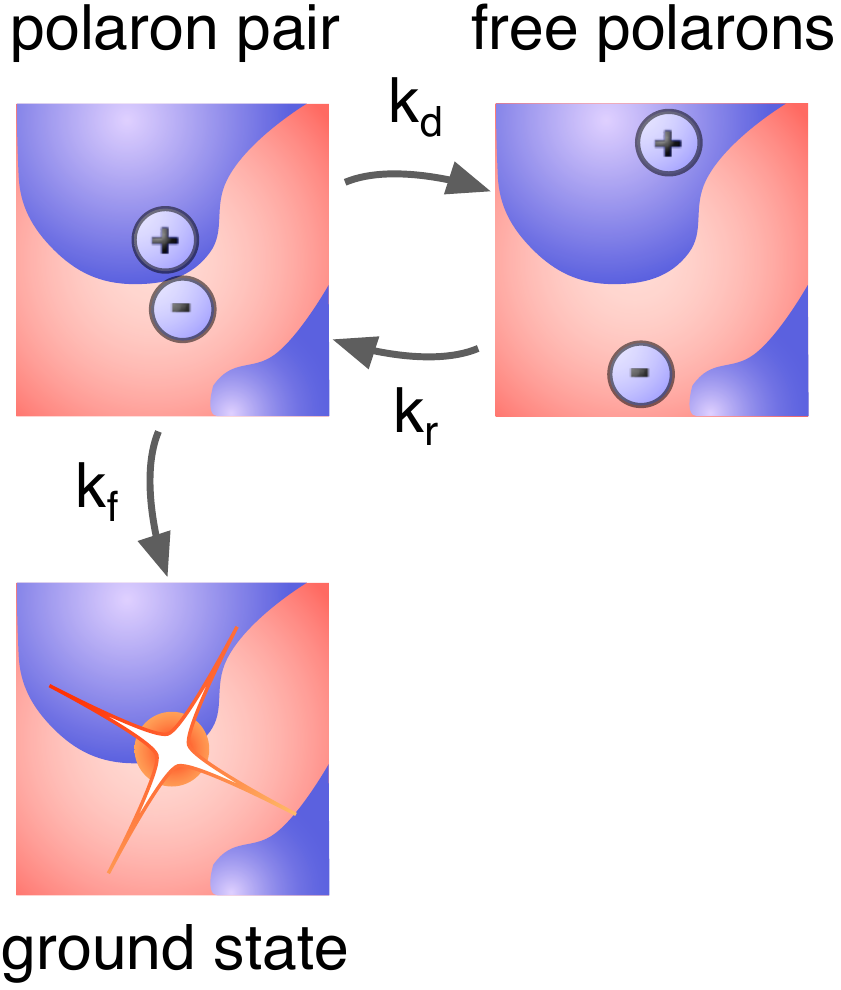}}\qquad
	\subfigure{\includegraphics[height=7cm]{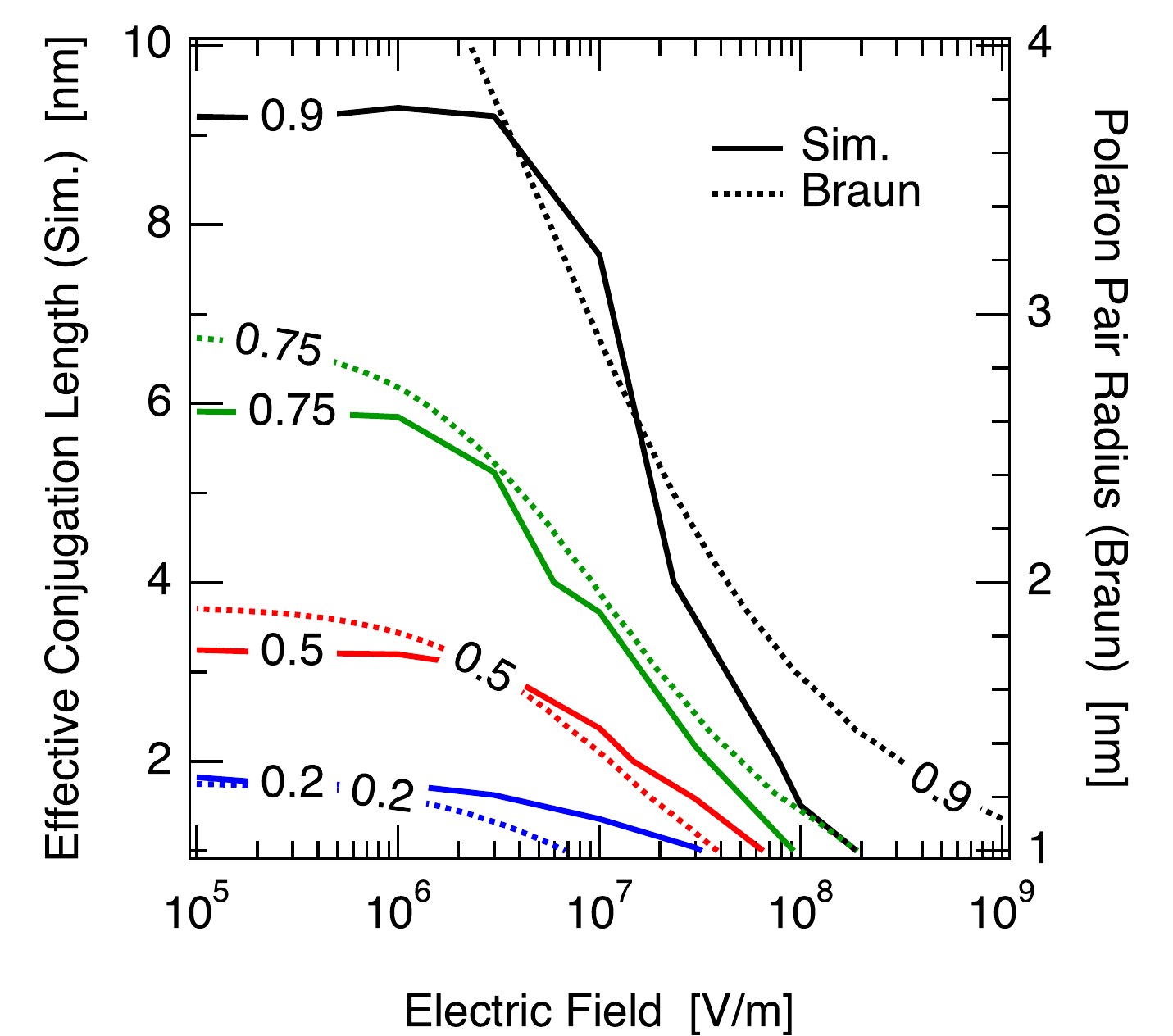}}
	\caption{Polaron pair dissociation in a donor--acceptor solar cell. (a) Schematic representation of the interplay between dissociation of polaron pairs to free charges with rate $k_d$---the reverse process of free polaron recombination, connected by detailed balance, going with rate $k_r$---or recombination of the polaron pair to the ground state with rate $k_f$. (b) Polaron pair dissociation yield with Monte Carlo simulation (solid lines) and the Onsager--Braun model (dashed lines) in $1:1$ donor--acceptor blends at $300$K  (dielectric constant $\epsilon=3.0$)~\cite{deibel2009a}. The contours are denoted with the yield values. Due to polymer chain entanglement, the polaron pair radius $r_{pp}$ is lower than the length of the  polymer chain segment (here, conjugation length $CL=2.5 r_{pp}$). The deviations between simulation and model are due to the Onsager--Braun model not considering energetic disorder and extended polymer chains. Thus, it is not suited perfectly well to describe polaron pair dissociation in disordered semiconductor blends. %
	\label{fig:pp-diss}}
\end{figure}

Once the singlet exciton has been dissociated with help of a suitable acceptor material, electron and hole reside on the acceptor resp.\ donor material, but are still Coulomb bound~\cite{deibel2010review2,clarke2010review}. This polaron pair also has to be separated in order to gain free charge carriers. The polaron pair dissociation is shown in Figure~\ref{fig:osc-bhj-morph} (iv) in the context of photogeneration in organic solar cells.

A commonly used model for considering the separation of two charges is the Onsager model, which has been applied to the dissociation of excitons as well as polaron pairs in organic semiconductors~\cite{onsager1938}. In his original publication from 1938, Onsager used it to calculate the probability to separate a Coulomb bound pair of ions of opposite charge with a given initial distance under assistance of an external electric field. 45 years later, Braun extended this model, accounting for the finite lifetime of the initial bound state, and applied it to the dissociation of charge transfer states in donor--acceptor systems~\cite{braun1984}. The polaron pair separation following the Braun--Onsager model is shown schematically in Figure~\ref{fig:pp-diss}(a). The polaron pair can either recombine to the ground state with a constant rate, given by its inverse lifetime $k_f=\tau_{PP}^{-1}$, or be dissociated to free charges with the rate $k_d$. Free charges can meet to generate bound polaron pairs again with the rate $k_r$. This recombination rate is partly based upon arguments of Langevin for nongeminate recombination~\cite{langevin1903,pope1999book}, and was used by Onsager to calculate the dissociation rate $k_d$ considering a detailed balance between recombination and dissociation. It is important to point out that within this framework, free charges never decay directly to the ground state: they always form polaron pairs first, implying that they have a finite chance to become free again (see Figure~\ref{fig:pp-diss}(a)).

The dissociation probability $P(F)$ is given by Onsager and Braun as
\begin{eqnarray}
	P(F) & = & \frac{k_d(F)}{k_d(F) + k_f}.
	\label{eqn:P}
\end{eqnarray}
Here, $F$ is the electric field and $k_f=\tau^{-1}$ the recombination rate of the polaron pair to the ground state. The field dependent dissociation rate $k_d(F)$ is given by
\begin{equation}
	k_d(F) = \frac{3\gamma}{4\pi r_{pp}^3} \exp\left( - \frac{E_b}{kT} \right) \frac{J_1\left( 2 \sqrt{-2b}  \right)}{\sqrt{-2b}}
	\label{eqn:kd}
\end{equation}
where $\gamma=q\mu/\epsilon\epsilon_0$ is the Langevin recombination factor~\cite{langevin1903},  $\mu$ the sum of electron and hole charge carrier mobility, $r_{pp}$ is the initial polaron pair radius, $E_b=e^2/(4\pi\epsilon\epsilon_0 r_{pp})$ is the Coulomb binding energy of the pair, $kT$ the thermal energy, $J_1$ the Bessel function of order one, and $b=e^3F/(8\pi\epsilon\epsilon_0(kT)^2)$. $e$ is the elementary charge, and $\epsilon\epsilon_0$ the effective dielectric constant of the organic semiconductor blend.  It is interesting to note that the polaron pair dissociation probability, Eqn.~(\ref{eqn:P}), contains mobility and polaron pair recombination rate in linear dependence, so that the $\mu\tau$-product is decisive for the dissociation yield.

The Braun--Onsager model is nowadays the most commonly applied model for describing the polaron pair dissociation in organic solar cells~\cite{mihailetchi2004a,koster2005b,buxton2006,deibel2008a}. 
Other approaches to describe the escape yield of charge pairs have been presented recently, including a review of Burshtein on non-Markovian theories of transfer reactions, including encounter theory~\cite{burshtein2004review}. Recently, Wojcik et al.~\cite{wojcik2009} presented an extension of Onsager theory to consider the finite lifetime of the bound state, claiming that the approach by Braun~\cite{braun1984} was based on the wrong assumption of dissociation and recombination following exponential kinetics. A comparison of experimental data to the so-called exact Onsager model remains to be done, which is partly in lack of meaningful data on the temperature and field dependence of the polaron pair dissociation.

A major drawback of Braun--Onsager and similar models is that the parameters $r_{pp}$ and $k_{f}$ are mostly used as fit factors in order to gain accordance with experimental data such as photocurrents~\cite{mihailetchi2004a,koster2005b,buxton2006,deibel2008a}. The reason is that the Braun--Onsager model does not account for energetic disorder or high local charge carrier mobilities due to polymer chains~\cite{hoofman1998,savenije2006,devizis2009} or nanocrystalline regions in either donor or acceptor phase. Recently, the effect of extended polymer chains on the dissociation yield was shown by Monte Carlo simulation~\cite{deibel2009a}, considering energetic disorder as well as effective conjugation lengths of ten monomer units~\cite{holdcroft1991}, in a donor--acceptor bulk heterojunction. The dissociation yields by Monte Carlo simulation and Onsager--Braun model are compared in Figure~\ref{fig:pp-diss}(b). Clearly, the increased conjugation length in the simulation has a profound impact on the separation yield. We found a qualitatively good agreement for dissociation yields between 20 and 75~\%. Beyond these separation efficiencies, the Onsager--Braun theory shows its limitations, as it cannot describe this behaviour. One reason might be that the theory does not account for hopping transport and energetic or spatial disorder, although these effects could be included implicitly in the (time and carrier concentration dependent) charge carrier mobility. We also point out that, to our knowledge, the strong temperature dependence predicted by the model has never been proven experimentally. 

In addition to delocalised charge carriers along polymer chains or nanocrystalline regions, also the properties of the donor--acceptor heterointerface can influence the dissociation yield. This effect will be most pronounced for bilayer solar cells, as in the planar geometry the surface dipoles do not cancel each other partly as in distributed heterojunctions. For pentacene/C$_{60}$ bilayers, it was recently shown by calculations of Verlaak et al.\ that the effective polaron pair binding energy is adjusted due to an additional potential, generated by interface multipoles, which decreases the Coulomb binding energy~\cite{verlaak2009,verlaak2007}. Already earlier, Arkhipov's dark dipole model had predicted this effect by accounting for partial charges, tranferred between molecules that are oriented parallel to the interface, thus forming a potential wall for a photogenerated polaron pair~\cite{arkhipov2003}. Static (partial) ground state charge transfer across the donor--acceptor interface has already been shown experimentally~\cite{benson-smith2007,osikowicz2007}. Verlaak et al.\ illustrated that depending on the molecular orientation of the pentacene relative to the fullerene, the surface dipole directly influences how easily polaron pairs are dissociated. Indeed, theoretical calculations predict that different orientations of molecules towards the heterojunction can result in attractive or repulsive interactions~\cite{sreearunothai2006}, thus being able to diminish or enhance the polaron pair dissociation process.

The fraction of polaron pairs which cannot be dissociated, can recombine geminately---a monomolecular process, being proportional to the polaron pair concentration, not the product of electron and hole densities. The loss process is either nonradiative or has a very low emission cross section~\cite{vandewal2008}. In the case that the polaron pair cannot escape the mutual influence, e.g. due to a too fine grained phase separation, one possible process is an electron back transfer from the acceptor to a triplet exciton state, as indicated in Figure~\ref{fig:ct}(c). This process is only possible if the energy of the triplet is favourable for the acceptor electron, i.e., if the donor triplet exciton energy is below the energy of the donor--acceptor charge transfer state~\cite{veldman2009}. This triplet generation in the donor material is independent of the intersystem crossing.

Another loss mechanism for polaron pairs is analog to the process found in literature in conjunction with electron or hole blocking layers~\cite{subbiah2010,yin2007a,irwin2008,hains2010,hains2010a}: extraction of a charge carrier at the wrong electrode. For polaron pairs, this process has been predicted by Monte Carlo simulations~\cite{strobel2010}. As polaron pairs dissociate only gradually, feeling the mutual Coulomb attraction even at longer distances---the Coulomb radius is 18~nm for a dielectric constant of 3 at room temperature---they can move along together in cases where the electric field is not much higher as compared to the influence of energetic disorder. As this process is hard to determine experimentally, it is difficult to estimate which relative importance the loss of polaron pairs at the interface holds.

An important question is whether or not the polaron pair separation is influenced by the excess energy available after exciton dissociation. The Durrant group investigated a range of polythiophenes blended with PCBM, and measured the polaron signal corresponding to the charge generation after the exciton dissociation by transient absorption~\cite{ohkita2008}. They found that the larger the Gibbs free energy $\Delta G_{cs}^{rel}$, the difference between the singlet exciton energy in the donor and the energy of the resulting polaron pair distributed over the donor--acceptor interface, the higher the polaron signal. Indeed, the polaron density is shown to depend exponentially on $\Delta G_{cs}^{rel}$. The best yield is achieved by P3HT due to having the highest excess energy. The authors have not determined the energy of the polaron pair state directly, although possible~\cite{veldman2008,vandewal2009a}, but have estimated it by considering the HOMO(donor)--LUMO(acceptor) energy gap. The strong dependence of the polaron density on the energy offset has two important implications: first, the excess energy after exciton dissociation provides the polaron pair with additional kinetic energy to overcome the Coulomb binding energy. It could maybe be considered as hot polaron pair, having a larger initial separation, thus making the mutual escape easier. Second, a larger $\Delta G_{cs}^{rel}$ might be related to a smaller HOMO(donor)--LUMO(acceptor) gap. The latter is directly proportional to the open circuit voltage, which would decrease accordingly. 

Thus, either photocurrent or open circuit voltage could be optimised in an organic solar cell, or a trade-off had to be found. However, these questions have not been answered conclusively yet: recently, an organic solar cell made from P3HT with an endohedral fullerene derivative was presented, which has its LUMO 300~meV closer to the P3HT LUMO than the reference acceptor PCBM~\cite{ross2009}. The open circuit voltage was increased accordingly by 300~mV, but the short circuit current---although somewhat lower as compared to the highly-optimised P3HT:PCBM solar cells---did not decrease exponentially, thus being much better than expected from the above-mentioned results~\cite{ohkita2008}. Indeed, recent results for a bulk heterojunction thin film with a copolymer PCPDTBT as donor material and PCBM as acceptor show that also other factors have an important influence: although having an excess energy hundreds of meV below the P3HT:PCBM reference system, the photogeneration yield was shown to be similarly good~\cite{clarke2009}. One reason might be that the positive and negative constituents of the donor excitation occupy different spatial parts of the copolymer, thus being already less strongly bound than in the case of P3HT. Also, we note that the studies~\cite{ohkita2008,clarke2009} where measured on optical films at zero electric field. 

Thus, kinetic excess energy seems to have an exponential influence on the polaron pair dissociation, but other factors also play an important role. Therefore, there is not necessarily a trade-off between high open circuit voltage and high photocurrent.

\subsection{Charge carrier transport} \label{sec:transport}

Following exciton dissociation and polaron pair separation, the charges can be transported towards the respective electrodes, as shown in Figure~\ref{fig:osc-bhj-morph} (v). Due to the lack of long-range order in the solution processed and evaporated organic semiconductors, the electrical transport mostly takes place by hopping from one localised state to the next, instead of band transport found in crystalline semiconductors.

In donor--acceptor solar cells, the donor material is mostly hole conducting, whereas the acceptor transports the electrons. In bilayer solar cells~\cite{cheyns2008,riede2008review}, the two semiconductors form separate layers with a defined interface (Figure~\ref{fig:osc-bilayer+bhj}(a)). Bulk heterojunctions instead consist of donor--acceptor blends with a certain phase separation (Figure~\ref{fig:osc-bilayer+bhj}(b)), which---from a macroscopic perspective---can be mostly described as an effective medium. Nevertheless, the charge carriers are transported predominantly on their respective phase.

In order to understand the hopping transport in the rather disordered organic semiconductors~\cite{arkhipov2006inbook,tessler2009review}, of which the solar cells are typically made, we will first consider what happens in a single material, before describing the bilayer or blend system.


Systems without long-range order consist of differently aligned molecules. If the wavefunctions for a given orbital between neighbouring sites do not overlap strongly, the localised charge carriers can only move by hopping from one site to the next, which leads to a very slow charge transport with low carrier mobilities as compared to crystalline materials.

Hopping transport is due to a combination of tunneling from one site to the next, a process depending on the transfer integral between the corresponding wave functions, and a thermally activated process. In the fifties, Marcus proposed a hopping rate from site $i$ to site $j$, which is suitable to describe the local charge transport~\cite{marcus1956,marcus1993},
\begin{eqnarray}
\nu_{ij} & = & \frac{\left| I_{ij} \right|^2}{\hbar} \sqrt{\frac{\pi}{\lambda kT}} \exp\left( -\frac{(\Delta G_{ij}+\lambda)^2}{4\lambda kT} \right) \quad .
\end{eqnarray}
Here, $I_{ij}$ is the transfer integral, i.e., the wave function overlap between sites $i$ and $j$, which is proportional to tunneling. $\lambda$ is the reorganisation energy, which is related to the polaron relaxation, $kT$ the thermal energy, and $\Delta G_{ij}$ is due to different energetic contributions, in particular the energy difference between the two sites. In disordered systems, the density of states is usually given by an exponential or gaussian distribution, so that $\Delta G_{ij}$ can be chosen as energy difference of two sites chosen from such a distribution. Thus, the polaronic self-trapping due to  the distortion of the molecule by the charge---which can be seen as a (lattice) polarisation, or a phonon cloud---as well as energetic disorder are considered. Experimentally, it is often difficult to distinguish between these two influences, as both have a similar impact on the charge transport properties. 

The Miller--Abrahams hopping rate~\cite{miller1960} is very similar,
\begin{equation}
	\nu_{ij} = \nu_0 \exp\left(-\gamma r_{ij} \right)\cases{\exp \left( -\frac{\Delta E_{ij}}{kT} \right) & $\Delta E_{ij}>0$ (uphop)\\1 & $\Delta E_{ij}\le0$ (downhop)} \quad ,
	\label{eqn:MA}
\end{equation}
although it shows the contributions of tunneling and thermal activation even more explicitly. $v_0$ is the maximum hopping rate, the so-called attempt-to-escape frequency. $\gamma$ is the inverse localisation radius, proportional to the transfer integral. $r_{ij}$ the distance of the sites $i$ and $j$: the first exponent describes the tunneling contribution. For hops upwards in energy, also a Boltzmann term analog to Marcus theory is used, although the energy difference between the sites $i$ and $j$, $\Delta E_{ij}$, usually considers only energetic disorder, but is sometimes enhanced to consider polaron relaxation as well.

Even in disordered systems, very high hopping rates can be achieved on a mesoscopic scale within regions of higher order. Experimentally, this is also observed: by using a contact-less method, the pulsed radiolysis microwave conductivity, the local charge carrier mobility within a material can be gained from the determined photoconductivity. Thus, instead of measuring the macroscopic charge transport, the local movement along a polymer chain or intrachain transport can for instance be observed~\cite{hoofman1998,warman2006}. The corresponding mobility values are not necessarily of direct relevance to electronic devices, they rather the intrinsic mobility which could in principal be achieved if structural defects, traps and other limitations were not present. Thus, these experimental techniques give important information concerning the optimisation potential of a given material. In contrast, the macroscopic mobility directly revant to devices can be determined by experimental transient photoconductivity~\cite{gill1972,tuladhar2005,baumann2008,tuladhar2009}, charge extraction~\cite{juska2000,mozer2005a,deibel2008b} or space charge limited current methods~\cite{blom2005,bohnenbuck2006}.

Information on the mesoscopic hopping transport, as given by the models after Marcus and Miller--Abrahams, can also be used to gain insight into the macroscopic properties of charge transport: The corresponding hopping rates can be applied in either Monte Carlo simulations or Hopping Master Equations. There, a larger piece of a material, made of maybe a million sites, each of them having its own energy chosen randomly from the density of states distribution, can be simulated. Master equations~\cite{pasveer2005,houili2006} as well as Monte Carlo simulations~\cite{schonherr1980,bassler1993} have been commonly used to describe charge transport in disordered materials, where analytic solutions are more difficult to be found. Single materials were considered~\cite{schonherr1980,bassler1993}, but also polaron pair dissociation and charge transport in donor--acceptor bilayers~\cite{peumans2004,houili2006,groves2008} and bulk heterojunctions~\cite{offermans2005a,deibel2009a} were investigated.

The results of these simulations can be fitted parametrically in order to make them more easily accessible by researchers to compare to experimental results. Thus, B{\"a}ssler~\cite{bassler1993} become famous for his gaussian disorder model, describing the charge transport in a gaussian density of states. From his Monte Carlo simulations, using the Miller--Abrahams hopping rate Eqn.~(\ref{eqn:MA}), he found the charge carrier mobility to depend on temperature $T$ and field $F$,
\begin{equation}
	\mu_{GDM} = \mu_\infty \exp\left( - \left( \frac{2\sigma}{3kT} \right)^2 + C \left( \left( \frac{\sigma}{kT} \right)^2 - \Sigma \right) F^{1/2} \right) \quad ,
	\label{eqn:mu_gdm}
\end{equation}
considering a gaussian density of states distribution with energetic width $\sigma$, also-called disorder parameter. $\Sigma$, corresponding to spatial disorder, as well $C$ and $2/3$ are scaling factors from the parametric fits. The experimental observation of $\ln \mu \propto F^{1/2}$, already described empirically by Gill~\cite{gill1972}, and $\ln \mu  \propto 1/T^2$ are both compatible with recent findings~\cite{mozer2004,tuladhar2009}. Indeed, this parametric model is commonly used. Later on, also a correlated disorder model was proposed~\cite{novikov1998}, where spatial and energy disorder depend on each other, with the motivation that a changing environment for a given molecule will have an influence on its energetic position. 

Similar to the approach of B{\"a}ssler, experimental findings of a carrier concentration depending mobility~\cite{tanase2003}---due to filling of the gaussian density of states, where lower states act as charge traps---were simulated by a Hopping Master Equation approach and fitted parametically, yielding an empirical description~\cite{pasveer2005} which is sometimes termed enhanced gaussian disorder model. 

\begin{figure}[bt]
	\centering
	\includegraphics[width=15cm]{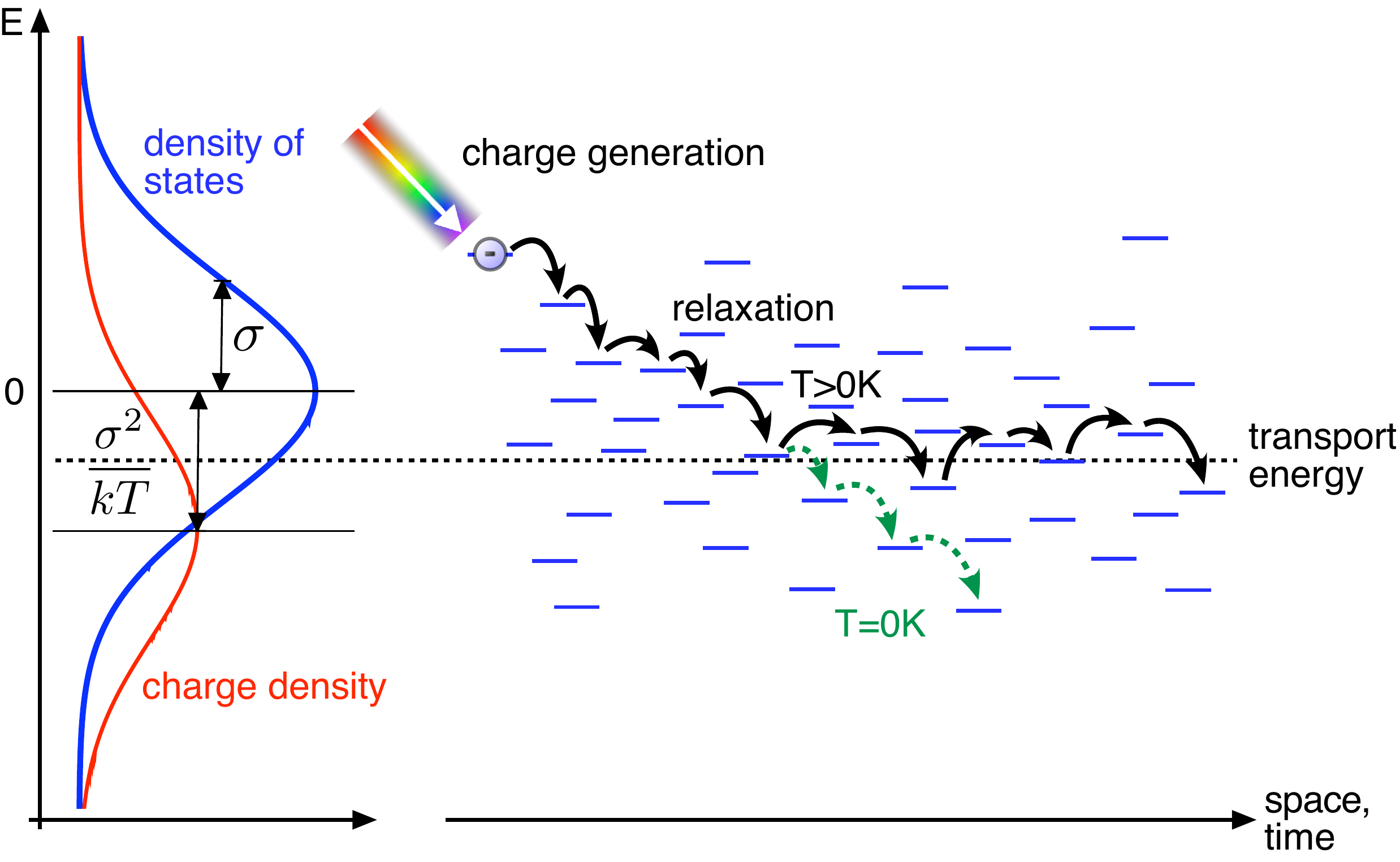}
	\caption{Charge transport by hopping between localised states with a gaussian energy distribution and disorder parameter $\sigma$ (blue solid line). The centre of the steady state charge distribution, also-called density of occupied states distribution (red solid line), is shifted by $\sigma^2/kT$ down as compared to the density of states centre. Photogenerated (and injected) charge carriers can be generated at higher energies, followed by a relaxation of the charge carriers to a quasi-equilibrium. The steady state charge transport takes place around the effective transport energy, which depends mainly on temperature and disorder. At $0$~K temperature, the charge carriers relax to the deepest states where they remain trapped. %
	\label{fig:transport-gDOS}}
\end{figure}

An important concept with respect to these investigations is the so-called transport energy~\cite{monroe1985,arkhipov2001a,tessler2009review}. It is analog to the mobility edge in amorphous inorganic semiconductors such as $\alpha$-Si. In a disordered organic device with a gaussian density of states of width $\sigma$, the charge density under steady state conditions thermalises on average to $-\sigma^2/kT$ below the center of the density of states distribution for a given temperature. The charge transport, however, is carried by hopping processes around the transport energy, as indicated in Figure~\ref{fig:transport-gDOS}. Charge carriers below this energy mostly do not contribute to the conductivity, they remain immobile. As the transport energy increases with higher temperature due to the thermal energy available to the charge carriers, the number of states accessible to the mobile charge carriers is also raised. The latter effect originates from  the shape of the density of states. Thus, the total carrier concentration $n$ can be described as the sum of mobile, conductive charge carriers $n_c$ around the transport energy, and immobile charge carriers $n_t$ trapped in the tails of the gaussian or exponential density of states. 

The usefulness of the concept of transport energy stems from the description of charge transport in these disordered systems by an equivalence model adapted from trap-rich inorganic semiconductors, the multiple trapping and release model. Though physically different---the latter assumes band transport---it is a valid approach to consider the charge transport of mobile charge carriers $n_c$ as band-like, and the tail states as traps. This approach can be used for analytic considerations~\cite{arkhipov2005,bohnenbuck2006} and macroscopic simulations~\cite{koster2005b,deibel2008a}, although in the latter case the tail states---important mainly for investigating the temperature or time dependence---have up to now always been omitted. Thus, the transport energy concept allows us to describe the charge transport through disordered semiconductors in equivalence to band transport plus trapping. Therefore, macroscopic device simulations of organic solar cells can be made with this concept, by either using the (enhanced) gaussian disorder model---the transport energy being implicitly included---or an analytic solution which is slower in terms of computation time. Nevertheless, the latter approach grasps more of the physics, and has made a lot of progress in recent years in describing the charge transport properties of hopping between localised states analytically~\cite{arkhipov2001a,rubel2004,fishchuk2007}.

The above mentioned models for charge transport consider energetic disorder on a mesoscopic scale, thus irrespective of ab-initio molecule--molecule interactions. The molecular interactions at planar donor--acceptor heterointerface in view of polaron pair dissociation were already studied by ab-initio methods~\cite{verlaak2009}. Still, a lot of work has to be done to understand the detailed process of charge generation and transport in organic semiconductors on multiscales. It is well-known that the local charge transport in organic semiconductors is often orders of magnitude faster than on a macroscopic scale~\cite{hoofman1998,savenije2006,devizis2009}. A complete multiscale model for charge transport, however, has not been presented for conjugated polymers yet, and in particular not for bilayer systems or blends. For highly ordered discotic liquid crystals in a homeotropic alignment, similar to one-dimensional nanowires, the first promising approaches have been performed~\cite{kirkpatrick2007,marcon2009}. In the near future, also ab-initio models connecting microscopic to macroscopic transport properties will be made using similar methods~\cite{cheung2008review,troisi2009,cheung2009a}.


Although microscopic charge transport to single organic layers is not completely described yet, charge transport in a bilayer or blend system such as an organic solar cell is more complicated to grasp. The dark current in a bilayer cell has to cross the donor--acceptor heterointerface. In a bulk heterojunction, percolation paths connect both material phases to both electrodes, the asymmetry given by the electrode work functions. The charge transport in organic solar cells is changed as compared to in single material devices, due to either the energy barrier (bilayer) or the higher spatial disorder (bulk heterojunction). In the latter case, hole transport through the polymer phase is for instance hindered by a fullerene nanocrystal blocking the transport path. These considerations for the dark conductivity are also valid for illumination. Corresponding transient photocurrent measurements have been performed, for instance on PPV--PCBM~\cite{tuladhar2005} and P3HT--PCBM bulk heterojunctions~\cite{baumann2008}. In both cases, the typical solar cell blend ratios, 1:4 resp.\ 1:1, show almost balanced electron--hole mobilities, which is favourable for the solar cell performance. The dependence of the charge carrier mobility on temperature and electric field given by the gaussian disorder model, Eqn.~(\ref{eqn:mu_gdm}), remains valid, although only on a mesoscopic scale~\cite{houili2006}: the parameters are changed effectively due to the different environment in the blend system, but up to now molecule--molecule interactions were not considered. Interface dipoles (Section~\ref{sec:pp-diss}) will also influence the charge transport across the heterojunctions. 

Generally, the microscopic description of charge transport in organic semiconductors leaves a lot to be desired as yet, even more so in donor--acceptor combinations, although a lot of progress has been made recently. Till then, the parametric models derived from Monte Carlo simulations---the most prominent still being the Gaussian disorder model---will have to be applied.

\subsection{Charge carrier recombination}\label{sec:recombination}

During the transport of the separated charges to their respective contacts (Figure~\ref{fig:osc-bhj-morph} (v)), charge carrier recombination can take place. This process is called nongeminate---in contrast to the geminate recombination during polaron pair dissociation (Section~\ref{sec:pp-diss})---as the participating positive and negative charges do not have a common precursor state, and thus occur independently. The recombination partners could either be injected into the organic semiconductor, or originate from different successfully dissociated polaron pairs. As in disordered organic semiconductors, charge transport takes place by hopping between localised states instead of band transport, the loss mechanisms cannot be described by a classical band--band or Shockley-Read-Hall recombination. 

The order of recombination is determined by the number of participants, and is thus crucial for the understanding of the origin of the loss mechanism. The charge carrier dynamics are given by the continuity equation,
\begin{equation}
	\frac{dn}{dt} = -\frac{1}{q} \frac{dj_n}{dx} + G - R \quad ,
	\label{eqn:continuity}
\end{equation}
shown here for the electron concentration $n$, with the spatial derivative of the electron current $j_n$, the elementary charge $q$, the optical generation rate $G$, and the recombination rate $R$. The latter includes the order of recombination. A monomolecular decay is given as $n/\tau$. Recombination of a mobile charge with a trapped carrier is considered to be a first order process, despite the participation of two constituents---if the density of trapped charges clearly exceeds the number of mobile ones, therefore seeming inexhaustible. If the concentration of trapped charges is similar or lower compared to the mobile ones, the decay will be bimolecular, as the trapped charges are depleted as well. Geminate recombination of a freshly generated polaron pair at the heterointerface is monomolecular, because the two constituents originate from the same precursor state. In contrast, if two mobile and independent charges are involved, the bimolecular recombination rate describes this nongeminate electron--hole recombination---it is thus a second order process. Higher order processes are well established from inorganic semiconductors, for example three-body recombination such as Auger or impact recombination~\cite{sze1995book}, although not for disordered inorganics. 

In bulk heterojunction and coevaporated small molecule solar cells, recombination of charges always occurs with the polaron pair as intermediate state (see Figure~\ref{fig:pp-diss}). Thus, free charges do not recombine directly upon meeting one another, but build a bound polaron pair first, which has a finite chance of dissociating again. Considering a coupled differential equation system of the continuity equations for free charges (Eqn.~\ref{eqn:continuity}) and polaron pairs (Eqn.~\ref{eqn:continuity-pp}), Koster et al.~\cite{koster2005b} were able to simplify the evaluation of charge losses in organic solar cells. The continuity equation for polaron pairs $\pp$ reads
\begin{equation}
	\frac{\pp}{dt} = G_{\pp} - k_f\pp - \underbrace{k_d \pp}_{G} + R,
	\label{eqn:continuity-pp}
\end{equation}
where $G_{\pp}$ is the polaron pair generation rate, for ideal exciton dissociation equalling the singlet generation rate, $k_f$ is the polaron pair recombination rate to the ground state (see Figure~\ref{fig:pp-diss}), $k_d$ is the dissociation rate from polaron pair to free polaron. The term in conjunction with the latter corresponds to the generation rate of free polarons, $G$, and is therefore a loss of polaron pairs. Similarly, a recombination of free polarons $R$ leads to the generation of polaron pairs and has a positive sign. Coupling Eqns.~(\ref{eqn:continuity}) and~(\ref{eqn:continuity-pp}) under steady state conditions, where the time derivatives become zero, leads to 
\begin{equation}
	\frac{dn}{dt} = -\frac{1}{q} \frac{dj_n}{dx} + PG_{\pp} - (1-P)R \quad .
	\label{eqn:continuity-p}
\end{equation}
Here, $P$ is the polaron pair dissociation probability, which---in terms of competing rates---corresponds to $k_d/(k_d+k_f)$ (Eqn.~(\ref{eqn:P})) and is usually given by the Braun--Onsager or exact Onsager model (Section~\ref{sec:pp-diss}). At flat band conditions~\cite{limpinsel2010}, approximately at the open circuit voltage, the  spatial current derivative becomes zero, thus further simplifying Eqn.~(\ref{eqn:continuity-p}).

\begin{figure}
	\begin{center}
		\subfigure{\includegraphics[width=6.5cm]{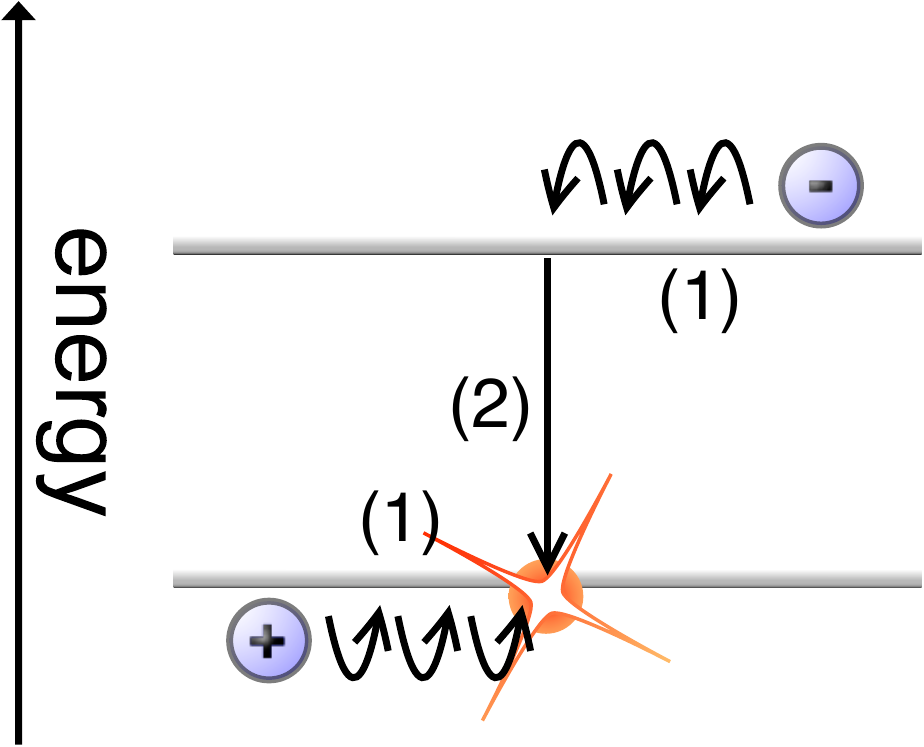}}\qquad
		\subfigure{\includegraphics[height=6cm]{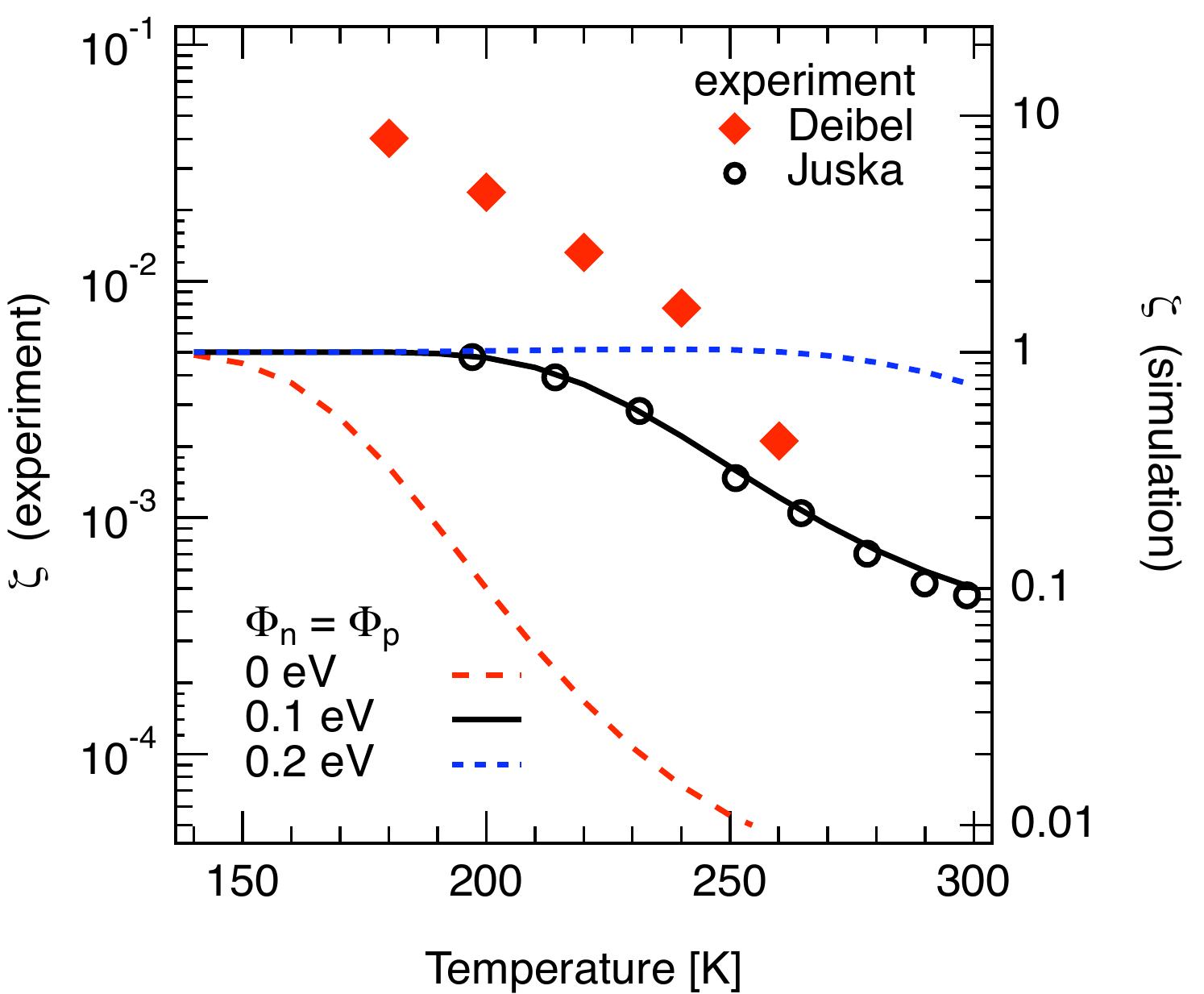}}%
		
		\caption{(a) The bimolecular Langevin recombination of electrons and holes. The recombination rate is limited by the slow approach of the recombination partners (1), not the actual annihilation process (2). (b) The experimentally found reduced prefactor $\zeta$ of the bimolecular recombination depending on temperature can partly be explained by the different gradients of electron and hole concentration in organic devices, which can not be distinguished and reconstructed by the mostly used charge extraction techniques. Macroscopic simulations (lines) allow to attain the concentration gradients and thus explain the experimental reduction factors (symbols)~\cite{deibel2009}.%
		\label{fig:recombination-Langevin}}
	\end{center}
\end{figure}

For free charges in low mobility materials, a nongeminate recombination of the second order (bimolecular)---following Langevin's theory~\cite{langevin1903,pope1999book}---is usually expected.
The Langevin recombination rate is given as
\begin{equation}
	R = \gamma (n p - n_i^2 ) ,
	\label{eqn:Langevin}
\end{equation}
with electron concentration $n$, hole concentration $p$, and $n_i$ the intrinsic carrier concentration. The bimolecular recombination prefactor $k_{br}$  is now expressed as Langevin recombination prefactor $\gamma$, the derivation of which is shown in the book by Pope and Swenberg~\cite{pope1999book}. The recombination process is indicated in Figure~\ref{fig:recombination-Langevin}(a). It is assumed that the rate limiting factor for recombination is the finding of the respective recombination partners (1), and \emph{not} the actual recombination rate (2). Neglecting process (2) as it is faster than (1), the finding of electron and hole depends on the sum of their diffusivities or---considering the Einstein relation~\cite{roichman2002,bisquert2008}---their mobilities. Thus, 
\begin{equation}
	\gamma=\frac{q}{\epsilon_r\epsilon_0}(\mu_e+\mu_h)	\quad ,
	\label{eqn:gamma}
\end{equation} 
where $q$ is the elementary charge, $\epsilon_r\epsilon_0$ the effective dielectric constant of the ambipolar semiconductor, $\mu_e$ and $\mu_h$ the electron and hole mobilities.

Several experimental investigations have shown that in bulk heterojunctions, an additional factor for reducing the Langevin recombination strength $\gamma$ is needed to describe the charge carrier dynamics appropriately~\cite{juska2006,deibel2008b}. Explanations have been attempted based on the notion that the charge carriers are spatially separated due to the blend morphology, where electrons and holes reside in the acceptor resp.\ donor material, and can only recombine at the heterointerface. It was suggested that either a thermally activated recombination~\cite{adriaenssens1997} could be responsible for the reduced recombination rate, or the minimum mobility of the two charges was relevant in Eqn.~(\ref{eqn:gamma}) instead of the sum~\cite{koster2006}. However, the temperature dependence of the correction factor could be described with neither approach~\cite{deibel2008b}. Only recently, a model was proposed which considers the organic solar cell as device with carrier concentration gradients formed due to the electrode/blend/electrode device configuration~\cite{deibel2009}. The resulting correction factor can explain the experimentally found temperature dependence, as shown in Figure~\ref{fig:recombination-Langevin}(b), although a residual temperature independent factor remains unresolved so far. An alternative explanation for the temperature dependence of the Langevin prefactor in form of a unified dissociation and recombination model was also given~\cite{hilczer2010}.

In bulk heterojunction devices, some researchers have attempted to gain the dominant recombination process by investigating the light dependence of the solar cell photocurrent. Mostly, a linear or slightly sublinear dependence is found, corresponding to a first order process~\cite{riedel2005}. This behaviour has been interpreted as predominant monomolecular recombination. However, other explanations for the same behaviours are either no recombination, which also gives a linear dependence, or a weak recombination of higher order, which could lead to the experimental observation of a  sublinear photocurrent. 

In contrast, recombination of mobile polarons with order larger than two have been reported by several researchers~\cite{shuttle2008,deibel2008b,juska2008}. Indeed, the apparent recombination order grows with decreasing temperature~\cite{foertig2009} and depends on the processing conditions~\cite{shuttle2008}, making trimolecular recombination improbable. A similar behaviour was observed in dye solar cells, where the variation to the expected monomolecular recombination was attributed to a delayed recombination due to trapping of charges in the tails of the density of states~\cite{zaban2003}. This observation has been found consistent with trapping in disordered materials~\cite{rudenko1982,arkhipov1982}, and its influence on nongeminate bimolecular recombination in bulk heterojunction solar cells~\cite{foertig2009,clarke2009a}. The recombination mechanism works on injected and photogenerated charge carriers identically~\cite{shuttle2008b,garcia-belmonte2009}. 

In bilayer solar cells, which do not contain a mixed donor--acceptor layer, the occurence of nongeminate---particularly bimolecular---recombination is fortunately unexpected~\cite{walzer2007review,riede2008review}. Once the bound electron--hole pairs have been separated, escaping the chance to recombine geminately, they usually do not meet again due to the device configuration of two separate layers, one electron, one hole conducting, with a planar heterointerface inbetween.

\subsection{Charge carrier extraction} \label{sec:extraction}

Free photogenerated charges that do not recombine can finally be extracted from the device in order to yield a photocurrent (Figure~\ref{fig:osc-bhj-morph} (vi)). The charge extraction process depends strongly on the device architecture, which determines the steady state carrier concentrations. Another important aspect is surface recombination at the metal--organic interface, which influences the carrier concentration at the interface, and thus the charge extraction.

\begin{figure}[bt]
	\centering
	\subfigure{\includegraphics[width=7cm]{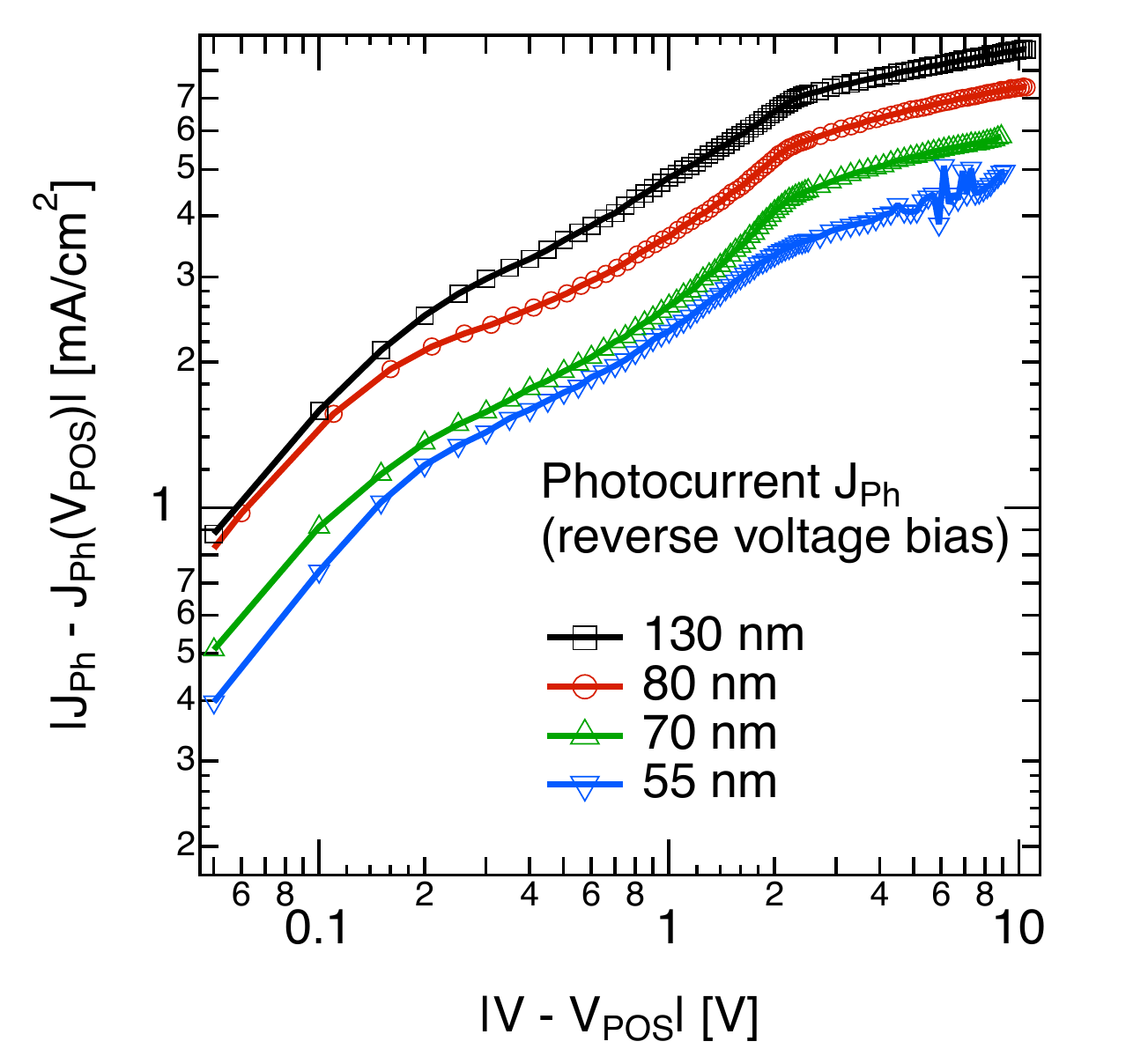}} \quad
	\subfigure{\includegraphics[width=7cm]{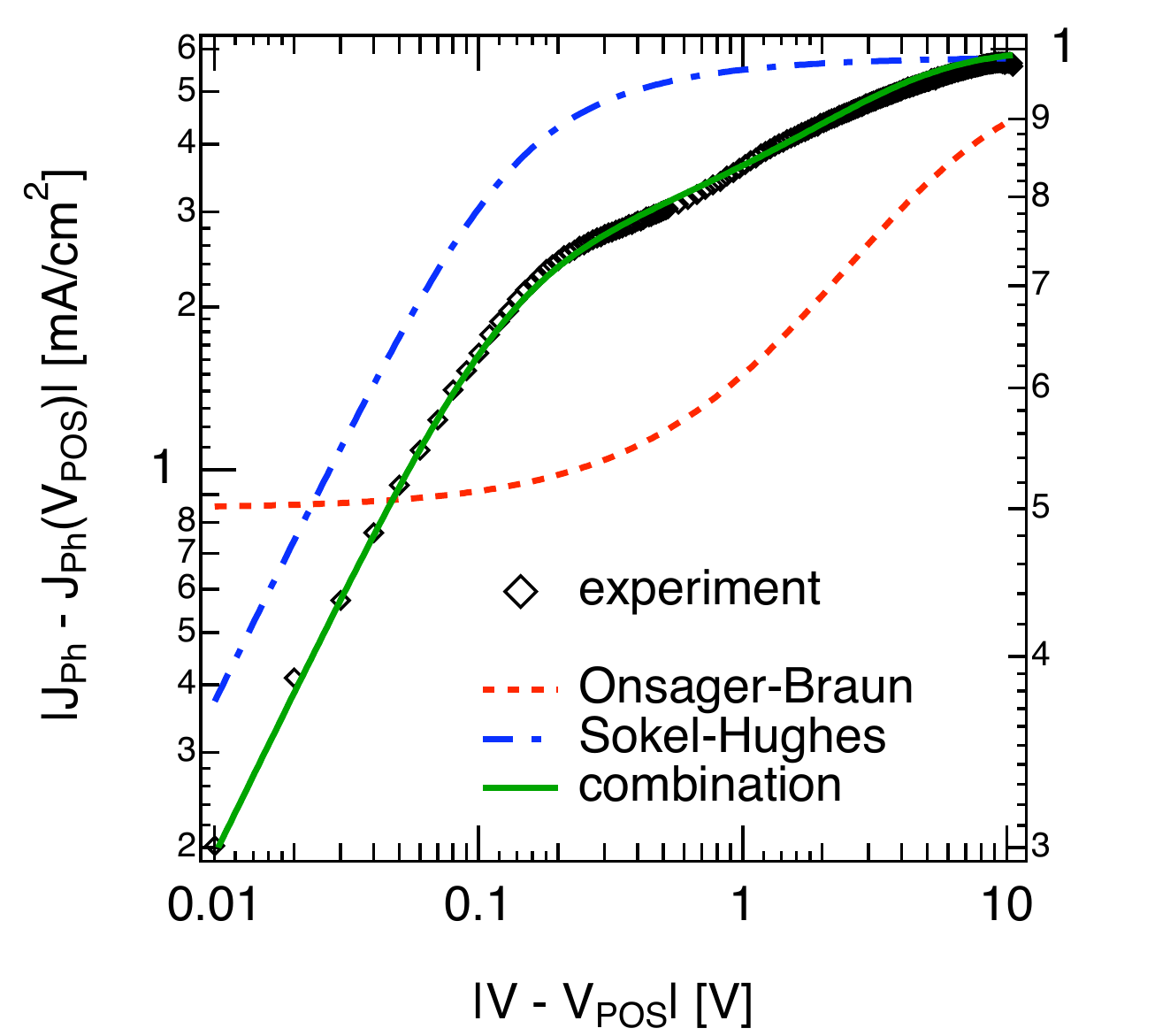}}
	\caption{Photocurrent of a bulk heterojunction organic solar cells: (a) The photocurrent is strongly voltage dependent. (b) Photocurrent described by a combination of Onsager--Braun model for polaron pair dissociation, and Sokel--Hughes model for charge extraction, in accordance with Eqn.~(\ref{eqn:OB+SH})~\cite{limpinsel2010}.%
	\label{fig:photocurrent}}
\end{figure}

\begin{figure}[bt]
	\centering
	\includegraphics[width=9cm]{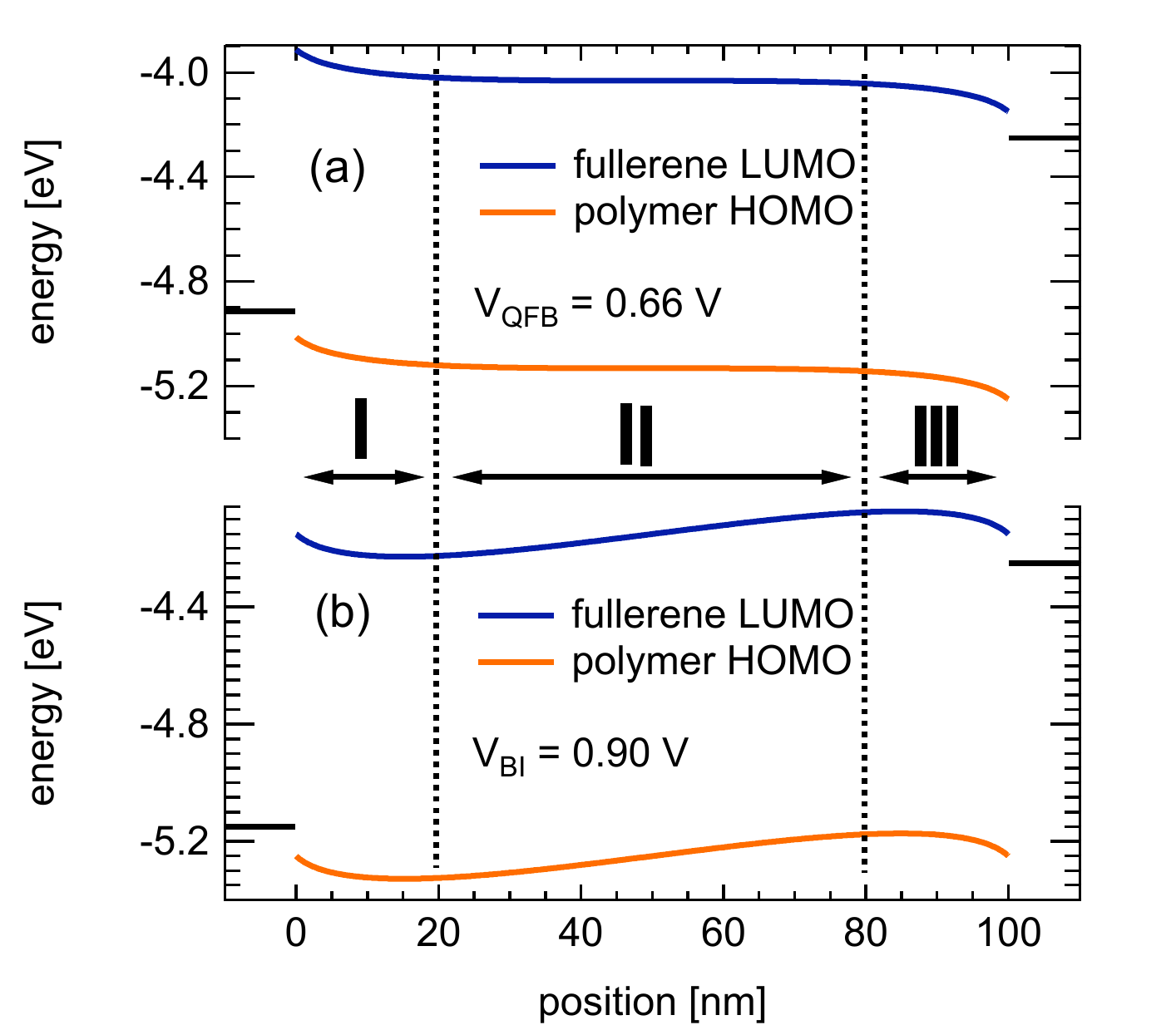}
	\caption{Potential energy scheme of an organic bulk heterojunction solar cell. By means of a macroscopic simulation, the HOMO and LUMO levels of a 100~nm thick device were calculated. The ITO contact is on the left, the metal anode on the right side. The top graph shows the quasi flat band case, in which the bulk region II shows zero electric field; only the contact regions I and III exhibit band bending. (Bottom) At an applied voltage equal to the built-in potential, the internal electric field is nonzero over the whole device. If the internal electric field is calculated by $F=V/L_{eff}$, a constant electric field over the extent of the device thickness is incorrectly expected. Thus, to decribe the internal field, an effective device thickness unequal to the physical thickness can be applied for modelling~\cite{limpinsel2010}.%
	\label{fig:band}}
\end{figure}

For bulk heterojunction solar cells, an actively debated question is the origin of the field dependence of the photocurrent: is the dominant current loss mechanism the polaron pair dissociation or the charge extraction process? In 2004, Mihailetchi et al.~\cite{mihailetchi2004a} was able to describe the experimental photocurrent of a bulk heterojunction solar cell by polaron pair dissociation, applying the Braun--Onsager model~\cite{braun1984,onsager1938}, and an additional diffusion term for polarons based on a model by Sokel and Hughes~\cite{sokel1982}. Recently, investigations of the photocurrent of polymer based solar cells by a pulsed technique were presented~\cite{ooi2008}. The photocurrent was found to be symmetric relative to a point of optimum symmetry. The field dependence was explained qualitatively by charge collection from a photoactive layer, whereas the polaron pair dissociation was assumed to be field independent. As explained in Section~\ref{sec:pp-diss}, despite the high yield of polaron pair dissociation already at low fields, this process does depend on the field. Therefore, it becomes clear that both dissociation and extraction determine the field dependence of the photocurrent in bulk heterojunction solar cells. Indeed, the polaron diffusion term in the photocurrent description of Mihailetchi et al.~\cite{mihailetchi2004a} already accounted for charge extraction by the Sokel and Hughes model~\cite{sokel1982}. This model is analytic, therefore with several simplifying assumptions such as a constant voltage drop across the extent of the device, the surface recombination velocity~\cite{scott1999a} supposedly infinite, and a negligence of trapping and recombination. Nevertheless, such models help us to gain insight into charge extraction. After Sokel and Hughes, the photocurrent in a single layer device is thus given as
\begin{equation}
	J_{ph,SH} = J_{ph,max} \left( \frac{\exp(qV/kT)+1}{\exp(qV/kT)-1} - \frac{2kT}{q} \right) ,
	\label{eqn:SH}
\end{equation}
where $J_{ph,max}=qGL$ denotes the maximum photocurrent. $q$ is the elementary charge, $G$ the generation rate of free charges in the bulk, $L$ the device thickness, $kT/q$ the thermal voltage, and $V$ the internal voltage. Most importantly, Eqn.~(\ref{eqn:SH}) does depend on the voltage across the device rather than the electric field---in contrast to the field dependent polaron pair dissociation. A recent study shows that the photocurrent is indeed strongly voltage dependent (Figure~\ref{fig:photocurrent}(a)), although the exact shape can only be reconstructed by a convolution of models concerning polaron pair dissociation and charge extraction~\cite{limpinsel2010}, as shown in Figure~\ref{fig:photocurrent}(b). Thus, the complete photocurrent can be written as
\begin{equation}
	J_{ph} = qP(F)G_{\pp}L  \left( \frac{\exp(qV/kT)+1}{\exp(qV/kT)-1} - \frac{2kT}{q} \right) ,
	\label{eqn:OB+SH}
\end{equation}
where $P(F)$ is the polaron pair dissociation yield given by Onsager--Braun, Eqn.~(\ref{eqn:P}), and $G_{\pp}$ is the generation rate of polaron pairs. Here, the free charge generation rate $G=P(F)G_{\pp}$. The experimental data can be described well with this model~\cite{mihailetchi2004a,foertig2009}, as shown in Figure~\ref{fig:photocurrent}(b), although an effective device thickness $L_{eff}$ has to be used to calculate the internal field $F=V/L_{eff}$. It approximates the influence of the inhomogenous voltage drop across the device, particularly close to the electrodes, as shown in Figure~\ref{fig:band}~\cite{limpinsel2010}. 

For bilayer solar cells, the consideration of polaron pair dissociation and charge extraction is also important~\cite{riede2008review}. The former process is similar, although due to the planar heterojunction interface a potential drop improves the polaron pair dissociation~\cite{peumans2004}. Therefore, polaron pair dissociation is efficient although bilayer devices are usually made of small molecules where delocalisation along polymer chains~\cite{deibel2009a} cannot assist in the process of charge separation. Delocalisation along nanocystals, however, could enhance the dissociation yield. The charge transport and extraction is in any case easier than in bulk heterojunction solar cells, as the spatial disorder as introduced by intermixing of donor and acceptor phases is much lower.

The resulting photocurrent is thus generated by the extraced charges: the holes leave the device through the anode, the electrons are extracted from the cathode.

In suboptimal photovoltaic devices, additional mechanisms influencing the charge extraction have been reported. Sometimes, a space-charge limited photocurrent is observed in less efficient solar cells, for instance due to imbalanced electron and hole mobilities~\cite{mihailetchi2005}. Another issue is the occurence of so-called double diodes, which corresponds to an inflexion point in the current--voltage characteristics~\cite{nelson2004,kumar2009,wagenpfahl2010}. The explanations for these s-shaped solar cell characteristics vary. For instance, they be modeled by a decrease of charge transfer rate over the interface~\cite{nelson2004} or by surface recombination~\cite{wagenpfahl2010, wagenpfahl2010a}. The latter concept is more general and includes also a low charge extraction rate. 
Another proposition was that interface dipoles~\cite{lee2009} are responsible for a voltage drop leading to double diodes~\cite{kumar2009}.

Before, the surface recombination velocity was not considered adequately, as it was never investigated directly in organic semiconductors. Therefore, in macroscopic simulations, the surface recombination velocity was usually unrealistically considered to be infinite~\cite{koster2005b,buxton2007,mandoc2007,deibel2008a}. Only recently, the finite values as postulated by Scott and Malliaras~\cite{scott1999a} for amorphous organic semiconductor devices were accounted for~\cite{kirchartz2009,wagenpfahl2010,wagenpfahl2010a}. Indeed, surface recombination does not always imply a loss of charges. Considering the surface recombination velocity of only one charge carrier, a better way of visualising this process is an extraction rate of charges from the device. Thus, this process is directly relevant to charge extration. If the surface recombination velocity of a given carrier type is too low, it cannot leave the device quickly enough, will therefore pile up and build a space charge region at the corresponding electrode. If this happens, the above-mentioned s-shaped current--voltage curve will appear. Only if both charge carrier types are extracted at the same electrode does this process correspond to a loss mechanism with zero net current.

Charge extraction concludes the steps from photogeneration of excitons to gain a photocurrent, shown in Figure~\ref{fig:osc-bhj-morph}. The photocurrent dominates two of the three basic solar cell parameters, short circuit current and fill factor. For understanding the open circuit voltage, additional considerations are necessary, as described in the following section.

\subsection{Energetics: the open circuit voltage} \label{sec:voc}

\begin{figure}[bt]
	\centering
	\includegraphics[width=13cm]{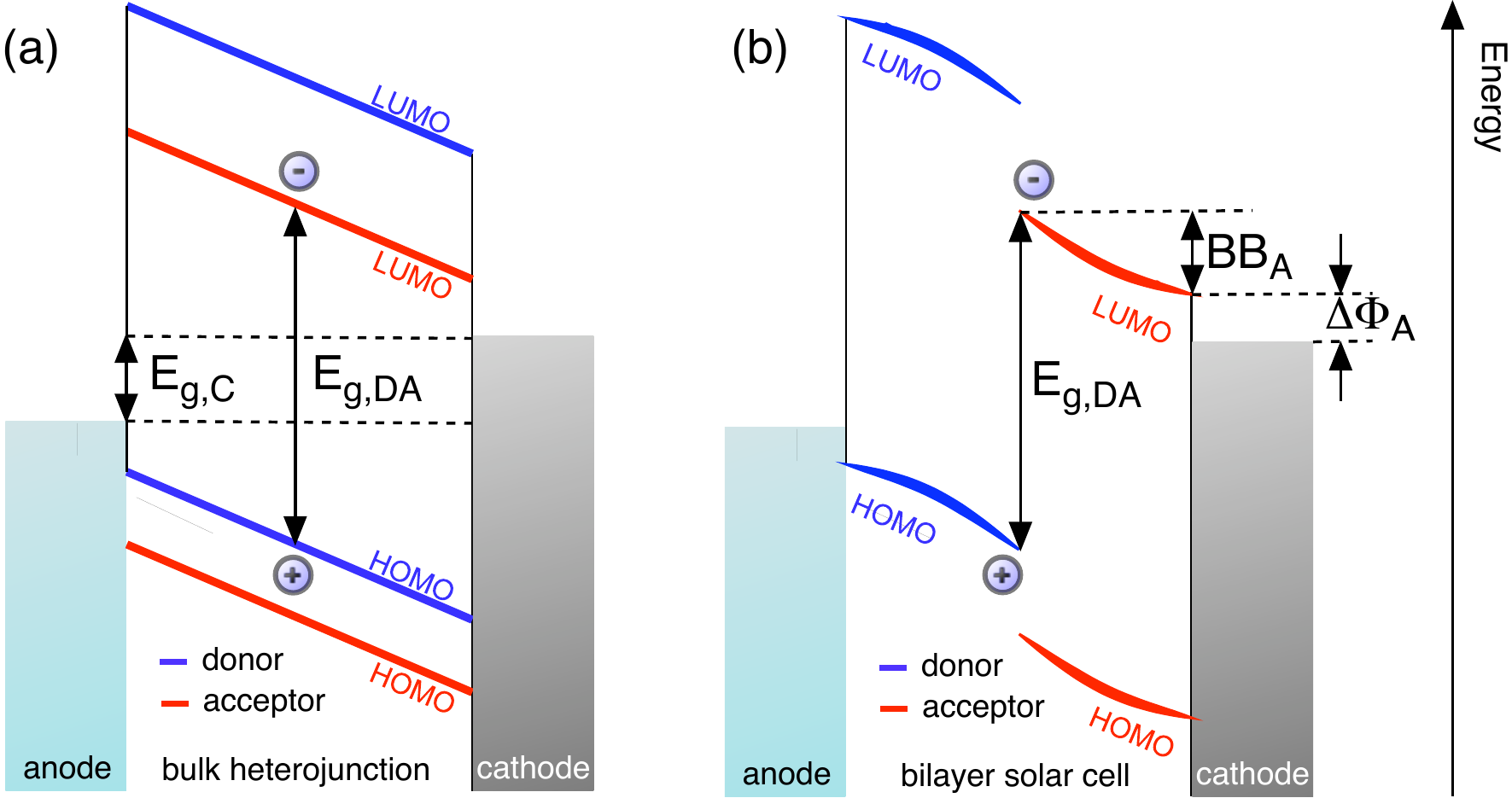}
	\caption{The open circuit voltage in organic solar cells. (a) Bulk heterojunction device with an effective band gap between donor and acceptor. The maximum open circuit voltage is given by the energy of the donor--acceptor polaron pair state $E_{g,DA}$~\cite{vandewal2009a} or, in some cases, the work function difference of the electrodes $E_{g,C}$. (b) In a bilayer device, band bending ($BB$, here indicated for the acceptor $A$) at the planar heterojunction can influence the open circuit voltage~\cite{cheyns2008} and compensate a higher injection barrier $\Delta \Phi$. The binding energies of excitons and polaron pairs were neglected in both schematic drawings.
	\label{fig:voc}}
\end{figure}

The open circuit voltage is a very important factor for understanding and optimising organic photovoltaic devices. The relevant band diagram of a bulk heterojunction solar cell is presented in Figure~\ref{fig:voc}(a). Two different energy levels which can govern the open circuit voltage are indicated: first, the effective bandgap $E_{g,DA}$, corresponding to the energy of the polaron pair after exciton dissociation, thus depending on donor and acceptor energetics. Second, the difference of the workfunctions of anode and cathode.  In this schematic drawing it is also indicated that under open circuit conditions, the potential is not necessarily flat. The bands of donor and acceptor are shown to be parallel, which is a good approximation for very fine intermixing of the blend constituents. Nevertheless, generally a weak band bending between donor and acceptor is possible. In contrast, bilayer solar cells can show a strong band bending, as donor and acceptor layer are only coupled at the planar heterojunction interface, as shown in Figure~\ref{fig:voc}(b).

In the bilayer solar cell, the open circuit voltage is determined by the energy difference of the quasi Fermi-levels at zero net current flow. The energy of the polaron pair or charge transfer complex is the maximum possible open circuit voltage in the device. The experimentally determined open circuit voltage, however, also includes contributions by the band bending. In case of Figure~\ref{fig:voc}(b) these are losses, but in principle band bending can also lead to a gain in open circuit voltage. Also, the quasi Fermi levels are influenced by the injection barriers at the anode and cathode. These influences can be summarised in the general equation
\begin{eqnarray}
	V_{oc} & = & \frac{E_{g,DA}}{q} + BB_D + BB_A - \Delta\Phi_D - \Delta\Phi_A ,
\end{eqnarray}
where $BB_D$ and $BB_A$ correspond to the band bending in the donor and acceptor layer, respectively, which can be positive as well as negative. $\Delta\Phi$ are the injection barriers from anode into donor or cathode into acceptor, respectively. Here, the injection barriers seem to reduce the open circuit voltage linearly. Cheyns et al.~\cite{cheyns2008} calculated the band bending, and considered thermionic emission as injection mechanism, and found that the band bending cancels the effect of the injection barriers,
\begin{eqnarray}
	V_{oc,Cheyns} & = & \frac{E_g}{q} - \frac{kT}{q} \ln \left( \frac{N_D N_A}{p_h n_h} \right) .
	\label{eqn:voc-cheyns}
\end{eqnarray}
Here, $kT/q$ is the thermal voltage, $N_D$ and $N_A$ are the donor and acceptor effective density of states, respectively. $p_h$ and $n_h$ are the hole resp.\ electron concentration at the donor--acceptor heterojunction. Although the open circuit voltage is now independent of the injection barriers, the slope of current vs.\ voltage at $V_{oc}$---the inverse of which can be expressed as diffusion resistance---is decreased, thus reducing the solar cell efficiency~\cite{cheyns2008}. These findings were also shown experimentally by the authors. Several earlier publications already pointed out the almost linear dependence of the open circuit voltage on the effective donor--acceptor bandgap~\cite{kietzke2006,rand2007}.

Concerning the open circuit voltage in  bulk heterojunction solar cells, the first impression yields similar dependencies. In 2001, Brabec et al.~\cite{brabec2001} presented experimental measurements showing an almost linear correlation of $V_{oc}$ with donor--acceptor effective bandgap for solar cells based on blends from a polyphenylenevinylene with different fullerene derivatives. Gadisa et al.~\cite{gadisa2004} drew similar conclusions from different donor polymers with PCBM as acceptor. A few years later, Koster et al.~\cite{koster2005} presented a description of $V_{oc}$ derived from the quasi Fermi level difference, neglecting injection barriers. The equation, derived in part by using the continuity equation Eqn.~(\ref{eqn:continuity-p}), directly shows the experimental dependence of the open circuit voltage on the effective bandgap, 
\begin{eqnarray}
	V_{oc,Koster} & = & \frac{E_g}{q} - \frac{kT}{q} \ln \left( \frac{(1-P)\gamma N^2}{PG} \right) .
\end{eqnarray}
Here, $P$ is the probability to dissociate excitons into free charges, for instance given by Eqn.~(\ref{eqn:P}), $G$ is the exciton generation rate, and $\gamma$ the Langevin recombination strength described in Section~\ref{sec:pp-diss}. $N$ is the effective density of states. This equation is analog to Eqn.~(\ref{eqn:voc-cheyns}) for bilayer solar cells: also in bulk heterojunctions does the open circuit voltage depend on the effective bandgap, the temperature, and the carrier concentration---although, here in the bulk, not at the planar heterojunction interface as in bilayers.

The investigations of Brabec et al.~\cite{brabec2001} and Gadisa et al.~\cite{gadisa2004} also showed the effect of the injection barriers to be negligible for a wide range of cathode metals on polyphenylenevinylene based solar cells. In contrast, the contact properties were found to be important for polyfluorene based photovoltaic devices~\cite{ramsdale2002}: for four out of five contact metals, a linear dependence of the workfunction difference to the open circuit voltage could be shown. Only for the anode--cathode combination ITO--Calcium a deviation was found. It may be of importance to point out that the open circuit voltage in polyfluorene based devices is much higher as for polyphenylenevinylene based solar cells. These results allow for the interpretation that two different regimes of contact and bulk limitation determine the experimental open circuit voltage, the minimum of $E_{g,C}$ resp.\ $E_{g,DA}$  (see Figure~\ref{fig:voc}). In other words, if the correct anode--cathode combination is chosen, the maximum open circuit voltage is determined by the effective bandgap of the donor--acceptor blend. 

Although open circuit voltage and effective bandgap are usually linearly proportional, a certain offset voltage is evident. Scharber et al.~\cite{scharber2006} determined $V_{oc}=1/q | $HOMO$_D - $LUMO$_A | - 0.3$~V for a vast range of donor--acceptor combinations. In 2008, Vandewal et al.~\cite{vandewal2008} presented very sensitive Fourier transform photocurrent spectroscopy measurements, in which the authors observe the weak quantum efficiency of the polaron pairs. Thus, they are able to directly determine the effective bandgap of the blend material, $E_{g,DA}$, which is an advance over using molecular orbital energies instead. Nevertheless, a discrepancy of several $100$~mV was observed for polyphenylenevinylene based solar cells, Vandewal et al.\ find $V_{oc}=E_{g,DA}/q-0.43$~V. Veldman et al.~\cite{veldman2009} considered the relation between the molecular orbitals, the polaron pair energy (although not measured directly in constrast to Vandewal et al.), and the open circuit voltage for several donor--acceptor combinations. The effective bandgap, given by the polaron pair energy---denoted as charge transfer state energy---are $0.3$~eV below the corresponding HOMO$_D - $LUMO$_A$ energy. The quasi-Fermi level splitting corresponding to the open circuit voltage turns out to be about $0.4$--$0.5$~eV below the effective bandgap. This discrepancy is assigned to approximately $0.4$~eV from band bending due to charge carrier diffusion, and about $0.1$~eV due to energetic disorder and the energy needed to dissociate polaron pairs.

Recent results by Vandewal et al.~\cite{vandewal2008,vandewal2009a} present in more detail the origin of the dependence of open circuit voltage on the effective bandgap. They find that the energy of the charge transfer state, or polaron pair, is directly proportional to the experimental open circuit voltage. The former, however, is larger by several hundred meV and corresponds to the theoretical maximum of the open circuit voltage---a quasi-Fermi level splitting cannot exceed the energy of the charge transfer state, but can be lower due to reasons discussed below. The polaron pair energy depends directly on the donor--acceptor properties and the morphology, such as the polymer crystallinity~\cite{vandewal2009}. By a complementary combination of experimental techniques and a data analysis based on theoretical guidance by Shockley--Queisser~\cite{shockley1961} and Rau~\cite{rau2007,kirchartz2008d} they could relate the energy of the charge transfer state~\cite{vandewal2009a} to the open circuit voltage and assign the energy difference to nonradiative recombination. However, considering the original paper of Shockley and Queisser~\cite{shockley1961}, it becomes evident that both, radiative and nonradiative recombination limit the open circuit voltage~\cite{deibel2010review2}. As the energy of the charge transfer state depends on the effective bandgap of the solar cell, i.e. HOMO$_D - $LUMO$_A$, the previous findings described above, which related the open circuit voltage to the HOMO--LUMO difference at the donor--acceptor interface, are explained as well. The details of recombination in organic solar cells (Section~\ref{sec:recombination}) and how to reduce them in order to minimise the loss in open circuit voltage, however, are not completely resolved yet.

In order to optimise the open circuit voltage, the difference between polaron pair energy and open circuit voltage will have to be reduced, while ensuring that the energy levels are still suitable for an efficient exciton dissociation without electron back transfer, and thus yielding an efficient photocurrent generation as well. This is for instance done by tuning the LUMO level of the acceptor molecules~\cite{riedel2005,ross2009}. Simultaneously, recombination will have to be minimised in order to best exploit the maximum possible open circuit voltage, given by the polaron pair energy.

\subsection{Morphology: a trade off between dissociation and transport} \label{sec:morphology}

The morphology of the active layer is of major importance in bulk heterojunction solar cells~\cite{chen2009review}. Due to the trade-off between exciton dissociation (Section~\ref{sec:exciton})---where a fine-grained phase segragation is ideal---and charge transport (Section~\ref{sec:transport})---where a bilayer configuration is optimum---the spatial dimensions of the donor--acceptor phases are crucial for the device performance~\cite{moule2009review}.

The influence on the device performance due to different donor--acceptor ratios or thermal annealing was assigned to changes of the nanomorphology of PPV:PCBM~\cite{hoppe2004}, P3HT--PCBM~\cite{chirvase2004,vanlaeke2006a}, or polymer--polymer solar cells~\cite{mcneill2008}. In particular, thermal annealing was found to improve the photocurrent by up to one order of magnitude~\cite{padinger2003,mcneill2008}. The molecular weight of the polymer components was shown to have a strong influence on crystalline packing~\cite{donley2005}, charge transport~\cite{zen2004,goh2005, ballantyne2008}, polaron pair dissociation~\cite{moet2009} as well as the solar cell performance~\cite{hiorns2006,ballantyne2008,sommer2009}. Some donor--acceptor combinations, for instance P3HT:PCBM~\cite{peet2006} and  the copolymer donor PCPDTBT in conjunction with PCBM~\cite{peet2007,coates2008}, show an improved power conversion efficiency when alkane dithiol additives are included in the blend solution for film casting. Being a selective solvent for PCBM, the additive seems to increase the degree of phase separation and to improve the local crystallinity in the films. Which constituent exactly exhibits this higher order is not yet completely understood~\cite{peet2009review}.

Insight into the internal structure of the bulk heterojunction solar cells was obtained by different methods, which give information on lateral, vertical, or threedimensional morphology changes. Lateral information was won by traditional~\cite{hoppe2004} and conductive atomic force microscopy~\cite{alexeev2008} as well as transmission electron microscopy~\cite{vanlaeke2006,ma2007,ma2007a,jo2009}. A chemical mapping was done by scanning transmission x-ray microscopy, although it is limited to domains larger than 30~nm due to the lateral resolution~\cite{mcneill2006a}. With the combination of x-ray diffraction techniques~\cite{vanlaeke2006a,chiu2008,jo2009}, insights into the partial order, or crystallinity, could be obtained. In P3HT:PCBM films, annealing was found to increase the size of PCBM aggregates from around 35~nm to 60~nm, whereas the crystalline P3HT regions also became more prominent~\cite{vanlaeke2006a,chiu2008}. Indeed, polymer fibre growth was found by transmission electron microscopy in P3HT:PCBM solar cells upon annealing~\cite{savenije2005,savenije2006,bertho2009}.

For the vertical device configuration, ellipsometry~\cite{campoy-quiles2008} as well as cross-sectional atomic force microscopy were applied~\cite{hoppe2004}. The chemical composition was investigated by time-of-flight secondary ion mass spectrometry~\cite{bulle-lieuwma2003}, even with lateral resolution~\cite{jo2009}. The growth of small molecule solar cells by organic vapour phase deposition was also investigated by scanning electron microscopy and atomic force microscopy (also cross-sectional)~\cite{yang2005}.

Most information, however, can be extracted from three-dimensional images, which can be investigated by nanotomography techniques. A copolymer:PCBM solar cell, typically giving 3-4~\% power conversion efficiency, was measured by means of electron tomography~\cite{andersson2009a}. From the data analysis, a model picture of PCBM rich regions surrounded by copolymer rich regions was extracted. The detailed influence of acceptor material, donor--acceptor ratio, solvent and substrate was investigated extensively~\cite{barrau2009}. Also, P3HT:PCBM was studied. Upon annealing, a threedimensional network on the nanometre-scale with crystalline order was found, which exhibits concentration gradients throughout the thickness of the photoactive layer, assisting the process of selective charge extraction~\cite{vanbavel2009,vanbavel2009a}. 

Indeed, a vertical phase segregation, with a higher donor (acceptor) concentration at the anode (cathode), might offer an improvement over a random donor--acceptor distribution. The reason is that photogenerated holes are transported to the anode within the donor phase, and are extracted at the anode. If, however, the concentration of the acceptor at the anode is higher than that of the donor, the probability of extraction of electrons at the anode is increased. This extraction at the wrong electrode is a loss mechanism, and thus unfavourable for the photovoltaic performance. Electron blocking layers might help~\cite{yin2007a,subbiah2010,hains2010,hains2010a}, but a favourable gradient of the donor--acceptor ratio throughout the sample could already reduce the losses. The vertical phase segregation was investigated in P3HT:PCBM samples~\cite{campoy-quiles2008}. On PEDOT/P3HT:PCBM/air layer configurations, the authors found that in pristine samples, the PCBM fraction at the anode is two-thirds, at the cathode only one-third. Annealing of the thin film reduces the gradient to an approximately constant donor--acceptor ratio vs.\ depth within the sample. For complete devices, the trend seemed similar. Thus, annealing reduces the formerly unfavourable distribution and enhances the selectivity of the device at the electrodes. These findings were recently confirmed by complementary techniques~\cite{jo2009}.

Recently, intercalation of fullerene acceptors between the side chains of donor polymers was found to play an important role in determining the donor--acceptor ratio optimum for charge generation. 
In blends where intercalation occurs, fullerenes first fill the space between the polymer side chains before larger fullerene phases, favourable for transport of electrons to the cathode, are formed~\cite{mayer2009}. Regions with pure fullerenes in blends of PPV derivatives with PCBM were already observed earlier~\cite{hoppe2004}. Whether or not intercalation occurs can be influenced by the size of the fullerene acceptor relative to the spacing of neighbouring side chains along the polymer backbone. In blends where intercalation occurs in combination with PC$_{70}$BM, bis-PC$_{70}$BM with two instead of one sidechain can---depending on the distribution of the donor side chains---already prevent the intercalation~\cite{cates2009}. If intercalation does not occur, an increased distance between adjacent polymer chains can be observed by x-ray diffraction. This has a direct effect on the optimum donor--acceptor ratio: using the thiophene derivative pBTTT, solar cell efficiencies were compared. The intercalated pBTTT:PC$_{70}$BM blend was clearly better with a 1:4 than with a 1:1 ratio. In contrast, the blend with bis-PC$_{70}$BM was somewhat better in 1:1 as compared to the 1:4 donor--acceptor ratio. However, the role of intercalation and the conditions under which it might be beneficial for the solar cell performance is not yet clarified.

A revived interest in solution-processed bilayer solar cell becomes evident lately. The restrictions of bilayers in disordered films for photovoltaic applications due to the limited exciton diffusion seem to be diminished as higher quality materials are synthesised nowadays. Solar cells of above 3~\% efficiency were solution-processed from P3HT/PCBM planar heterojunction solar cells~\cite{ayzner2009}, and also chemical fixing of layers was considered~\cite{hoven2010a}. More work, however, is needed to reevaluate the potential of solution-processed bilayer solar cells.

Knowing the requirements for optimum exciton separation and charge transport, the  challenge is to control the phase segregation accordingly by direct means. This issue is addressed in Section~\ref{sec:config}.

\subsection{Instrinsic lifetime of organic solar cells}

The lifetime of organic solar cells---in other words the stability of the solar cell output performance against influences such as exposure to air and humidity---is a very important topic for the commercialisation of photovoltaic devices. Usually, encapsulated solar cell modules have to withstand certified test procedures such as the damp heat test, an exposure to moist air at 85\degree{}C at 85~\% relative humidity. The corresponding extrinsic stability, referring to encapsulated devices, is mainly determine by the quality of the encapsulation layers~\cite{krebs2005}. It was shown by the company Konarka, that encapsulated photovoltaic devices were able to show a high performance on a roof-top for almost 2 years. 

For an intrinsic lifetime of organic photovoltaic devices, materials and electrodes need to be stable even without any encapsulation. In order to find an optimum performance--cost relation in terms of stability, also the intrinsic stability needs to be enhanced, and has therefore recently become a focus of research efforts.

Indeed, the fundamental degradation mechanisms in organic semiconductors under exposure to air and humidity have been understood only in part. The degradation of organic solar cells, consisting of at least two active materials, is even more complex. The usual multilayer structure of a bulk heterojunction solar cell consists of transparent anode (usually indium tin oxide), a polymeric interlayer (usually PEDOT:PSS), the photoactive blend (often the  P3HT:PCBM composite) and a cathode (e.g., LiF/Aluminium). Reactions with oxygen and water in the volume~\cite{seemann2009,schafferhans2010} and at the interfaces of these layers~\cite{reese2008} have to be considered. The oxygen induced mechanisms can be reversible~\cite{seemann2009}, such as hole doping by oxygen~\cite{meijer2003a,liao2008,schafferhans2008}, as well as irreversible~\cite{kondakov2005}.

It was found that oxidative degradation of conjugated polymers is connected to the photoexcited states, and that oxygen as well as humid air assist in this process. In solution, polymer degradation takes place via the photosensitised generation of singlet oxygen, which then reacts with the polymer chains~\cite{holdcroft1991}. This reaction induces polymerchain breaking und leads to aldehydes, aromatic and aliphatic ketones and carbonylic structures as products~\cite{neugebauer2000}. 

For thin films, it was reported that molecular oxygen leads to the formation of weak charge transfer complexes (CTC) with polythiophenes~\cite{abdou1997}. According to this study, polymers with low ionisation potential and amorphous structure were most susceptible to oxygen diffusion into the bulk, which is the prerequisite for the CTC formation. Indeed, regiorandom P3HT degrades significantly faster than its regioregular counterpart, which is probably due to the higher singlet oxygen generation yield. The interaction with oxygen leads to doping~\cite{liao2008,schafferhans2010}, , which was also investigated theoretically by band-structure calculations~\cite{lu2007}. Under additional photoexcitation, excitation quenching was observed~\cite{luer2004}. The fluorescence quenching kinetics consist of a fast, reversible decay component, as well as a slow component. The latter is partly reversible, which is assigned to the above-mentioned formation of charge-transfer complexes between excited singlet states of polythiophene and oxygen.
Also, deep traps are formed in polymers~\cite{schafferhans2008} as well as small molecules~\cite{kondakov2005}. These effects result in a higher conductivity due to doping~\cite{meijer2003a,schafferhans2010} and a decrease of the charge carrier mobility~\cite{abdou1997,schafferhans2008}. The role of water in the degradation process is also not completely clear yet. In the presence of water, the P3HT photooxidation rate is increased. If no oxygen is present, water does not seem to play a role~\cite{luer2004}. 

In the case of the fullerene C$_{60}$, oxygen also causes decreased electron mobilities~\cite{taponnier2005,konenkamp1999} and increased trap densities~\cite{matsushima2007}, as found by investigations of field effect transistors. Little is known of other acceptor materials.

In donor--acceptor blends, even more degradation paths exist due to the more complex interplay of components. Efficiency losses due to light~\cite{reese2008,schafferhans2010}, oxygen~\cite{seemann2009,neugebauer2000} and water~\cite{norrman2009} are reported. The details of the degradation processes, however, are still not completely understood. On the one hand, the stability of a PPV donor seemed to be enhanced due to the shorter lifetime of the primary excited state~\cite{neugebauer2000} due to the ultrafast electron transfer to the fullerene acceptor, thus decreasing the reactivity against oxygen. On the other hand, a higher triplet generation---populated via the donor--acceptor polaron pair state---can lead to a more efficient generation of singlet oxygen. For PPV:PCBM bulk heterojunction solar cells, photooxidation was found to mainly reduce the short circuit current, which was explained by a reduction of the charge carrier mobility (measured in transistor configuration) due to trapping~\cite{pacios2006}. Polymer bleaching was found to be too small to account for the losses, and recombination was actually reported to decrease. Recently, two different degradation conditions were investigated in P3HT:PCBM solar cells: in the dark and under simultaneous illumination. Exposure of the solar cells to synthetic air in the dark results in a loss of Jsc of about 60~\% within 120~h. In contrast, bias light during oxygen exposure results in an acceleration of the degradation and a loss of all solar cell parameters. The doping resulting from the exposure to molecular oxygen is responsible for the loss in short circuit current, as shown by macroscopic device simulations. The charge carrier mobility in the blend was shown to decrease only slightly, in contrast to pure P3HT~\cite{schafferhans2008} Although a decreased mobility may result in a loss of the fill factor, as shown by macroscopic simulations, the experimentally observed mobility decrease is too small to be the origin of the FF drop in the case of photodegradation. Oxygen induced photodegradation was found to also result in an increase of the density of deeper traps, which could be the origin of the decreased fill factor and open circuit voltage~\cite{schafferhans2010}.

Another important issue is the stability of the morphological donor--acceptor phase separation~\cite{jorgensen2008}. As many processes, described in the previous sections, depend on the molecular order as well as the spatial dimensions of the polymer--fullerene demixing, a change of these parameters with time can have an impact on the performance. Although organic solar cells do show a high stability when being properly encapsulated, as outlined above, there are studies which point out changes of the donor--acceptor microstructure with temperature and time. In solar cells based on P3HT or MDMO-PPV blended with PCBM, the annealing  at 110\degree{}C was reported to lead to the formation of PCBM clusters, in addition increasing the order of the P3HT phase in the respective cells~\cite{bertho2008}. Indeed, pure PCBM had previously been reported to have a tendency to crystallise~\cite{yang2004,klimov2006}. Up to a certain degree, the demixing of donor and acceptor has a positive impact on the solar cell efficiency, until the trade-off between exciton dissociation and charge transport is optimum. Further phase separation, observed on longer times scales, leads to a spatial donor--acceptor distance exceeding the diffusion length of  photogenerated excitons---which results in losses of the photocurrent~\cite{schuller2004}. Another study reported on a rather stable performance of annealed P3HT:PCBM solar cells~\cite{paci2008a}, attributing the loss in efficiency to oxidation of the Ca cathode due to residual water in the PEDOT:PSS interlayer. Nevertheless, solar cells based on a PPV derivative donor with a higher glass transition temperature of almost 140\degree{}C---as compared to 50\degree{}C for MDMO-PPV and even less for P3HT---were found to inhibit a higher thermal stability concerning the donor--PCBM phase separation~\cite{bertho2008}. Also by changing the functionalisation of P3HT it was possible to improve the thermal stability of the donor--acceptor blend without increasing the glass transition temperature~\cite{campo2009}. These two approaches may be the keys to improve the morphological long-term stability. 

Although the very low processing cost to mass production techniques offer a good energy balance despite the limited intrinsic stability of the organic compounds used today for organic solar cells, a deeper understanding and newly designed materials are needed to be competetive in the market.

\section{Novel concepts for higher efficiencies} \label{sec:concepts}

In this section, the factors limiting state-of-the-art organic solar cells will be summarised on basis of the current knowledge about the elementary processes of photogeneration in these devices (Section~\ref{sec:limits}). In Section~\ref{sec:eff}, the fundamental limits of energy conversion in solar cells will be discussed. Corresponding optimisation routes as well as novel concepts for device configurations addressing these limitations will be discussed subsequently, such as material development (Section~\ref{sec:material}), multijunction solar cells (Section~\ref{sec:multijunction}), optical approaches to enhance light absorption (Section~\ref{sec:optical}) and advanced device configurations (Section~\ref{sec:config}). These different approaches show the high potential for performance improvements of organic photovoltaics by research and development.

\subsection{What limits state-of-the-art organic solar cells?} \label{sec:limits}

\begin{figure}[bt]
	\centering
	\includegraphics[height=8cm]{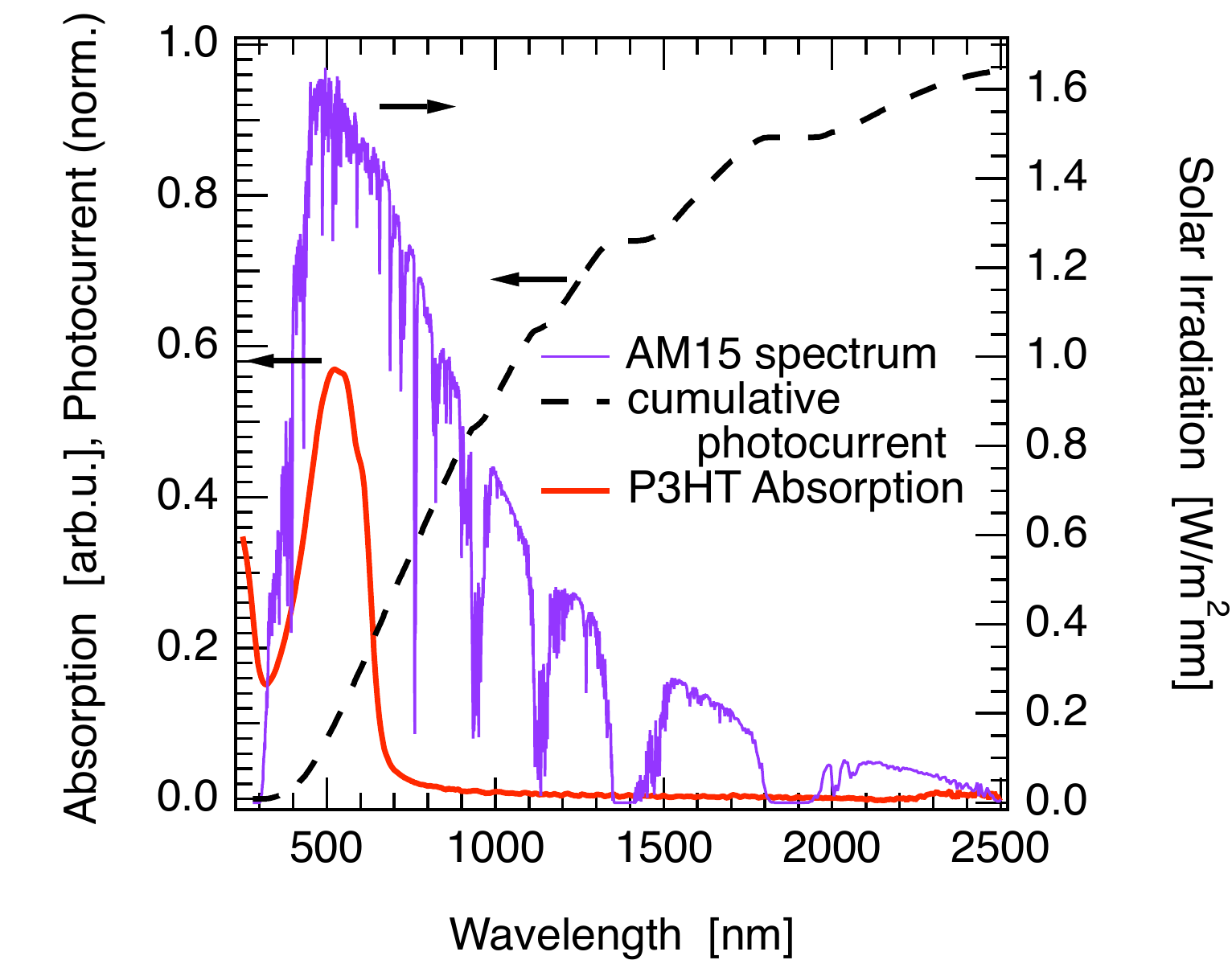}
	\caption{AM1.5 solar spectrum (Direct + Circumsolar, after ASTMG173), and the resulting cumulative, normalised photocurrent, which corresponds to the conversion of all photons below the given wavelength. The absorption spectrum of the commonly used absorber material for bulk heterojunction solar cells, poly(3-hexylthiophene) (P3HT), is included for comparison. Improving the absorption spectrum in organic solar cells is a major optimisation route.}. %
	\label{fig:am15}
\end{figure}

Considering that the energy conversion by organic solar cells relies on several steps from light absorption to charge extraction---discussed in detail in Section~\ref{sec:device}---it is evident that not all of these mechanisms have a high efficiency. Thus, the processes limiting the device performance are listed below, and approaches for avoiding the limitation or improving the process are given as well.

\begin{description}
	\item[Light Absorption] (Section~\ref{sec:exciton}) Organic semiconductors exhibit high absorption coefficients. Limiting is mainly the typically high bandgap and the narrow absorption width of few hundred ~nm, which is lower as compared to inorganic semiconductors. Figure~\ref{fig:am15} shows that an organic solar cell with an absorption edge at 650~nm, typical for the common P3HT absorber material, can generate up to 20~\% of the maximum available photocurrent, whereas silicon and Cu(In,Ga)Se$_2$ solar cells achieve up to 60~\%. Low bandgap organic semiconductors have already been synthesised, but yield lower open circuit voltages as single junction cells.\\
	Approaches for potential improvements: optical concepts (Section~\ref{sec:optical}), multijunction solar cells (Section~\ref{sec:multijunction}), novel materials with lower bandgap and wider absorption range (Section~\ref{sec:material}).
	
	\item[Exciton Dissociation] (Section~\ref{sec:exciton}) The dissociation of excitons is ultrafast and of almost 100~\% efficiency, if (and only if) the diffusion length is sufficient to reach the interface. Nevertheless, the generation of a polaron pair from an exciton is only possible when energy can be gained for the charge carriers. This energy is lost for the open circuit voltage.\\
	Approaches for potential improvements:  novel organic materials with higher dielectric constant to reduce the exciton binding energy (Section~\ref{sec:material}), and advanced device configurations for an optimum phase separation of the donor--acceptor domains (Section~\ref{sec:config}).
	
	\item[Polaron Pair Dissociation] (Section~\ref{sec:pp-diss}) The yield of polaron pair separation is not optimum. The fill factor and the short circuit current are both influenced by this process. Also, the energy of the charge transfer state determines the maximum open circuit voltage, which is reduced to its experimental value by recombination.\\
	Approaches for potential improvements: novel materials with higher dielectric constant (Section~\ref{sec:material}), optimum phase separation and/or nanocrystallinity of the donor and acceptor domains (Section~\ref{sec:config}).
	
	\item[Charge Transport and Recombination] (Section~\ref{sec:transport} and~\ref{sec:recombination}) The charge mobility limits the power conversion efficiency in bulk heterojunction solar cells to a certain degree (Sections~\ref{sec:transport} and~\ref{sec:eff}). As the recombination rate at short circuit current is rather low, the extraction depth is sufficient for state-of-the-art devices, but is limited for devices further beyond 200~nm thickness. At voltages approaching the open circuit voltage, charge recombination is also critical, limiting the maximum open circuit voltage.\\
	Approaches for potential improvements: novel materials with higher charge carrier mobility (Section~\ref{sec:material}) for improved transport. Unfortunately, in the regime of Langevin recombination, the loss rate increases linearly with mobility. Thus, such efforts need to be combined with, e.g., nanostructed material phases to reduce the recombination probability (Section~\ref{sec:config}).
	
	\item[Charge Extraction] (Section~\ref{sec:extraction}) The charge extraction in bulk heterojunction solar cells is more limited in comparison to bilayer or p--i--n solar cells. However, in general terms the low but sufficient surface recombination velocity ensures an efficient charge extraction in state-of-the-art solar cells.\\
	Approaches for potential improvements: it is unclear wether or not the charge extraction in bulk heterojunctions can be improved further. If yes, then by using a more advanced device layouts, such as donor/donor--acceptor/acceptor configurations or blocking layers.

\end{description}

In all cases, more research is needed on the fundamental processes governing energy conversion by organic solar cells, in order to find the most promising optimisation routes.

\subsection{Efficiency estimates} \label{sec:eff}

\begin{figure}[bt]
	\centering
	\includegraphics[height=7cm]{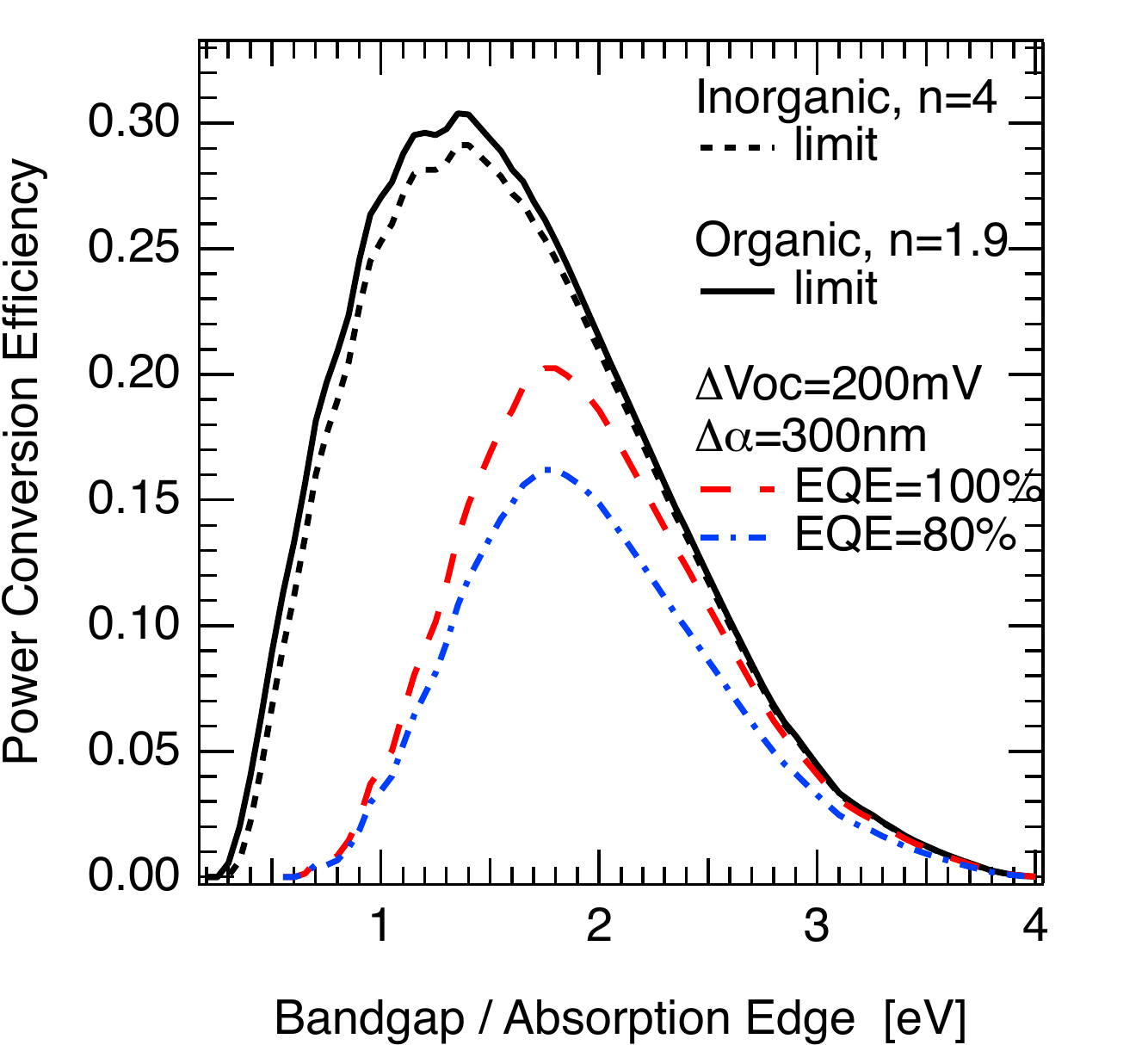}%
	\caption{Efficiency limit for solar cells after the calculations of Shockley and Queisser~\cite{shockley1961}. Different cases are compared: the limit for inorganic semiconductors (black dashed line) and organic semiconductors (black solid). the latter, however, usually have a limited absorption width (here, 300~nm) and a loss in the open circuit voltage due to charge separation (here, 0.2V). The cases of 100~\% (red long-dashed line) and 80~\% (blue dash-dot) quantum efficiency are compared. The donor polymer P3HT, the absorption range of which was shown in Figure~\ref{fig:am15}, has an optical bandgap of 1.85~eV. %
	\label{fig:sq}}
\end{figure}


The well-known detailed balance calculation for inorganic single gap solar cells after Shockley and Queisser gives a theoretical maximum of about 30~\% power conversion efficiency~\cite{shockley1961}. The upper limit for organic solar cells is in principle slightly higher due to their lower refractive index, as shown in Figure~\ref{fig:sq}. However, despite the favourable absorption coefficient in organic materials, the absorption width is usually much lower as compared to their inorganic couterparts. The Shockley--Queisser limit for two exemplary absorption widths of 300~nm is shown in Figure~\ref{fig:sq}. Interestingly, not only the maximum efficiency is lowered, but also the optimum absorption edge at which this maximum is reached. As in organic solar cells the light is usually absorbed in one material, but the respective charges transported in two, absorption gap and effective transport gap do not coincide anymore. In order to illustrate the loss of open circuit voltage $\Delta V_{oc}$, corresponding to the loss in energy due to exciton and polaron pair dissociation, we have assumed here a value of 0.2~V. In these calculations, quantum efficiencies of 80 and 100~\% were assumed, respectively. Disregarding advanced concepts for increasing the power conversion efficiency at the moment, these calculations allow to draw some basic conclusions. First, particularly the absorption widths of organic materials should by increased, bearing in mind that a lower bandgap material also lowers the open circuit voltage. Second, the reduction of the open circuit voltage by dissociation of excitons and polaron pairs should be minimised, for instance by increasing the dielectric constant or by choosing a favourable energy offset between donor and acceptor. It is unclear by how much $\Delta V_{oc}$ can be reduced, but probably not below 0.1V~\cite{veldman2009}. However, even with the constraints of open circuit voltage loss and narrow absorption width, above 15~\% power conversion efficiency are predicted, as shown in Figure~\ref{fig:sq}.


\begin{figure}[bt]
	\centering
	\subfigure[]{\includegraphics[height=8cm]{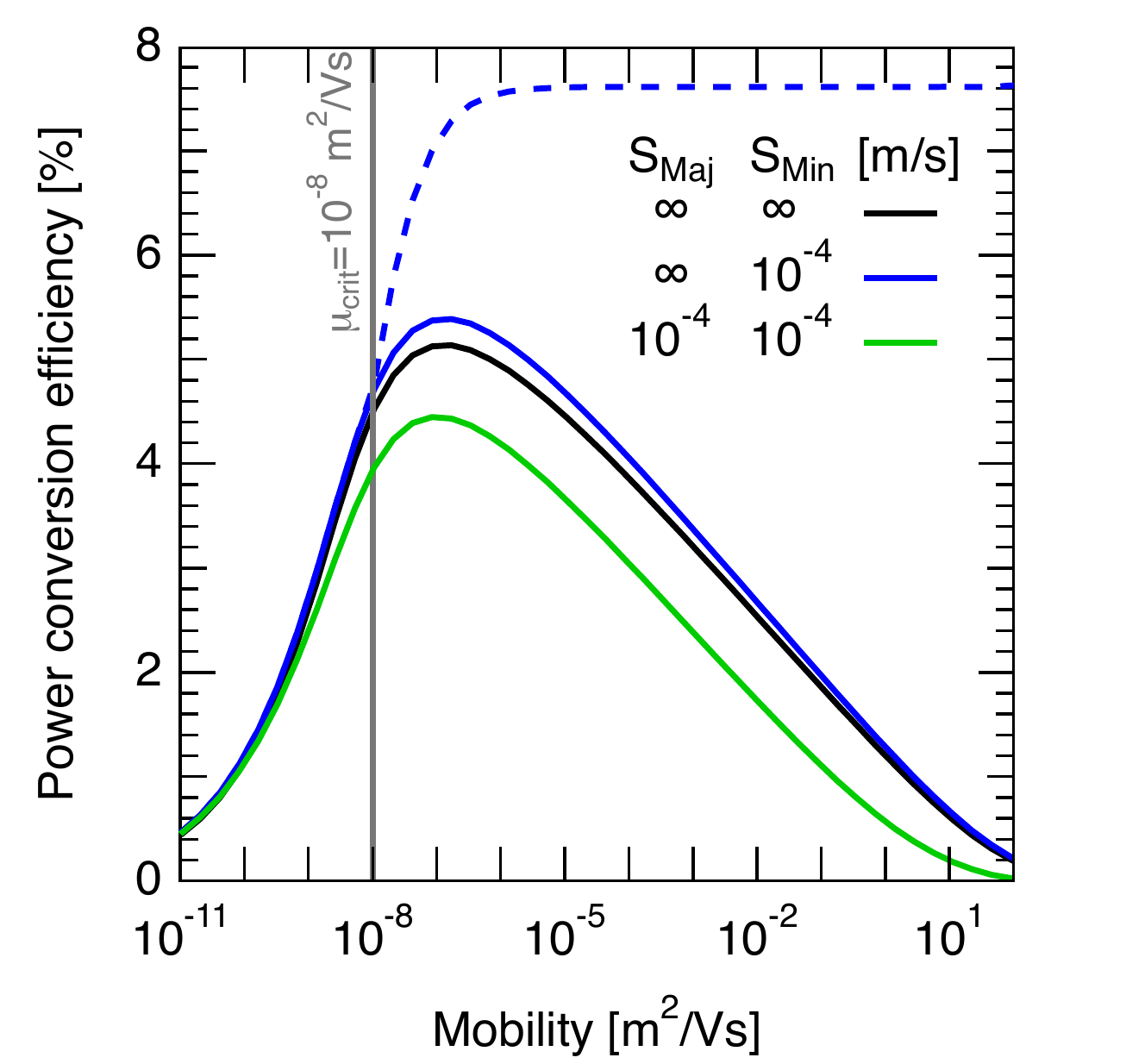}} \quad
	\subfigure[]{\includegraphics[height=7.8cm]{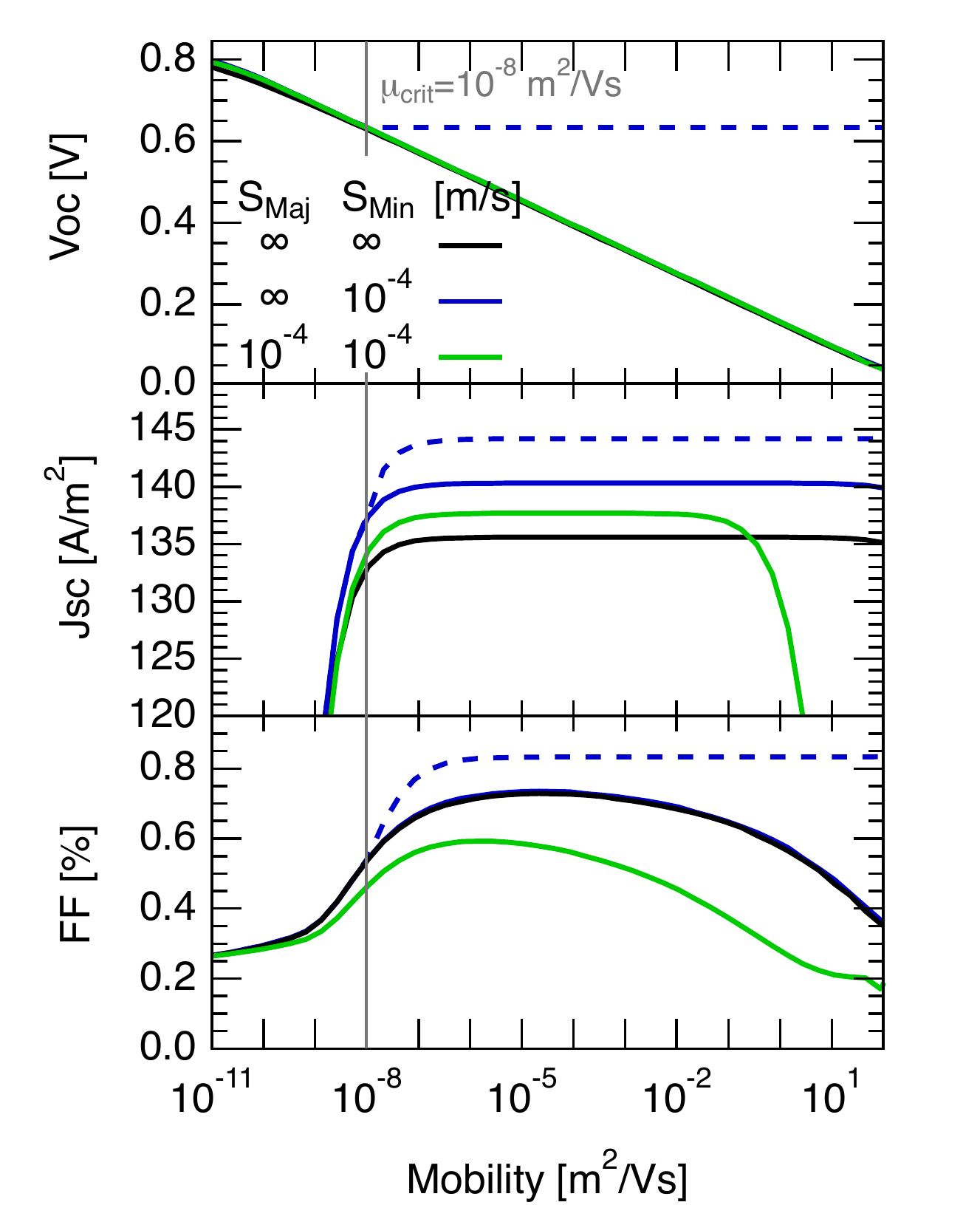}}
	\caption{Simulated performance of a bulk heterojunction solar cell in dependence on the electron and hole mobility~\cite{wagenpfahl2010a}. (a) The power conversion efficiency shows a distinct maximum at a mobility of $10^{-7}$~m$^2/$Vs for the given simulation parameters, if a reduced Langevin recombination is assumed for all mobilities (solid lines). Shown is a set of three solid lines for different combinations of surface recombination velocities of majority $S_{maj}$ and minority carriers $S_{min}$ (see text). However, Langevin recombination occurs only for low mobility materials. Therefore, in the simulation a critical mobility $\mu_{crit}= 10^{-8}$~m$^2/$Vs was defined additionally, above which a fixed recombination rate is used. In this case, the power conversion efficiency vs.\ mobility does not decrease at high mobilities, but saturates. (b) The solar cell parameters open circuit voltage, fill factor, and short ciruit current vs.\ mobility for the simulation parameters used in (a). The vertical lines indicate the critical mobility defined above. Both Figures are adapted from Ref.~\cite{wagenpfahl2010a} with permission; (c) 2010 IEEE.%
	\label{fig:musim}}
\end{figure}

The Shockley--Queisser limit~\cite{shockley1961} is an upper limit for the power conversion efficiency, but is by far not reached by today's best organic solar cells. A major part of the discrepancy is due to charge transport and recombination, as will be outlined in the following. 

Although the disordered organic semiconductors considered here exhibit charge carrier mobilities which are orders of magnitude lower than for inorganic crystals, the resulting solar cells show adequate mobilities. As discussed in the previous sections, the mobility has direct influence on several processes important for efficient organic solar cells: polaron pair dissociation, charge transport, and charge extraction---for instance by surface recombination. The band bending, responsible for a major part of the difference between polaron pair energy and open circuit voltage, depends on the charge carrier mobility. Thus, the important question is, how much does a weak charge transport limit the performance of organic solar cells?

This issue has been addressed by several publications for bulk heterojunction solar cells~\cite{mandoc2007,kirchartz2009,deibel2008a,wagenpfahl2010a} by using macroscopic simulations. A typical result is shown in Figure~\ref{fig:musim}(a). The simulated power conversion efficiency is shown in dependence of charge carrier mobility, here assumed to be balanced for electrons and holes. The corresponding solar cell parameters short circuit current, open circuit voltage, and fill factor are shown in Figure~\ref{fig:musim}(b). Due to polaron pair dissociation, the short circuit current increases with mobility in accordance with the Braun--Onsager Model until the separation yield is 100~\%. The open circuit voltage, unfortunately, decreases steadily if the Langevin recombination model is assumed for all mobilities. The power conversion efficiency shows a maximum at rather low mobilities, which are just somewhat higher than the ones determined for typically used materials~\cite{tuladhar2005,baumann2008}. In Figure~\ref{fig:musim}(a), an increase of about 20~\% in efficiency could be gained by raising the mobility from a typical experimental value of $10^{-8}$~m$^2/$Vs to the one yielding the maximum efficiency, here between $10^{-7}$ and $10^{-6}$~m$^2/$Vs. The latter of course depends on the simulation parameters. Thus, the performance of state-of-the-art bulk heterojunction solar cells seems to be limited to some extent by the charge carrier mobility.

This, however, is unrealistic as Langevin recombination is also found for low mobility materials. Therefore, we assume for comparison that above a critical mobility, the recombination rate is not defined by the rate of the two recombination partners finding each other (as done by Langevin), but is considered to be constant. This leads to higher internal steady-state concentrations of charge carriers which increase the open circuit voltage (compare Eqn.~(\ref{eqn:voc-cheyns})). If, in addition, a finite minority recombination velocity is assumed, which prevents charges to be extracted at the wrong contact, a  higher and indeed saturated open circuit voltage is seen for high mobilities above the critical value (i.e. in the non-Langevin bulk recombination regime assumed here). Here, minority (majority) charge means electrons (holes) at the anode (cathode) and holes (electrons) at the cathode (anode). In contrast, if---although unrealistic---an infinite minority surface recombination is assumed, the minority charges are extracted at both electrodes, leading to a reduced minority charge carrier density. Accordingly, the band bending is increased and the internal electric field enhanced. This leads to a reduction of the quasi-Fermi level splitting---thus, the open circuit voltage would decrease if the  surface recombination velocities were infinite~\cite{kirchartz2009,wagenpfahl2010a}.

Consequently, the power conversion efficiency in organic solar cells is only partly limited by the charge carrier mobility. More severe is the impact of recombination by both, bulk and surface losses.


\begin{figure}[bt]
	\centering
	\includegraphics[height=7cm]{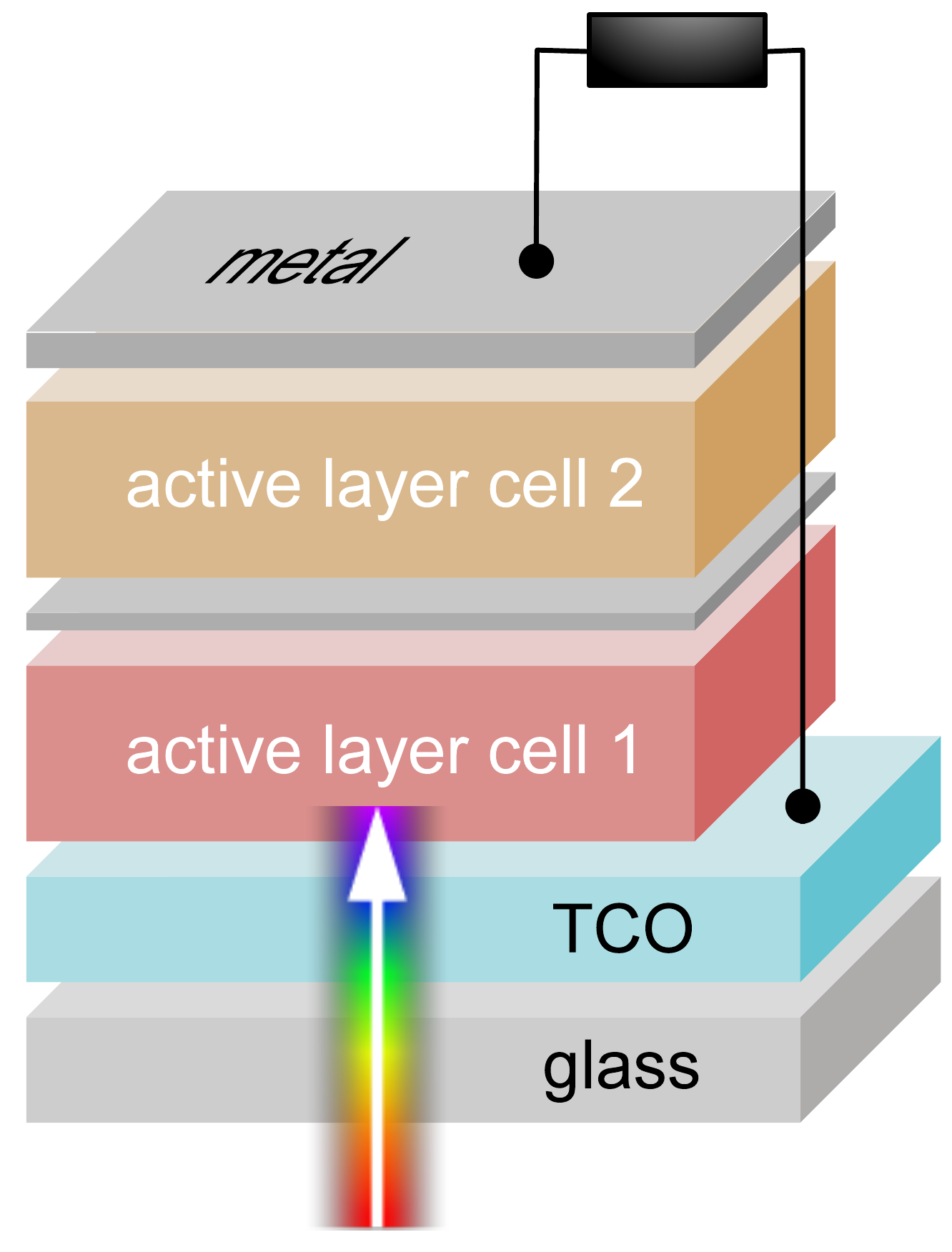}%
	\caption{Schematic device configuration of an organic tandem solar cell, where two subcells are connected in series. The incident light first meets cell 1, which we therefore call top cell as it faces the sun. Ideally, the high energy light (for instance blue) is absorbed in cell 1. The low energy light (red, infrared) is transmitted through cell 1 and can be absorbed by cell 2. The voltages of both subcells ideally add up, whereas the photocurrent has to be as balanced as possible.  %
	\label{fig:tandem}}
\end{figure}

Even higher efficiencies as compared to the single junction devices described above can in principal be achieved by multijunction concepts, the simplest instance of which is a tandem solar cell. Multijunctions are promising as the absorption regimes of the subcells can complement one another, for instance by a combination of low-bandgap and high-bandgap polymers. Ideally, a tandem solar cell made of a series connection of two subcells, illustrated in Figure~\ref{fig:tandem}, works as follows. Both subcells generate their own photocurrent by absorbing light and generating charges (as described for single layer cells before), each having its own open circuit voltage and photocurrent. Of course, as the two cells are connected in series, they influence each other. The photogenerated holes of cell 1---or top cell, as it faces the light source---are extracted by the ITO. The electrons have to recombine with photogenerated holes from cell 2 (the bottom cell): that is what the intermediate recombination layer is for. If the photocurrent of top and bottom cell is initially unbalanced, the electric field is redistributed, such that the photocurrent becomes balanced---at a lower value, approximately determined by the worse of the two cells, at the same time reducing the open circuit voltage of the corresponding sub cell. The open circuit voltage of the tandem stack is determined by the sum of both subcell's open circuit voltages. Thus, for an ideal tandem solar cell, complementary absorption ranges, and balanced photocurrents are needed. The experimental progress with multijunction solar cells will be discussed in Section~\ref{sec:multijunction}.

Dennler et al.~\cite{dennler2008} presented optical simulations by considering thin film interference effects. They verify that the 6.5~\% efficiency record tandem solar cell by Kim et al.~\cite{kim2007b} indeed used an advantageous thickness combination for both subcells (approx.\ 180 and 130~nm) in order to optimise the photocurrent by balancing it. They continued to consider other thickness combinations with balanced photocurrent, and found higher charge carrier generation rates by increasing the thickness of both cells. Finally, Dennler et al.\ claimed that by going as far as 565~nm for the P3HT based cell, and 225~nm for the PCPDTBT based sub cell, the efficiency can be as high as 9~\% power conversion efficiency. However, bear in mind that the authors performed only optical, not electrical simulations. Thicknesses of several hundred nanometres would certainly lead to very high losses: the internal field decreases for thicker devices, leading to weaker charge separation and extraction as well as higher recombination.

\begin{figure}[bt]
	\centering
	\subfigure{\includegraphics[height=7cm]{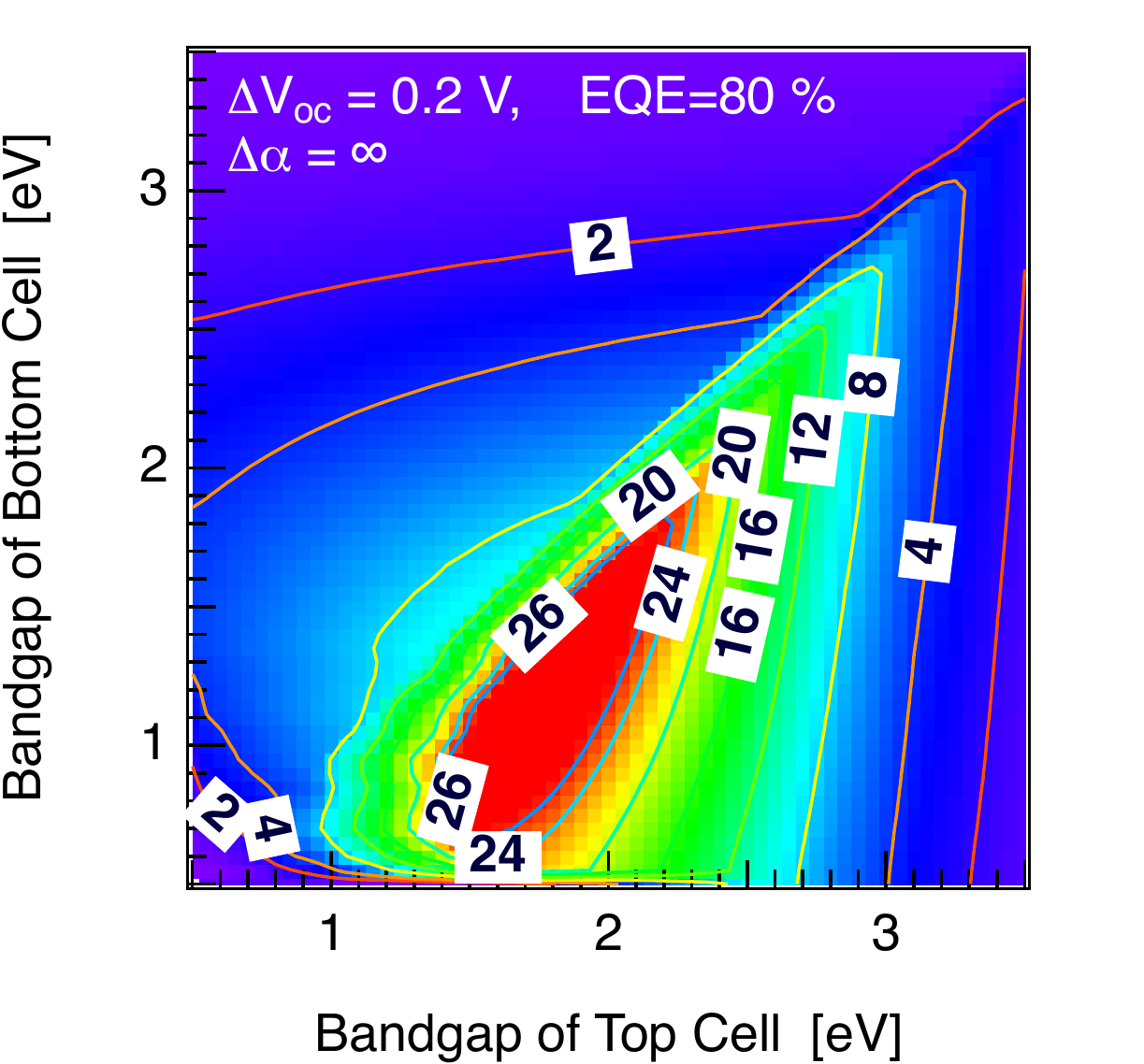}}\quad
	\subfigure{\includegraphics[height=7cm]{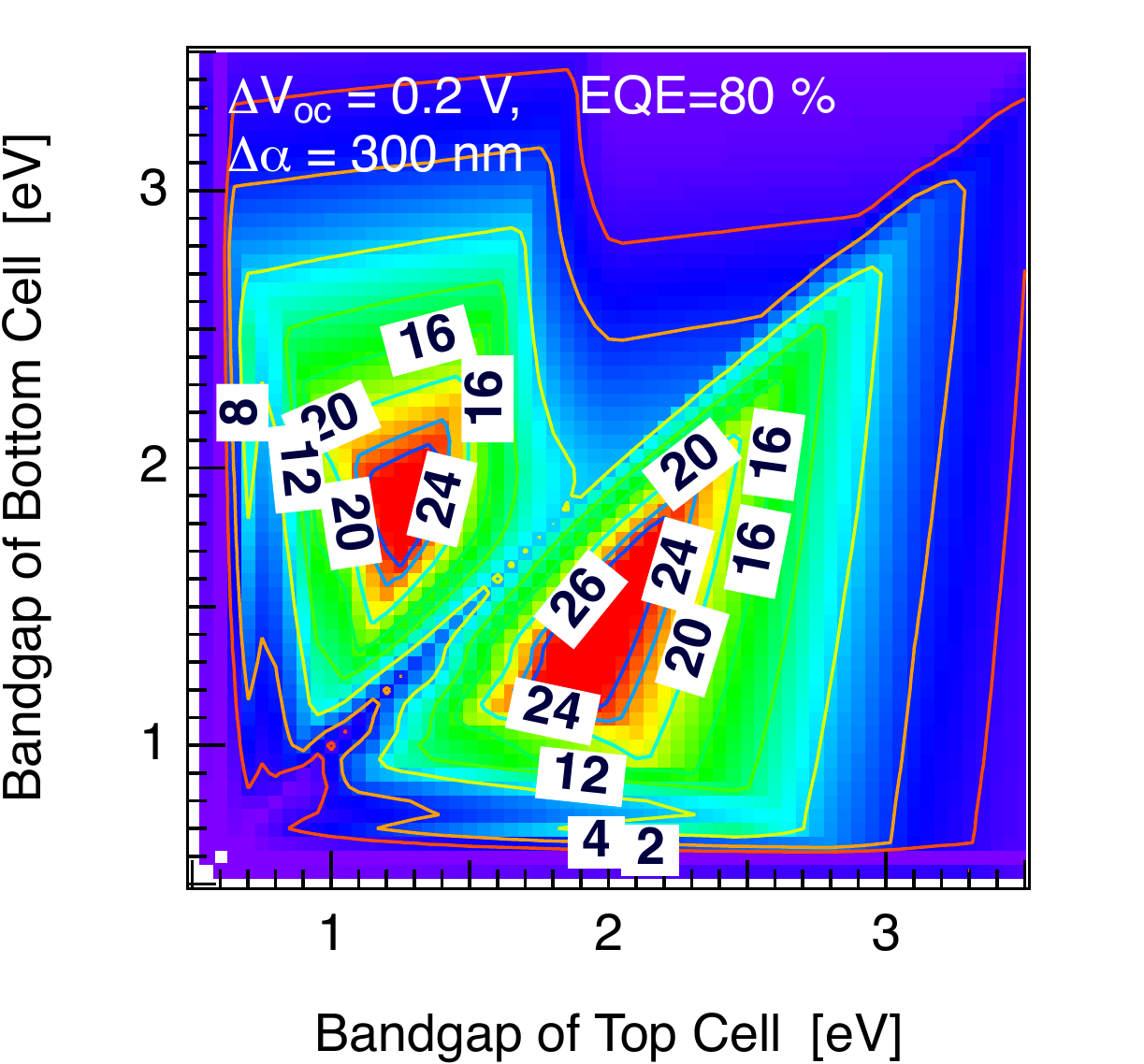}}%
	\caption{Efficiency limit of organic tandem solar cells after the calculations of Shockley and Queisser~\cite{shockley1961}. The open circuit voltage was reduced by 0.2V in order to account for losses in the solar cell, the quantum efficiency was set to 80~\%. The response of a corresponding single junction is shown in Figure~\ref{fig:sq}. (a) wide absorption bands (here: infinite), with results similar to Ref.~\cite{dennler2009review}. (b) Organic semiconductors usually have a limited absorption width, which has a strong impact on the efficiency calculations. Here, again 300~nm absorption width were assumed.%
	\label{fig:sq-tandem}}
\end{figure}

We also consider the maximum efficiency of tandem solar cells within the framework of Shockley and Queisser~\cite{shockley1961}, in analogy to treatment above for single junctions. The calculations are shown in Figure~\ref{fig:sq-tandem} for (a) infinite and (b) finite and thus more realistic absorption width of the active layers. The representation is asymmetric, as a low bandgap material as first cell would absorb a larger fraction of the high energy photons, which then were not available for the bottom cell with high bandgap material. This behaviour is particularly problematic for wide absorption bands. Therefore, the low bandgap materials are placed below the high bandgap material in tandem configurations. In case of finite absorption width, under the same conditions as for the single cell estimated above (external quantum efficiency of 80~\%, voltage drop of 0.2~V, absorption width 300~nm), a maximum of 26~\% of the tandem stack can in principal be achieved. Practically, this value can only be approached and not achieved, as discussed in Section~\ref{sec:multijunction}.

\subsection{Material engineering} \label{sec:material}

The recent records in power conversion efficiency of organic solar cells, 3.5~\% for modules (208~cm$^2$ area) and 7.9~\% for small devices~\cite{green2010review}, were not achieved by using the well-researched P3HT:PCBM compound system. Instead, novel materials yielding almost 50~\% higher open circuit voltage and still high photocurrents were used. Indeed, several aspects of organic solar cells can potentially be optimised on the molecular level by synthetic design. Properties looked for are the ability to self-organise---enhancing order and thus charge transport---and an absorption spectrum as wide as possible, being one of the major limiting factors yet. Nowadays, in most cases only the donor material absorbs light efficiently; an absorbing acceptor has a large potential for increasing the photocurrent. Additionally, by a variation of the relative energy levels of donor and acceptor material, thus optimising the energetic position of the charge transfer complexes, the energy loss due to the electron transfer can be minimised.

An alternative approach is to use low-bandgap polymers. These have an enhanced absorption in the red part of the spectrum, in order to shift the absorption gap closer to the values predicted after the Shockley--Queisser framework discussed above (Section~\ref{sec:eff} and in particular Figure~\ref{fig:sq}). Recently, low bandgap novel donor polymers with an absorption up to 1300~nm have been presented. Devices derived from blends with PCBM, however, showed that the open circuit voltage and  the photocurrent are too low to yield a good solar cell~\cite{gong2009}. Many of the novel copolymer low bandgap systems show either an overall lower charge generation efficiency---such as the still promising copolymer donor poly[2,6-(4,4-bis- (2-ethylhexyl)-4$H$-cyclopenta[2,1-b;3,4-b']-dithiophene)-\emph{alt}-4,7-(2,1,3-benzothiadiazole)] (PCPDTBT)~\cite{peet2007}---or have absorption bands as narrow as P3HT and thus absorb little in the blue part of the visible spectrum~\cite{voelker2010}. Nevertheless, the efficiency records in the last months~\cite{park2009,green2010review} were also achieved with low-bandgap donor materials. Indeed, recently a trend towards using copolymers can be observed in order to address the issue of narrow absorption width. These conjugated polymers can have low bandgaps resulting from the intrachain coupling between the electron donating and accepting units on each monomer~\cite{kim2009d}. Some of these concepts embed metals~\cite{wong2007} or elemental (inorganic) semiconductors~\cite{morana2010} into the chemical structure, thus showing that further enhancements of  the absorption width as well as the charge generation and transport properties are to be expected. Alternatively (or additionally), nanoparticle sensitisers (Section~\ref{sec:optical}) and the multijunction approach (Section~\ref{sec:multijunction}) can be used to improve the photon harvesting. 

A rather well performing polyfluorene copolymer was presented in 2007~\cite{zhang2005,zhang2008c}, and is based on a promising derivative already published earlier~\cite{svensson2003}. 4.2~\% power conversion efficiency for a blend of  poly(9,9-didecanefluorene-alt-(bis-thienylene) benzothiadiazole) (PF10TB) with PCBM were published~\cite{sloof2007}. The internal quantum efficiency amounted to almost 80~\%, and the open circuit voltage was 1~V. Another class of alternating copolymer donors, based on poly(2,7-carbazole) derivatives~\cite{blouin2008}, was optimised to show even lower photocurrent losses. Recently, a poly[$N$-9''-hepta-decanyl-2,7-carbazole-alt-5,5-(4',7'-di-2-thienyl-2',1',3'-benzothiadiazole) (PCDTBT):PC$_{70}$BM bulk heterojunction solar cell with almost 100~\% internal quantum efficiency was presented~\cite{park2009}. Although this device absorbed only light with wavelengths below 650~nm, power conversion efficiencies of 6~\% could be achieved. A third class of donor copolymers which have recently attracted a lot of interest where the cyclopentadithiophene-based compounds; one of them, PCPDTTBT, was applied as the low-bandgap donor in an efficient tandem solar cell~\cite{kim2007b}. Interesting in these examples of copolymer donor classes is that by using the different electron deficient moieties, the effective gap of the donor material can be tuned, while keeping the HOMO low to yield high open circuit voltages. For donor-acceptor type copolymers, the effectively lowered exciton binding energy leads to a more efficient photogeneration~\cite{clarke2009a}. 

A more classical approach to an effective lowering of the binding energy of bound charges was done by adjusting the side chains of PPV derivatives in order to increase the dielectric constant by almost a factor of two~\cite{breselge2006}. Although the charge separation efficiency was enhanced, the resulting solar cells had efficiencies below 1~\%~\cite{lenes2008a}.

Recently, small molecules based on merocyanines were solution processed with PCBM to bulk heterojunction solar cells, yielding power conversion efficiencies of 1.7~\%~\cite{kronenberg2008}. For evaporated solar cells, subphthalocyanine and subnaphthalocyanine donor materials have shown promising properties and efficiencies of 3~\% in bilayer configuration~\cite{mutolo2006,gommans2007, heremans2009review}.

Other promising donor materials can be found in two recent reviews~\cite{dennler2009review,chen2009review1}.

With respect to efficient electron acceptors, until now the fullerenes~\cite{smalley1999review} have remained the most efficient option as constituent in bulk heterojunction solar cells~\cite{yu1995,hummelen1995,imahori2007}. Acceptors based on conjugated polymers have exceeded the one percent efficiency limit~\cite{mcneill2009review}, but still remain comparably low. For evaporated solar cells, fluorenated sub-phthalocyanines were presented, yielding efficiencies of about 1~\%~\cite{gommans2009a}.

The acceptor material can also contribute to the photocurent if its absorption is high, and ideally complementary to the donor absorption range. A prerequisite is that exciton dissociation by hole transfer to the donor is energetically possible and favourable. For instance, this has been realised by blending a PPV derivative with PC$_{70}$BM~\cite{wienk2003}. 

Enhancing the energetics via tuning the combination of donor and acceptor materials is also a viable option for performance enhancements. Some donor--acceptor combinations are energetically not optimal, since the excess kinetic energy after exciton dissociation dissipates. An optimised donor--acceptor gap thus can reduce the loss in open circuit voltage without sacrificing much photocurrent, as recently shown by use of fullerene derivatives with less acceptor strength relative to the donor materials~\cite{riedel2005,lenes2008,ross2009,ross2009a,liedtke2010}. Ross et al.\cite{ross2009} blended P3HT with an endohedral C$_{80}$-fullerene derivative, with HOMO and LUMO levels shifted by 300~meV upwards as compared to PCBM. The resulting open circuit voltage is accordingly higher by that amount in comparison with P3HT:PCBM reference cells.

\subsection{Multijunction solar cells} \label{sec:multijunction}

In Section~\ref{sec:eff}, tandem solar cells~\cite{hadipour2008} were already introduced as very effective way to increase the power conversion efficiency of organic photovoltaics. Clearly, multijunction devices work best if the subcells have complementary absorption ranges, and their thickness can be optimised by optical modeling in view of current matching, as described in Section~\ref{sec:eff}. In addition to enhancing the effective absorption width, tandem solar cells conserve the photon energy more carefully as single junction cells with broadband absorbers. The voltage at which charge collection occurs is close to the absorbed photon energy absorbed each subcell. Thus, less energy is lost by charge relaxation towards the transport gap in single junction cells.

Recently, Kim et al. presented 6.5~\% efficient organic tandem solar cells~\cite{kim2007b} made of two series connected subcells of polymer:fullerene blends. The anode was indium tin oxide on a glass substrate. The top cell (in terms of incident light) was PCPDTBT:PC$_{60}$BM. The recombination layer consisted of TiO$_\mathrm{x}$ and a special PEDOT:PSS, separating both subcells and connecting them in series. The bottom cell made of a P3HT:PC$_{70}$M blend was connected to the TiO$_\mathrm{x}$/Aluminium electrode. It was argued that the photocurrent of this device configuration was possibly overestimated due to the recombination layer covering the whole device, not only the active area.

Already in 2004, tandem solar cells with around 4~\% efficiency were presented based on small molecular phthalocyanine donors with C$_{60}$ acceptors~\cite{xue2004,rand2004}. Recently, the company Heliatek has boosted this architecture to above 6~\% efficiency~\cite{green2010review}, showing the potential of this technology.

As proof-of-concept, triple~\cite{gilot2007} and even six-fold junctions could be realised by processing from solution. Novel material combinations of low and wide bandgap materials were also tried, and excellent current matching when measuring similarly prepared front and back cells separately was found. The tandem cells yielded efficiencies of almost 5~\%~\cite{gilot2010}. A similar performance was achieved in tandem cells with parallel connection of the subcells. In the so-called three-terminal tandem cells, the short circuit current was doubled, reaching values above 15~mA/cm$^2$ and a power conversion efficieny of 4.8~\%.  An advantage of this device configuration is that the two subcells can also  be characterised independently~\cite{sista2010}. 

Potential drawbacks of multijunction solar cells such as the effect of the angular response~\cite{cheyns2008a,forberich2008} should be considered adequately, although an increased active layer thickness will minimise its influence. Thus, multijunction solar cells are viable options to enhance the effective absorption range in organic photovoltaics. The calculations described in Section~\ref{sec:eff} certainly emphasise the potential for efficiency gains.

\subsection{Optical concepts} \label{sec:optical}

The absorption of light can be enhanced by different concepts. The design of novel photoactive organic semiconductors with enhanced absorption properties was already discussed in Section~\ref{sec:material}.

\begin{figure}[bt]
	\centering
	\includegraphics[width=11cm]{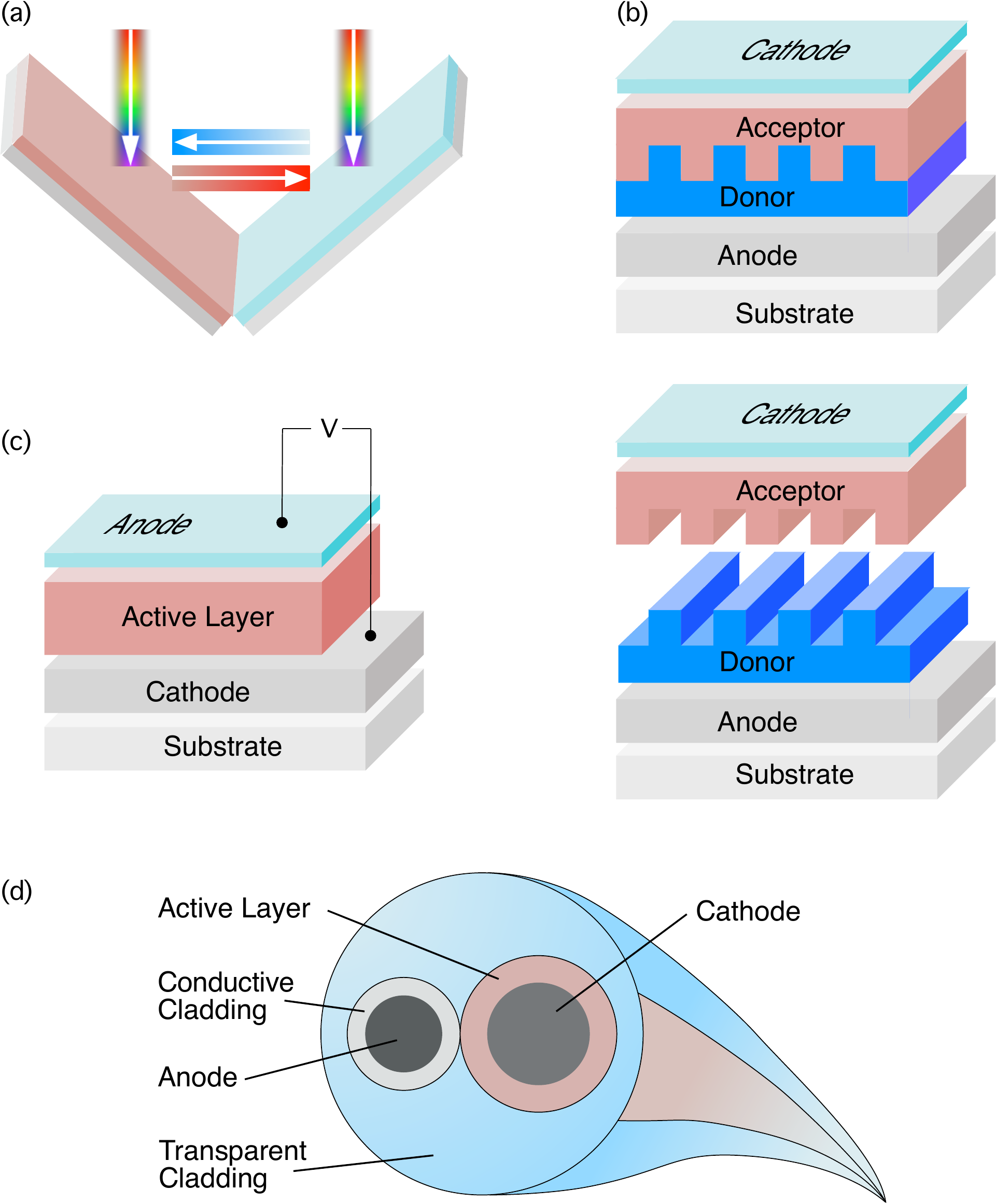}
	\caption{Exemplary optical concepts and advanced device configurations. (a) V shaped solar module arrangement~\cite{tvingstedt2007}, ideally with two photovoltaic cells with complementary absorption ranges facing one another. (b) Nano imprint lithography in order to control the donor--acceptor separation~\cite{cheyns2008b}. The ideal width of the donor trenches corresponds to twice the exciton diffusion length. (c) Inverted solar cell structure, which is allows illumination from either side, and is thus compatible with opaque substrates. (d) Photovoltaic nanowires with a diameter of approximately 0.25~mm~\cite{lee2009a}.} %
	\label{fig:alternative-device-configurations}
\end{figure}

Optical spacers were discussed and used as additional layers to shift the absorption maximum, resulting from thin film intereference effects~\cite{pettersson1999}, into the optimum position within the photoactive layer~\cite{kim2006b,gilot2007b}. If, however, the thickness of the different layers of the solar cell are  chosen optimally, for instance according to the guidelines of optical simulations, no additional gain can be expected from the spacer concept~\cite{andersson2009}. Indeed, in terms of absorption enhancements, spacers can only compensate for imperfect layer configurations. 

An alternative approach for organic solar cell modules might be the V shaped arrangement of the modules for mutual light reflection~\cite{tvingstedt2007}. In this concept, illustrated in Figure~\ref{fig:alternative-device-configurations}(a), solar cell modules are arranged next to each other within a certain angle, so that the incoming sun light is partly absorbed and partly reflected by one module. The other module, while also directly absorbing incoming light, increasing the light path due to the tilting angle, also absorbs the reflected photons partly. This configuration is a good way of optimising the light incoupling for state-of-the-art devices. Once thicker photoactive layers can be considered for organic solar cells due to improved organic semiconductors and better process control, the V shaped arrangement will not increase the absorption any more, unless solar cells with complementary absorption spectra are used.

One way to enhance the absorption within the photoactive layer is often described by the term plasmonics. Nanoscale objects such as metal quantum dots or nanoparticles are brought in contact with or embedded into the active layer in order to increase the photocurrent by enhancing the absorption width or coefficient~\cite{atwater2010review}. Either light scattering~\cite{catchpole2008review} or near-field concentration~\cite{schuller2010review} are applied in order to enhance the absorption witdh or coefficient.  Such concepts have already been tried in conjunction with solution processed~\cite{kim2008a} as well as evaporated~\cite{rand2004} organic solar cells. In another study, PbS nanocrystalline quantum dotes were used as sensitisers of P3HT:PCBM photodiodes, enhancing the absorption spectrum into the near infrared region~\cite{rauch2009}.

The absorption width can also be enhanced by downconversion and upconversion, concepts which---among others---are referred to by the term third generation photovoltaics~\cite{green2006book}. Upconversion by triplet--triplet annihilation on a dopant to generate otherwise high energy singlet excitons on the donor from low energy photons has already been presented~\cite{baluschev2006}, although the yield of such process is still very low. 

A different concept, which might be relevant to bilayer and bulk heterojunction device configurations, is the enhancement of the exciton diffusion length. By sensitising donor polymers with phosphorescent molecules, the diffusion length could be increased from 4 to 9~nm~\cite{rand2009}. If at the same time the phase segregation is increased accordingly, the charge transport of the separated polarons will be more efficient. 

A combination of these different approaches will be able to further enhance the photon-to-electron conversion efficiency of organic solar cells.

\subsection{Advanced device configurations} \label{sec:config}

The influence of the active layer morphology has a profound impact on the device performance, as discussed already in Section~\ref{sec:morphology} about the interplay between contradictory optimum spatial dimensions of the donor--acceptor dimensions for charge pair dissociation and charge transport. A method to control the spatial dimensions could help to adjust the optimum morphology directly. Different approaches have already been presented.

The optimum structure for an organic solar cell is anode, followed by a donor/donor--acceptor/acceptor sequence, and concluded by the cathode. Here, the mixed layer should ideally be ordered into separate donor and acceptor channels normal to the layer surface. The light absorbing channel should have dimensions not exceeding twice the exciton diffusion length for optimum light generation. The subsequent charge transport would not be hindered by spatial disorder due to the separation of donor and acceptor transport paths. The single donor and acceptor layers close to the respective electrodes could act as electron or hole blocking layers, respectively, as already done in evaporated small molecule solar cells.

One method to achieve such a configuration is nanoimprint lithography. The principle is shown in Figure~\ref{fig:alternative-device-configurations}(b). It has already been used to directly pattern the donor polymer P3HT~\cite{cheyns2008b}. Although the width of the trenches was 50~nm, and thus clearly too large for optimum exciton dissociation, the proof-of-concept could be shown. As acceptor, a perylenetetracarboxylic diimide was vacuum-deposited. The performance of the resulting devices was only around 0.1~\%, although an enhancement of the photocurrent due to the increased interface area was observed. In principle, nanoimprint lithography with resolutions below 10~nm was presented~\cite{chou1997}, and could possibly be combined with the recent solution-processed bilayer approaches~\cite{ayzner2009,hoven2010a}.

The commonly used device configuration in organic solar cells is essentially transparent substrate/transparent anode/active layer/metal cathode. In inverted device configurations, anode and cathode are exchanged (see Figure~\ref{fig:alternative-device-configurations}(c)). Instead of an ITO/PEDOT:PSS anode, which might be an unstable combination under harsh conditions~\cite{dejong2000}, for instance silver, ZnO or TiO$_x$ can be deposited as cathode on top of the substrate. The latter two options can even be solution-cast at low temperature, paving the way to solution-only processing of organic solar cells~\cite{white2006,waldauf2006a} The typical cathode, for instance the LiF/Al, which is prone to oxidation, can be replaced by either gold or a (semi)transparent highly conductive PEDOT:PSS/gold grid layer as anode on top. In the latter case, the illumination can also be done from top, thus allowing for opaque substrates if required. Power conversion efficiencies of above 3~\% were achieved~\cite{waldauf2006a,mor2007,ameri2008}.


Instead of using solution-cast, inorganic compounds as replacements for the classically used electrodes, they can also act as electron acceptors within the photoactive layer. Such solar cells, where organic and inorganic components are combined, ideally marrying the advantages of both material classes, are called hybrid solar cells~\cite{huynh2002}. The inorganic compounds investigated for this purpose are usually based on solution-processable nanoparticles. Their conductivity can principally be tuned by doping~\cite{hammer2008}. Although a lot of effort has been put into the optimisation of these devices~\cite{beek2006,goh2007,gunes2008}, efficiencies have reached only up to 3~\%~\cite{skompska2010review}. Recently, the major limiting factors of a P3HT:ZnO hybrid solar cell with 2~\% efficiency have been identified~\cite{oosterhout2009} as a combination of inefficient charge generation as a result of the low ZnO content, a too large-grained phase separation for favourable exciton dissociation, as well as the exciton losses at the electrodes. Thus, similar to organic bulk heterojunctions, a better control of the donor--acceptor phase segregation is required to gain in efficiency.

Finally, an interesting approach was implemented by photovoltaic wires~\cite{lee2009a}. Two parallel wires, one corresponding to the anode, one to the cathode, are enclosed by a transparent cladding. The photoactive blend is coated on top of the surface of one of these wires, encapsulated by a hole transport layer. This final layer is in contact with the second wire, thus being able to have a circuit for charge generation. A simplified device configuration is shown schematically in Figure~\ref{fig:alternative-device-configurations}(d). Efficiencies of around 3~\% were achieved; the area was determined as wire length multiplied by the diameter (equals the cross-sectional area). The feasibility of upscaling of such a device even to the 0.1-1 cm$^2$ range still has to be presented.

\section{Market potential}

The annual photovoltaics module production is currently around 3~G\Wp (gigawatt peak). Despite its impressive growth rate beyond 30~\% per year during the last decade, the fraction of photovoltaics in the electricity generation is still low~\cite{aberle2009review}. Today, crystalline and multicrystalline silicon solar cells dominate the commercial market by far: the market share of silicon-based photovoltaics technology is above 90~\%. 
Although the production cost of silicon based modules has decreased significantly during the last decade, it is still not competitive with other sources of electricity generation without public subsidies~\cite{poortmans2007book}. The prospects of substantially lowering the cost of the classical silicon solar cell technology is limited. Organic solar cells manufactured by mass production techniques such as roll-to-roll printing are a very promising alternative to achieve this goal.

In the following, we will give a brief overview over the processing techniques for the mass production of organic solar cells. Also, the prospects of commercialising organic photovoltaic modules---where low cost has to make up for the potential disadvantages of lower efficiency and lifetime as compared to other thin film technologies---will be considered.

\subsection{Cost}

In the past, the module manufacturing costs, which dominated the system costs, have declined by about 50~\% per decade~\cite{cameron2007}. This trend can only be kept if thin film technologies manufactured by mass production techniques are used. 

The desired cost is usually considered as cost in US-Dollar or Euro, per unit area, over peak power \Wp of the module, also per unit area. The peak power is defined for AM1.5 conditions with a power density of 1000\Wp/m$^2$ at a module temperature of 25\degree{}C. This temperature definition is a weak point of the definition of peak power, as different photovoltaic technologies exhibit different temperature coefficients: the power output of crystalline silicon modules declines as the temperature rises, in contrast to amorphous silicon or organic solar cells. Nevertheless, \Wp implicitly accounts for the efficiency of the device. This is important in as far as it determines the area for achieving a certain output power, as well as the area influencing the cost to manufacture and install the modules. Thus, this figure of merit is widely used, and the temperature coefficient is usually neglected. 

To be commercially competitive, module costs significantly below 1~\euro{}/\Wp need to be achieved. Therefore, thin film technologies with much lower usage of material due to increased absorption properties are called for~\cite{poortmans2007book}, ideally processed by using mass production technologies.

The estimation of component costs for different thin film technologies was performed by Zweibel~\cite{zweibel2007inbook}. The module costs are usually separated into the balance of module (BOM) and balance of system (BOS) costs.

BOM costs include processing equipment, materials and overhead. Costs related to the photoactive semiconductors can be considered separately, and may be termed nonBOM. The BOM costs for glass encapsulated modules including a transparent conductive oxide layer currently are at around 50 US-\$/m$^2$ for small production volumes of 25~M\Wp per year---this corresponds to about 0.8~US-\$/\Wp at 6~\% module efficiency. This could be lowered to 20-40~US-\$/m$^2$ in the future, and presumably even further when using foil as substrate. The nonBOM cost estimates for organic photovoltaics are subject to large uncertainties, as many details about technological requirements and upscaling issues are not known in detail yet . Zweibel calculates 9~US-\$/m$^2$ for organic modules at yearly production volumes of 1~G\Wp with efficiencies of 8~\%, a performance which is not reached today. These figures correspond to about 0.1~US-\$/\Wp. The complete cost of such a module just achieves below 1~US-\$/\Wp (mainly BOM and nonBOM). As compared to other thin film technologies these estimates seem rather conservative, however. Generally, the production volume plays a major role on lowering the module cost, which is due to scaling as well as technical progress---the learning curve. It holds true for material related costs as well as processing.

The BOS costs comprise for instance module installation as well as cables and inverters, and have been considered previously~\cite{zweibel2007inbook,dennler2008inbook}. As the BOS goes beyond the scope of this review, we will not consider it here.

The overall cost of an organic photovoltaic installation depends on the application type: on the roof top a much higher cost is expected than for a mobile application. Dennler et al.~\cite{dennler2009review} calculated a grid-connected residential roof-top installation with 1~k\Wp, under the assumption of 25~years installation lifetime and module lifetime of between 3 and 10~years, also considering the cost incurred by the replacement of degraded modules.  With a BOS cost assumed as 70~\euro{}/m$^2$ and a BOM cost of 50~\euro{}/m$^2$, it was concluded that a low-cost technology such as organic photovoltaics can become competitive at an efficiency of 7~\% and 7 years lifetime. The cost per peak power under these conditions was calculated to about 3~\euro{}/\Wp. This cost estimate is clearly higher than the one by Zweibel stated above. This is mainly due to the different perspective, which lies here on the application on a time frame of 25~years instead module production. 

The anticipated cost of generated electricity was also calculated by Dennler et al.~\cite{dennler2009review} for middle europe (for instance, Germany) with 1000 hours of sun per year. With BOS cost of 70~\euro{}/m$^2$ and a BOM cost of 50~\euro{}/m$^2$ as above, 25~\euro{}~cent/kWh could be achieved with modules of 4~\% efficiency and 5~years lifetime. In order to lower the energy-cost to only 10~cent/kWh, the BOM cost needs either to be reduced to 10~\euro{}/m$^2$ or the efficiency is to be raised to 12~\%. It has been shown in Section~\ref{sec:eff} that organic solar cells have the potential for even higher efficiencies, although the current technology has not reached that stage yet.

Concerning mobile applications, the break-even point for organic solar cells is easier to achieve. If for instance integrated into bags as charger for laptops or mobile phones, lifetimes of only a few years are required. Direct integration into electronic appliances, often designed with usability lifetimes of 3-5 years in mind, also limits the requirements for organic mini modules accordingly~\cite{brabec2005review}.

The currently lower efficiency and lifetime of organic photovoltaics as compared to some of their inorganic counterparts can be compensated by a much lower price. Thus, the module costs have the potential to become comparably small as compared to the system costs. Work is ongoing to make organic photovoltaics technology even more attractive by enhancing performance as well as lifetime, and decreasing the cost by applying the processing technologies discussed in the following section.

\subsection{Processing}

As discussed in the previous section, a mass production of solar cells is desirable to lower the cost. Organic solar cells are highly attractive as they can be processed on large areas in high volumes by roll-to-roll techniques. Those allow to achieve 10000~m$^2$ per hour, which is orders of magnitude faster than the processing of crystalline silicon~\cite{brabec2008review}.


Using roll-to-roll techniques, material can be deposited onto a flexible substrate from the gas phase in a vacuum process as well as the liquid phase by printing. Although roll-to-roll vacuum deposition is used industrially in the packaging industry and for processing organic light-emitting diodes, the vacuum steps lead to a low process speed~\cite{huebler2007inbook}, thus limiting the cost reduction potential. The option of using organic vapour phase deposition techniques~\cite{yang2004b}, which is performed at pressures much closer to 1~atmosphere, might be promising in this respect, but has not been investigated in detail. From today's perspective, liquid state processing seems to be the best choice for high volume roll-to-roll production of organic solar cells. We will therefore focus on solution-based coating and printing techniques.


Solar cell module production usually requires to process several interconnected solar cells on one substrate. In terms of a continuous substrate material from a roll, the width of the substrate will be separated into several stripes. Each of these stripes corresponds to an independent solar cell, although the layer configuration and patterning is chosen as such that the back contact of one cell is connected to the front contact of its neighbour. By this way, the desired voltage of the solar cell module can be adjusted. Thus, the processing of organic solar cell modules comprises the subsequent deposition of several patterned layers subsequently. 

Printing from the liquid state can directly yield patterned films. Techniques which yield unpatterned films are called coating methods. In principle, with coating several stripes of material can be deposited next to each other, but the resolution is too low to be able to reproducibly interconnect the neighbouring solar cells. Therefore, an additional postprocessing step for patterning is used instead, such as laser scribing, nanoimprinting, lithography or shadow masks (for evaporation).


The most commonly used deposition technique for organic solar cells is spin-coating. It is easily done and well controllable with high film uniformity. However, it is more suitable for research and development, as it is only appropriate for the production of single samples and incompatible with high volume roll-to-roll processing. Also, more than 90~\% of the material solution dropped onto the sample is lost due to the spinning of the sample at several hundred or thousand rotations per minute, which is required for the uniform film formation.

Doctor blading is a coating technique as well, but it is suitable for large area substrates, and can be integrated into a roll-to-roll production. The solution is dispensed by a blade, which is moved with up to several hundred microns distance across the substrate. The organic ink formulation satisfies the same requirements of viscosity as for the spin coating technique, and is thus directly compatible to processes published by scientific groups. Indeed, solar cells with efficiencies comparable to those processed by spin coating were presented~\cite{schilinsky2006}.

Different requirements for the organic ink formulation are needed for spray coating. As picoliter to femtoliter droplets of ink are transferred onto the substrate through tiny nozzles, highly diluted solutions are necessary to avoid clogging. Recent investigations showed that despite the film formation by individual droplets, the mean roughness for single pass spraycoating is only slightly above that of spin coated layers~\cite{girotto2009a}. This coating technique was already used to process solar cells~\cite{girotto2009a,vak2007,hoth2009} of around 3~\% efficiency. In principal, this approach allows to control the donor--acceptor ratio throughout the film: the droplets dry almost instantaneously, so that subsequent layers with slightly different mixture could be deposited, thus adjusting the vertical phase segregation~\cite{girotto2009a}. Nanoparticle-based silver top cathodes were also deposited by spray coating~\cite{girotto2009}, and an all-solution processed solar cell with 2.7~\% was made~\cite{hoth2009}.

Recently, P3HT:PCBM solar cells were processed by slot-die coating, another continuous wet coating method. In one report, power conversion efficiencies of up to 1.7~\% were obtained on a coater for 5~cm wide substrate foil~\cite{blankenburg2009}. Wetting and adhesion were improved by corona discharge on the PET foil/ITO/PEDOT:PSS substrates. The PEDOT:PSS layer could also be deposited by slot-die coating. The coater was suitable for 1-4~m/min advance,  leading to 10~m$^2$/h for this proof-of-concept investigation.
All-solution roll-to-roll processed polymer solar cells were also presented, yielding 0.3~\% efficiency~\cite{krebs2009}.

These coating techniques have potential for the production of organic solar cells on commercial scale only, if patterning is done. Although long stripes can be coated, the resolution is too low to yield the required electrical connection between neighbouring cells. Postpatterning at high throughput, for instance by laser, is already established from the inorganic photovoltaic production.


Even better suited are printing techniques, which can directly pattern during the deposition. One side effect is that the ink formulations for patterned films usually need to be more viscous to avoid that the patterns blur during solvent evaporation. This also implies that new ink formulations have to be developed when transferring a process from spin coating to printing.

In screen printing, a stencil is formed by selectively covering parts of a screen. During printing, ink is guided only through the open parts of the screen, and thus transferred as pattern onto the substrate. This technique requires high-viscosity inks. Although the throughput exceeds 1~m/s~\cite{li2007book}, only proof-of-concept solar cells with very low efficiencies could be produces yet~\cite{krebs2004}. Also, anodes were produced by screen printing~\cite{aernouts2004}. It remains open to which degree this technique can be envisioned for application in large scale processing of organic solar cells.

A very flexible method for smaller scale production is inkjet printing~\cite{singh2010review}. Ink droplets from a low viscosity solution are ejected onto the sample, and need to coalesce on the substrate to form a homogeneous film. Organic solar cells with 1.4~\%~\cite{aernouts2008} and even 3.5~\% were presented~\cite{hoth2008}. This technique has some potential for upscaling, although clogging of the print head may happen due to gelation. A throughput of 50~m$^2$/s can be achieved~\cite{huebler2007inbook}.

The printing methods going furthest into the low-viscosity regime are inket, flexography, and gravure printing. They have the highest suitability and potential regarding the production of organic solar cells. The lateral resolution of these techniques goes down to around 20~microns (flexography 40~microns), and ink viscosities in the range of 0.05-0.5~Pa~s (offset printing: 30-100~Pa~s) are ideal~\cite{huebler2007inbook}. The minimum layer thicknesses are in the range of several hundred nanometres, thus sufficiently thin for producing organic solar cells.

A print cylinder---engraved with elevated (flexography) or lowered patterns (gravure printing) to take the ink---is rotated in a trough that deposits ink on the cylinder. In the offset process, a printing plate cylinder is used to offset the image onto a compressible blank cylinder. The ink is then transferred from the print cylinder onto the substrate. Clogging, which can occur for screens and ink jet printheads, is avoided thus. 

Offset, flexography, and gravure printing are ideal for roll-to-roll processing. The substrate material, often referred to as web, needs to be flexible. It is continuously rolled up from a roll, enters the printing machines for the several layers, and is wound onto another roll---ideally as finished solar cell. Intermediate steps such as annealing or drying cycles are needed.

Organic photodiodes have been gravure printed~\cite{kopola2009}, as well as P3HT:PCBM solar cells (PEDOT and photoactive layer) with reported efficiencies of 1.68~\%~\cite{ding2009}. Offset printing has been used to print organic transistors~\cite{zielke2005} and ring oscillators~\cite{huebler2007} on flexible substrates. Although these techniques are in principal very well suited for the task of producing organic solar modules at high throughput, the actual potential has to be investigated. Indeed, dedicated efforts will be necessary to implement a full production process and suitable ink formulations for the active layer as well as the electrodes. A detailed review of the different coating and printing techniques in view of polymer solar cell processing was recently published~\cite{krebs2009review}.

\section{Summary and Outlook}

Organic solar cells have the potential to show high efficiencies at low cost, as they exhibit favourable optical and charge generation properties, and  are compatible with mass printing techniques. State-of-the-art record solar cells have almost reached 8~\% power conversion efficiency, and modules of above 200~cm$^2$ with 3.5~\% were presented. Both are based on newly synthesised organic semiconductors. 

In this review, we have briefly presented the historical development of organic solar cells. The record efficiencies stated above were made possible based on the advances of fundamental research and material development during the last two decades. The multistep process from photon absorption to charge extraction has been understood largely, although many unresolved questions remain concerning the detailed mechanisms of generation, recombination and transport of bound or separated charges. The factors limiting the performance of state-of-the-art solar cells were highlighted, such as the limited absorption width of the organic semiconductors. For almost all limitations, concepts to solve or circumvent these restrictions of the performance were discussed. For instance, in case of the absorption width, multijunction solar cells as well as optical concepts to enhance light absorption were presented. Nevertheless, the implementation of these concepts will require considerable efforts. The cost of organic solar modules can be competetive even on the roof-top. This, however, depends strongly on the advances to come in the following years in terms of production techniques as well as fundamental research to overcome stability issues and enhance the lifetime. For applications without the need for infrastructure such as inverters, low cost organic solar cells will be easier to realise. Indeed, already today organic photovoltaics have entered the market for niche products. This important step implies more efforts also in the direction of mass production compatible process optimisation and upscaling, being the prerequisites to enter the mass market.

\subsection*{Acknowledgments}

We are grateful to A. F{\"o}rtig, D. Rauh and J. Schafferhans for reading the manuscript. C.D. gratefully acknowledges the support of the Bavarian Academy of Sciences and Humanities. V.D.'s work at the ZAE Bayern is financed by the Bavarian Ministry of Economic Affairs, Infrastructure, Transport and Technology.

\end{document}